\documentclass[useAMS,usenatbib]{mn2e}
\usepackage{eufrak,graphicx,natbib}
\usepackage{epsfig}
\DeclareGraphicsExtensions{.pdf,.eps,.png,.jpg,.mps,.ps}
\newcommand{\Msun}{\mbox{$M_{\odot}$}}
\newcommand{\Lsun}{\mbox{$L_{\odot}$}}

\newcommand{\Mcl}{\mbox{$M$}}
\newcommand{\Mig}{\mbox{$M_{\rm i}^{\gamma}$}}
\newcommand{\be}{\mbox{\begin{equation}}}
\newcommand{\ee}{\mbox{\end{equation}}}
\newcommand{\Mi}{\mbox{$M_{\rm i}$}}
\newcommand{\Ni}{\mbox{$N_{\rm i}$}}
\newcommand{\tdis}{\mbox{$t_{\rm dis}$}}
\newcommand{\tzero}{\mbox{$t_0$}}
\newcommand{\tzeropostcc}{\mbox{$t_0^{\rm cc}$}}
\newcommand{\tzerocc}{\mbox{$t_0^{\rm cc}$}}
\newcommand{\tzerocctwo}{\mbox{$t_0^{\rm cc2}$}}
\newcommand{\tzerouf}{\mbox{$t_0^{\rm uf}$}}
\newcommand{\tzeroccuf}{\mbox{$t_0^{\rm cc, uf}$}}
\newcommand{\tdistot}{\mbox{$t_{\rm tot}$}}
\newcommand{\ttot}{\mbox{$t_{\rm tot}$}}

\newcommand{\tone}{\mbox{$t_{\rm 1\%}$}}

\newcommand{\jumpcc}{\mbox{$j_{\rm cc}$}}

\newcommand{\trel}{\mbox{$t_{\rm rh}$}}
\newcommand{\rh}{\mbox{$r_{\rm h}$}}

\newcommand{\muev}{\mbox{$\mu_{\rm ev}$}}
\newcommand{\muevt}{\mbox{$\mu_{\rm ev}(t)$}}

\newcommand{\mulumevt}{\mbox{$\mu^{\rm ev}_{\rm lum}(t)$}}

\newcommand{\nbody}{\mbox{$N$-body}}
\newcommand{\Nbody}{\mbox{$N$-body}}
\newcommand{\dndt}{\mbox{${\rm d}N/{\rm d}t$}}

\newcommand{\rj}{\mbox{$r_{\rm J}$}}
\newcommand{\rt}{\mbox{$r_{\rm t}$}}

\newcommand{\fdel}{\mbox{$f_{\rm delay}$}}
\newcommand{\fdelay}{\mbox{$f_{\rm delay}$}}

\newcommand{\find}{\mbox{$f_{\rm ind}$}}
\newcommand{\findmax}{\mbox{$f_{\rm ind}^{\rm max}$}}

\newcommand{\tcc}{\mbox{$t_{\rm cc}$}}
\newcommand{\tdel}{\mbox{$t_{\rm delay}$}}
\newcommand{\tdelay}{\mbox{$t_{\rm delay}$}}

\newcommand{\ndelay}{\mbox{$n_{\rm delay}$}}
\newcommand{\trh}{\mbox{$t_{\rm rh}$}}
\newcommand{\tcrh}{\mbox{$t_{\rm cr}(r_{\rm h})$}}
\newcommand{\tcrt}{\mbox{$t_{\rm cr}(r_{\rm t})$}}
\newcommand{\dNdt}{\mbox{$({\rm d}N/{\rm d}t)$}}
\newcommand{\dmdt}{\mbox{$({\rm d}M/{\rm d}t)$}}
\newcommand{\dmdtev}{\mbox{$({\rm d}M/{\rm d}t)_{\rm ev}$}}
\newcommand{\dmuevdt}{\mbox{$({\rm d} \mu/{\rm d}t)_{\rm ev}$}}

\newcommand{\dmdtind}{\mbox{$({\rm d}M/{\rm d}t)_{\rm ind}^{\rm ev}$}}
\newcommand{\dmdtdis}{\mbox{$({\rm d}M/{\rm d}t)_{\rm dis}$}}
\newcommand{\dmdtdisnorm}{\mbox{$({\rm d}M/{\rm d}t)_{\rm dis}^{\rm norm}$}}
\newcommand{\dmdteq}{\mbox{$\frac{{\rm d}M}{{\rm d}t}$}}
\newcommand{\dmdteveq}{\mbox{$\left(\frac{{\rm d}M}{{\rm d}t}\right)_{\rm ev}$}}
\newcommand{\dmdtremneq}{\mbox{$\left(\frac{{\rm d}M}{{\rm d}t}\right)_{\rm remn}$}}

\newcommand{\dmdtindeq}{\mbox{$\left(\frac{{\rm d}M}{{\rm d}t}\right)_{\rm ind}^{\rm ev}$}}
\newcommand{\dmdtdiseq}{\mbox{$\left(\frac{{\rm d}M}{{\rm d}t}\right)_{\rm dis}$}}

\newcommand{\dmuevdteq}{\mbox{$\frac{ {\rm d}\mu_{\rm ev}}{{\rm d}t}$}}

\newcommand{\mmean}{\mbox{$\overline{m}$}}

\newcommand{\mmeancc}{\mbox{$\overline{m}_{\rm cc}$}}
\newcommand{\rgal}{\mbox{$R_{\rm Gal}$}}
\newcommand{\Rgal}{\mbox{$R_{\rm Gal}$}}
\newcommand{\vgal}{\mbox{$v_{\rm Gal}$}}

\newcommand{\tmig}{\mbox{$t_0M_{\rm i}^{\gamma}$}}

\newcommand{\tnref}{\mbox{$t_{\rm ref}^N$}}
\newcommand{\tnrefcc}{\mbox{$t_{\rm ref,cc}^N$}}
\newcommand{\tnrefuf}{\mbox{$t_{\rm ref,uf}^N$}}
\newcommand{\tnrefccuf}{\mbox{$t_{\rm ref,cc,uf}^N$}}
\newcommand{\gammacc}{\mbox{$\gamma_{\rm cc}$}}
\newcommand{\gammacctwo}{\mbox{$\gamma_{\rm cc2}$}}
\newcommand{\gammauf}{\mbox{$\gamma_{\rm uf}$}}
\newcommand{\kms}{\mbox{${\rm km~s}^{-1}$}}
\newcommand{\rhoh}{\mbox{$\rho_{\rm h}$}}
\newcommand{\fkickBH}{\mbox{$f_{\rm kick}^{\rm BH}$}}
\newcommand{\fkickNS}{\mbox{$f_{\rm kick}^{\rm NS}$}}
\newcommand{\fkickWD}{\mbox{$f_{\rm kick}^{\rm WD}$}}
\newcommand{\muBHt}{\mbox{$\mu_{\rm BH}(t)$}}
\newcommand{\muNSt}{\mbox{$\mu_{\rm NS}(t)$}}
\newcommand{\muWDt}{\mbox{$\mu_{\rm WD}(t)$}}
\newcommand{\muBH}{\mbox{$\mu_{\rm BH}$}}
\newcommand{\muNS}{\mbox{$\mu_{\rm NS}$}}
\newcommand{\muWD}{\mbox{$\mu_{\rm WD}$}}

\newcommand{\mmax}{\mbox{$m_{\rm max}$}}
\newcommand{\mmin}{\mbox{$m_{\rm min}$}}

\newbox\grsign \setbox\grsign=\hbox{$>$} \newdimen\grdimen \grdimen=\ht\grsign
\newbox\simlessbox \newbox\simgreatbox \newbox\simpropbox \newbox\wtildebox 
\setbox\simgreatbox=\hbox{\raise.5ex\hbox{$>$}\llap
     {\lower.5ex\hbox{$\sim$}}}\ht1=\grdimen\dp1=0pt
\setbox\simlessbox=\hbox{\raise.5ex\hbox{$<$}\llap
     {\lower.5ex\hbox{$\sim$}}}\ht2=\grdimen\dp2=0pt
\def\simgreat{\mathrel{\copy\simgreatbox}}

\title[Mass loss and the evolution of star clusters]
{Mass loss rates and the mass evolution of star clusters}

\author[Henny J.G.L.M. Lamers, Holger Baumgardt and Mark Gieles]
{
 Henny J.G.L.M. Lamers$^{1}$\thanks{Email: lamers@astro.uu.nl},  
 Holger Baumgardt$^{2,3}$\thanks{Email: h.baumgardt@uq.edu.au} and 
 Mark Gieles$^{4,5}$\thanks{Email: mgieles@ast.cam.ac.uk}\\
 $^{1}$ Astronomical Institute, Utrecht University, 
       Princetonplein 5, NL-3584CC Utrecht, the Netherlands\\ 
 $^{2}$ Argelander Astronomical Institute, University of Bonn, Bonn, Germany\\
 $^{3}$ School of Mathematics and Physics, University of Queensland, QLD 4702, Brisbane, Australia \\
 $^{4}$ European Southern Observatory, Casilla 19001, Santiago 19, Chile\\  
 $^{5}$ Institute of Astronomy, University of Cambridge, Madingley Road, Cambridge, CB3 0HA, UK             
 }

\begin{document}

\date{Received date / accepted date}

\pagerange{\pageref{firstpage}--\pageref{lastpage}}
\pubyear{2009}

\maketitle

\begin{abstract}
We describe the interplay between stellar evolution and dynamical mass loss 
of evolving star clusters, based on the principles of stellar evolution and cluster dynamics
and on the details of a grid of $N$-body simulations
of Galactic cluster models. The cluster models have different initial masses, different orbits, including
elliptical ones, and 
different initial density profiles.
 We use two sets of cluster models: one set of Roche-lobe filling
models and a new set of cluster models that are initially underfilling
their tidal radius.
  
We identify four distinct mass loss effects: (1) mass loss by stellar evolution, (2) loss of stars 
induced by stellar evolution and (3) relaxation-driven mass loss before and (4) after core collapse.
At young ages the mass loss is dominated by stellar evolution, followed by the evolution-induced  
loss of stars. 
This evolution-induced mass loss is important if a cluster is strongly emersed in the tidal field.
Both the evolution-induced loss of stars and the relaxation-driven mass loss  need time to
build up.
 This is described by a delay-function that has a characteristic time scale
of a few crossing times for Roche-lobe filling clusters and a few half mass relaxation times for
initially Roche-lobe underfilling clusters. 
The relaxation-driven mass loss (called ``dissolution'' in this paper), 
can be described by a simple power law dependence
of the mass ${\rm d}(M/\Msun)/{\rm d}t =-(M/\Msun)^{1-\gamma}/t_0$, where $t_0$ depends on the 
orbit and environment of the cluster.
The index $\gamma$ is 0.65 for clusters with a 
King-parameter $W_0=5$ for the initial density distribution, and 0.80 for more concentrated
clusters with $W_0=7$. For initially Roche-lobe underfilling clusters the dissolution is described by
the same $\gamma=0.80$, independent of the initial density distribution.
The values of the constant $t_0$ are derived for the models and described by simple formulae
that depend on the orbit of the cluster.
The mass loss rate increases by about a factor two at core collapse and the 
mass dependence of the relaxation driven mass loss changes to $\gamma=0.70$ after core collapse.

We also present a simple recipe for predicting the mass evolution of individual star clusters with various
metallicities and in different environments, with an accuracy of a few percent in most cases.
This can be used to predict the mass evolution of cluster systems.

\end{abstract}



\section{Introduction}
\label{sec:1}

In this paper we study the mass loss and the mass history of star clusters
in a tidal field, based on a grid of $N$-body simulations.
The purpose of this study is three-fold: (a) to understand and quantitatively
describe the different effects that are responsible for mass loss, (b) to study the
interplay between the different mass loss mechanisms and (c) to develop a method
for predicting the mass history of individual clusters of different initial conditions 
and in different environments. This information is needed if one wants to analyse 
observed star cluster systems in different galaxies. 
 Therefore we describe in this paper in detail the different mass loss effects;
how much each one contributes to the mass loss rate; how it depends on cluster parameters
and environment and how these effects determine the mass history and total lifetime of a cluster.
In particular we will point out the importance of the loss of stars that is induced by stellar
evolution. This mass loss is proportional to the evolutionary mass loss and therefore
adds to the mass loss rates of clusters at young ages. 
We will show that both the evolution-induced mass loss and the relaxation-driven mass loss 
start slowly with a delay time on the order of a few crossing times at the tidal radius.
We will also show that 
the mass loss rate after core collapse is about a factor two higher than before core collapse,
and that the dependence of the relaxation-driven  mass loss on mass is different 
before and after core collapse.

The mass of star clusters decreases during their lifetime, until they are 
finally completely dissolved. The stars that are lost from clusters add to the
population of field stars. The mass loss is due to stellar evolution and to 
several dynamical effects such as two-body relaxation, tidal stripping and shocks. 
These effects have all been extensively studied individually. However, 
to understand and describe the combination and interplay of the effects one has to 
rely on dynamical simulations.

The effects of stellar evolution on star clusters can be studied 
by means of stellar evolution tracks for a large range of masses and metallicities
(e.g. Anders \& Fritze-v. Alvensleben 2003; Bruzual \& Charlot 2003; Fioc \& Rocca-Volmerange
1997; Leitherer et al. 1999; Maraston 2005).
The dynamical effects of cluster evolution have been described in a large number of
theoretical studies starting with Ambartsumian (1938), Spitzer (1940), Chandrasekhar (1943) 
and a series of papers by King, e.g. King (1958). 
This was followed by the seminal works of Spitzer (1958 and 1987) and by many other studies, e.g.
Chernoff \& Weinberg (1990); Gnedin \& Ostriker (1997);  Aarseth (1999); Fukushige \& Heggie (2000).
The first $N$-body simulations of clusters were done by von Hoerner (1960). For a review of early
\Nbody\ simulations of star clusters, see Aarseth \& Lecar (1973).

The recent advancement of computational power, in particular the development of the 
$GRAPE$-computers (Makino et al. 2003) 
and the use of Graphics Processing Units (GPU)
 has allowed the improvement and verification of these 
theoretical models by means of direct $N$-body simulations 
(Vesperini \& Heggie 1997; Portegies Zwart et al. 1998;
Baumgardt \& Makino 2003, hereafter BM03; Gieles \& Baumgardt 2008).
For the purpose of the present study
the following results are particularly important: \\
(i) The realization that mass loss by tidal effects 
does not only scale with the half-mass relaxation time (as was assumed in earlier studies), 
but by
a combination of the half-mass relaxation time and the crossing time (Fukushige \& Heggie 2000; 
Baumgardt 2001; BM03). 
This implies that the lifetime due to evaporation in a tidal field does not scale
linearly with the cluster mass $M$, but with $M^{\gamma}$ with $\gamma \simeq 0.6$ to 0.7. \\
(ii) The realization that mass loss by shocks due to the passage of spiral arms and giant
molecular clouds scales with the density $M/r^3$ of the clusters 
  (Spitzer 1958, Gnedin \& Ostriker 1997).
Adopting the observed mean mass-radius relation of $r \propto M^{0.1}$ for clusters 
in spiral galaxies
(Larsen 2004, Scheepmaker et al. 2007) then also results in a mass loss rate that 
scales approximately as $M^{\gamma}$, with $\gamma$ similar to the value of evaporation in 
a tidal field (Gieles et al. 2006, 2007). This is in agreement with empirical determinations of 
$\gamma \simeq 0.6$ from studies of cluster samples in different galaxies (Boutloukos \& Lamers 2003;
Gieles et al. 2005; Gieles 2009).\\
(iii) A grid of cluster evolution models with different initial masses,
different initial concentration factors and in different Galactic orbits by means of $N$-body
simulations (BM03) allows a study of the interplay between stellar evolution and 
dynamical mass loss, that is not easily done by theoretical studies. 
In particular it shows how the 
mass loss depends on mass, age and external conditions, and how the stellar mass function
evolves during the life of the cluster.
   
In this paper we will use a grid of \nbody\ simulations of Roche-lobe filling models  (BM03),
supplemented with a new grid for Roche-lobe underfilling models, of Galactic clusters of different
initial mass, different initial concentrations and in different orbits to describe the process of 
mass loss from clusters and the interplay between the different effects. We also
derive a method for calculating the mass loss and mass history for clusters of different metallicity 
and in different environments. 
This results in an improvement of the analytical description of the
mass history of clusters that was based on a combination of stellar evolution and dynamical effects.
(Lamers et al. 2005).

The paper is arranged as follows.
In Sect. 2 we describe the mass loss processes of star clusters: stellar evolution and
dynamical effects.
In Sect. 3 we describe the results of N-body simulations of BM03 used in this study.
Sect. 4 deals with the mass loss due to stellar evolution, i.e. both the direct mass loss
and the evolution-induced loss of stars.
In Sections 5 and 6 we describe the relaxation-driven mass loss 
respectively before and after core collapse.
Section 7 deals with the mass evolution 
of clusters in elliptical orbits around the galaxy and Sect. 8 deals with initially Roche-lobe
underfilling clusters. In Sect. 9 we study the relation between the total age of a cluster
and the initial parameters.
In Sect. 10 we predict the mass loss history of clusters and its main contributions.
Sections 11 and 12 contain a discussion and the summary plus conclusions of this study.
In Appendix A we present a recipe to predict the mass history of star clusters in
different environments and with different metallicities. In Appendix B we tabulate 
numerical coefficients to calculate the mass loss of clusters by stellar evolution.


\section{Mass loss processes}
\label{sec:2}

Clusters lose mass by stellar evolution and by dynamical effects, such as two-body relaxation
and tidal stripping of stars in a cluster that is emersed in a steady tidal field and shocks. 
The mass loss by stellar evolution
is in the form of gas ejected by stellar winds and by supernovae, but also in the form of 
compact remnants that may be ejected if they get a kick velocity at birth. Mass loss by dynamical effects
is always in the form of stars. Throughout this paper we will refer to these two effects 
respectively as
``mass loss by stellar evolution'' and ``dissolution'', either in a steady potential field 
or due to tidal perturbation (shocks).

\subsection{Mass loss by stellar evolution}  
\label{sec:2.1}

The mass fraction that is lost by stellar evolution depends on the metallicity and
 on the adopted stellar initial mass function. We have calculated these for 
clusters with a Kroupa IMF, using the evolutionary calculations
of Hurley et al. (2000) by assuming no dynamical mass loss. The data are
provided by Pols (2007, Private Communication).

The various contributions to the evolutionary mass loss for (non dissolving) 
clusters with metallicities of
Z=0.0004, 0.001, 0.004, 0.008 and 0.02 can be expressed with very high accuracy  (better than $\sim$ 1\%)
by 3rd order polynomials as function of time. We have calculated these fit formulae for
models with a Kroupa (2001) IMF in the range of 0.1 to 100 \Msun.
These models have an initial mean stellar mass of 0.638 \Msun.
The fit formulae for clusters are listed in Appendix B for the following parameters:\\
(i) the remaining mass fraction $\mu(t)= M(t)/\Mi$,\\
(ii) the mass fractions of black holes $\mu_{\rm BH}=M_{\rm BH}/\Mi$, 
neutron stars $\mu_{\rm NS}$ and white dwarfs $\mu_{\rm WD}$,\\
(iii) the mean mass of all stars $<m>$ and of the black holes $<m>_{\rm BH}$, neutron stars, $<m>_{\rm BNS}$ 
and white dwarf $<m>_{\rm WD}$, \\
(iv) the luminosity $L(t)/\Lsun$ of a cluster with an initial mass of 1 \Msun.

The mass fraction that is lost by winds and supernova ejecta is $1-\mu(t)$.
If compact remnants are ejected with a kick velocity then the 
remaining mass fraction due to stellar evolution  is 

\begin{equation}
\muevt \equiv \mu(t)- \fkickBH \muBHt - \fkickNS \muNSt - \fkickWD \muWDt
\label{eq:muevt}
\end{equation}
where \fkickBH, \fkickNS\ and \fkickWD\ are the fractions of these stellar remnants that
are ejected out of the cluster by their kick velocity. If all BHs are kicked out then $\fkickBH=1$
and if all WDs are retained then $\fkickWD=0$.
The fraction of the {\it luminous mass} that is left by stellar evolution is

\begin{equation}
\mulumevt = \mu(t) - \muBHt - \muNSt - \muWDt. 
\label{eq:mulumevt}
\end{equation}
All fraction $\mu$ are expressed relative to the {\it initial} cluster mass \Mi.

 The mass loss rate of a cluster due to stellar evolution can now be expressed as

\begin{equation}
\left(\frac{{\rm d}\Mcl}{{\rm d}t}\right)_{\rm ev} = M_{\rm lum}(t) \cdot \frac{{\rm d}\muevt}{{\rm d}t}
\label{eq:dmdtev}
\end{equation}
which is negative, since $\muev$ decreases with time.
This expression is strictly valid for the early phases of cluster lifetime before the preferential 
loss of low mass stars by dynamical effects has changed the shape of the mass function. 
Since stellar evolution
dominates the mass loss only in the early phase of the clusters lifetime, equation
\ref{eq:dmdtev} is a good approximation. 
(For a description of evolutionary mass loss in a cluster with preferential loss of low mass stars
see Kruijssen \& Lamers 2008, Kruijssen 2009 and Trenti et al. 2010.)

In the description of the mass loss of the cluster models studied by \nbody\ simulations
in this paper the effect of the changing  mass function
due to evolution and the preferential loss of low mass stars is properly taken into account, 
as it is in the output of the simulations.

\subsection{Mass loss by dynamical effects or ``dissolution''}  
\label{sec:2.2}

The time-dependent mass loss by dissolution can be described by 
$\dmdtdis = - M/t_{\rm dis}$, where {\it $t_{\rm dis}\equiv ({\rm d} \ln (M)/{\rm d} t)^{-1}$ is the dissolution 
time scale} that depends on the actual cluster mass and on the environment of the cluster.
Let us assume that we can describe $t_{\rm dis}$ as a power-law function of mass, 
as $t_{\rm dis} = t_0 (M/\Msun)^{\gamma}$, with the constant  
$t_0$ being the {\it dissolution parameter} (which is the hypothetical dissolution time scale
of a cluster of 1 \Msun).
The changes in the cluster mass due to dissolution is then described by\footnote{
Throughout the rest of this paper all masses \Mi, $M$ and $m$ are in units of \Msun\ and all ages are in Myrs.}

\begin{equation}
\left(\frac{{\rm d}M}{{\rm d}t}\right)_{\rm dis} = - \frac{M(t)}{t_{\rm dis}(t)} = - \frac{M(t)^{1-\gamma}}{t_0}
\label{eq:dmdtdis}
\end{equation}
We stress that the dissolution time scale \tdis\  is not the same
as the {\it total life time of the cluster}, \tdistot, although these are related.  
Integration of Eq. \ref{eq:dmdtdis} shows that
$\tdistot = \tmig / \gamma$ in the absence of stellar evolution, 
where $\Mi$ is the initial mass. 
In reality stellar evolution also removes part of the cluster mass. 
This implies that this simple estimate of \tdistot\ 
overestimates the real cluster lifetime.

Theoretical considerations suggest that $\gamma \simeq 0.65$ to 0.85.
This follows from the following dynamical arguments.

\subsubsection{Dissolution  in a steady tidal field}   
\label{sec:2.2.1}

Spitzer (1987)  has argued that a fraction of the stars $\xi$ escapes each $\trel$, such that

\begin{equation}
\left(\frac{{\rm d}N}{{\rm d}t}\right)_{\rm dis} = -\frac{\xi N}{\trel}
\label{eq:dmdtspitzer}
\end{equation}
where  $\trel \propto (N^{1/2}/ \ln \Lambda) \rh^{3/2}~ \mmean ^{-1/2}$
is the half-mass relaxation time, \rh\ is the half-mass radius, $\mmean $ is the mean stellar mass,
$N=M/\mmean $ is the number of stars and $\ln \Lambda$ is the Coulomb logarithm.
The value of $\xi$ is larger for Roche-lobe filling clusters than for 
clusters in isolation. 

In analytical studies of cluster 
dissolution a single value for $\xi$ is usually assumed, implying 
that the cluster lifetime is a constant times $\trel$ (see e.g. Spitzer 1987).  
However, a recent theoretical study
by Lee (2002) and Gieles \& Baumgardt (2008) has shown that $\xi$ is a strong 
function of the Roche-lobe filling factor $\rh/\rj$, (where $\rj$ is the 
Jacobi radius, i.e. the Roche-lobe radius for clusters). 
These authors found  
for clusters with $\rh/\rj \geq  0.05$ that $\xi$ scales roughly as $(\rh/\rj)^{3/2}$.
Since $\rj\propto M^{1/3}\omega^{-2/3}$, where $\omega$ is the angular frequency of the cluster
 orbit, $\xi\propto t_{\rm cross}\times \omega$. 
Here $t_{\rm cross}$ is the mean crossing time of stars in a cluster:
$t_{\rm cross} \propto \rh^{3/2}/ \sqrt{GM}. $
 This dependence of $\xi$ on $\rh^{3/2}$ cancels the $\rh^{3/2}$ 
dependence of $\trel$ such that the radius becomes an unimportant 
parameter in \dndt. This can be understood intuitively as follows: 
for a smaller (larger) radius, relaxation becomes more (less) important, 
while the escape criterion due to the tidal field becomes less (more) important.

On top of this, Baumgardt (2001) and BM03  showed that the dissolution
time scale does not scale linearly with $\trel$, but rather with 
 a combination of $\trel$ and
the crossing time, $t_{\rm cross}$, because even unbound stars need time to
 leave a cluster (see Fukushige \& Heggie 2000 for details).
 
The relevant time scale is 

\begin{eqnarray}
t_{\rm dyn} &\propto& \trel^{x} t_{\rm cross}^{1-x}\\
	           &\propto& \left(\frac{N}{\ln\Lambda}\right)^x\,t_{\rm cross}
\label{eq:treltcr}
\end{eqnarray}  
This implies that {dissolution time scale in terms of number of stars is}

\begin{eqnarray}
\tdis^N~&\equiv& ~ -\frac{N}{{\rm d}N/{\rm d}t}~ = ~\frac{t_{\rm dyn}}{\xi} \\
	          &\propto& \left(\frac{N}{\ln\Lambda}\right)^x\,\omega^{-1}
\label{eq:tdisN}
\end{eqnarray} 
where $\tdis^N$ is the dissolution time scale if dissolution is expressed in terms of 
$N$ instead of $M$.
In the range of $ 10^4 < N < 10^6$ we can approximate $\Lambda\simeq 0.02 N$ 
(Giersz \& Heggie 1994) and $N/\ln \Lambda \propto N^{0.80}$ and so $\tdis ^N \propto N^{p}$
with $p =0.80 x$. \footnote{
Using the total lifetime as an indicator of the dynamical time BM03 found that
$x=0.75$ for Roche-lobe filling models with an initial concentration factor of the density King-profile 
$W_0=5$ and
$x=0.82$ for the more centrally concentrated $W_0=7$ models. This would imply $p \simeq 0.60$
and 0.66 for Roche-lobe filling models of  $W_0=5$ and 7 respectively.}

In this paper we describe the dissolution time as a function of $M$ instead of $N$.
Comparing the two expressions $\tdis \propto M^\gamma$ and  $\tdis ^N \propto N^p $ we see that
$\gamma \ne p$ if the mean stellar mass \mmean\ changes during the clusters lifetime.
If \mmean\ decreases as function of time, i.e. increases as function of $M(t)$, then $\gamma<p$,
whereas $\gamma > p$ if \mmean\ increases with time. We will see below that after an 
initial phase, dominated by stellar evolution, \mmean\ increases with time due to the preferential
loss of low mass star by tidal stripping. So the values of $\gamma$ are expected to be slightly
higher than the values of $p$.

\subsubsection{Dissolution due to shocks}  
\label{sec:2.2.2}

Clusters can also be destroyed by shocks (e.g. Ostriker et al. 1972, Spitzer 1987, 
Chernoff \& Weinberg 1990, Gnedin \& Ostriker 1997)
due to encounters with spiral arms or giant molecular clouds in the disk of a galaxy,
by disk shocking for clusters in orbits inclined with respect to the Galactic plane
and by bulge shocking for clusters in highly elliptical orbits. 

The mass loss rate due to shocks depends on the cluster half-mass radius, \rh, as
$\dmdt \propto M/\rhoh\propto \rh^3$, with $\rhoh$ the density within \rh. 
This implies that the mass loss time scale depends on the cluster properties as
$\tdis \equiv  -M/\dmdt \propto M/\rh^3$, which is proportional to the cluster density. 
The observed mean mass-radius relation of clusters is not well defined, but 
Larsen (2004) and  Scheepmaker et al. (2007) find a mean relation of 
$\rh \propto M^{\lambda} $ with $\lambda \simeq 0.13$.
So the time scale for mass loss due to shocks is $\tdis \propto M^{0.61}$ for a 
constant mean stellar mass (Gieles et al. 2006, 2007).
This dependence is almost the same as that for tidal dissolution.  

\subsubsection{The expected values of $\tzero$ and $\gamma$}
\label{sec:2.2.3}

Based on the arguments of the previous subsections we expect that the combined mass loss by tidal 
dissolution and shocks can be described by a function of the form

\begin{equation}
\left( \frac{{\rm d}N}{{\rm d}t}\right)_{\rm dis} = -\frac{N}{\tdis^N} = -\frac{N^{1-p}}{\tzero^N} ~~~{\rm or} ~~~
\left( \frac{{\rm d}M}{{\rm d}t}\right)_{\rm dis} = - \frac{M^{1-\gamma}}{t_0} 
\label{eq:dndtdmdt}
\end{equation}
with $p \simeq 0.65$ and $\gamma \simgreat p$. We will use the second expression
with the value of $\gamma$ derived from the \Nbody\ simulations of BM03.

In an environment where cluster  
dissolution is only due to stellar evolution, internal dynamical effects
and tidal stripping, \tzero\ depends on the potential field in which the
cluster moves. If clusters move in elliptical orbits in a logarithmic potential field, i.e. 
a constant galactic rotation velocity $v_{\rm Gal}$, the dissolution time is reduced by
a factor $1-\epsilon$, compared to a circular orbit at $R_A$,
where the eccentricity $\epsilon = (R_A-R_P)/(R_A+R_P)$
and $R_A$ and $R_P$ are the apogalactic and perigalactic distances respectively.  
This implies that we expect the value of $\tzero$ to vary as

\begin{equation}
\tzero = \tnref \times \left(\frac{1-\epsilon}{\mmean^{\gamma}}\right) \left( \frac{\rgal}{8.5 {\rm kpc}}\right)
  \left(\frac{\vgal}{220 {\rm \kms}}\right)^{-1} 
\label{eq:tnrefdef}
\end{equation}
where \tnref\ is a constant, whose value will be derived from the \nbody\ simulations of BM03.

The factors $\Rgal/8.5{\rm kpc}$ and $\vgal / 220\, {\rm km~s}^{-1}$
provide a scaling for calculating the dissolution of clusters at different galactocentric distances 
in other galaxies  with constant rotation velocity.
  The factor $1/ \mmean^\gamma$ is a result of the conversion
of $N$ to $M$ if the mean stellar mass \mmean\ is about constant.
We will see below that this yields a very good description
of the mass loss rate and $M(t)$ for all models of BM03, if \mmean\ is chosen 
appropriately.\footnote
{If other processes, such as encounters with GMCs or spiral density waves, are important, 
  then the value of \tzero\
will be smaller than predicted by Eq. \ref{eq:tnrefdef}. Wielen (1985) and Lamers \& Gieles (2006) found that
clusters in the solar neighbourhood are mainly destroyed by encounters with GMCs which reduces 
\tzero\ by a factor 4 compared to Eq. \ref{eq:tnrefdef}.  Kruijssen \& Mieske (2009) have derived the values of
\tzero\ for a number of galactic globular clusters in elliptical orbits, assuming $\gamma=0.70$.
}

\section{The models of Roche-lobe filling clusters}   
\label{sec:3}

\begin{table*}
\caption{The N-body models of BM03 used in this study}
\centering

\resizebox{\textwidth}{!}{

\begin{tabular}{r r r r r r r r | r  r  r r r | r r r r r r r}
\multicolumn{8}{l}{Input parameters} & \multicolumn{5}{l}{Timescales} & \multicolumn{7}{l}{Output parameters} \\
 $\#$  & Mass      & nr & $W_0$ & $R_{\rm Gal}$ & Orbit & $\rt$ & $\rh$ & $t_{\rm 1\%}$ & $t_{\rm cc}$ & $t_{\rm rh}$ & $t_{\rm cr}(r_{\rm h})$ & $t_{\rm cr}(r_{\rm t})$ &
 $\gamma$ &  $t_0$ & $\tmig$ & $t_{\rm del}$ & $f_{\rm ind}^{\rm max}$  & $\tzeropostcc$ & $j_{\rm cc}$     \\ 
     & $M_{\odot}$ & stars &    & kpc        &      & pc &  pc       & Gyr       & Gyr     & Gyr        & Myr
     & Myr  &     &  Myr  & Gyr     &  Myr       &    &  Myr  &    \\ \hline

  1 & 71952 & 128k & 5&  15& circ & 89.6 &16.75 & 45.3 & 36.42 & 7.20 &15.24 & 133 & 0.65 & 40.0 & 57.4 &  400 & 1.1 & 16.1 & 1.60 \\ 
  2 & 35915 &  64k & 5&  15& circ & 71.0 &13.28 & 26.9 & 22.51 & 3.88 &15.24 & 133 & 0.65 & 42.0 & 38.4 &  400 & 1.0 & 17.2 & 1.61 \\ 
  3 & 18205 &  32k & 5&  15& circ & 56.7 &10.59 & 19.8 & 14.24 & 2.13 &15.24 & 133 & 0.65 & 41.2 & 24.2 &  400 & 1.0 & 20.0 & 1.40 \\ 
  4 &  8808 &  16k & 5&  15& circ & 44.5 & 8.32 & 13.4 &  8.72 & 1.13 &15.24 & 133 & 0.65 & 37.9 & 13.7 &  400 & 0.7 & 20.4 & 1.29 \\
  5 &  4489 &   8k & 5&  15& circ & 35.5 & 6.64 &  9.0 &  5.92 & 0.63 &15.24 & 133 & 0.65 & 36.0 &  8.5 &  400 & 0.7 & 19.0 & 1.37 \\ \hline

  6 & 71236 & 128k & 5& 8.5& circ & 61.1 &11.43 & 26.5 & 21.34 & 4.05 & 8.63 &  76 & 0.65 & 21.5 & 32.1 &  228 & 0.8 &  8.5 & 1.70 \\ 
  7 & 36334 &  64k & 5& 8.5& circ & 48.8 & 9.13 & 17.2 & 13.91 & 2.22 & 8.63 &  76 & 0.65 & 22.0 & 20.3 &  228 & 0.8 & 10.4 & 1.40 \\ 
  8 & 18408 &  32k & 5& 8.5& circ & 39.0 & 7.28 & 11.1 &  8.41 & 1.22 & 8.63 &  76 & 0.65 & 21.7 & 12.8 &  228 & 0.8 &  9.9 & 1.50 \\
  9 &  9003 &  16k & 5& 8.5& circ & 30.7 & 5.74 &  7.5 &  5.06 & 0.65 & 8.63 &  76 & 0.65 & 20.5 &  7.4 &  228 & 0.7 & 10.7 & 1.34 \\ 
 10 &  4497 &   8k & 5& 8.5& circ & 24.3 & 4.55 &  4.9 &  3.30 & 0.36 & 8.63 &  76 & 0.65 & 20.0 &  4.7 &  228 & 0.8 & 10.2 & 1.42 \\ \hline 

 11 & 71218 & 128k & 5& 2.8& circ & 29.4 & 5.50 &  9.3 &  7.66 & 1.35 & 2.87 &  25 & 0.65 &  7.5 & 10.7 &   75 & 0.7 &  3.0 & 1.62 \\ 
 12 & 35863 &  64k & 5& 2.8& circ & 23.4 & 4.37 &  5.9 &  4.63 & 0.73 & 2.87 &  25 & 0.65 &  6.7 &  5.8 &   75 & 0.7 &  3.3 & 1.36 \\ 
 13 & 18274 &  32k & 5& 2.8& circ & 18.7 & 3.49 &  3.6 &  2.85 & 0.40 & 2.87 &  25 & 0.65 &  6.0 &  3.5 &   75 & 0.6 &  3.3 & 1.26 \\ 
 14 &  9024 &  16k & 5& 2.8& circ & 14.8 & 2.76 &  2.3 &  1.58 & 0.22 & 2.87 &  25 & 0.65 &  5.3 &  2.0 &   75 & 0.4 &  3.2 & 1.16 \\ 
 15 &  4442 &   8k & 5& 2.8& circ & 11.7 & 2.18 &  1.3 &  0.85 & 0.12 & 2.87 &  25 & 0.65 &  4.4 &  1.0 &   75 & 0.4 &  2.8 & 1.13 \\ \hline
 
 16 & 71699 & 128k & 7& 8.5& circ & 61.3 & 7.11 & 28.5 & 12.62 & 1.99 & 4.22 &  76 & 0.80 &  6.4 & 46.1 &  228 & 0.8 & 11.5 & 1.52 \\
 17 & 35611 &  64k & 7& 8.5& circ & 48.5 & 5.63 & 17.2 &  7.87 & 1.07 & 4.22 &  76 & 0.80 &  6.5 & 28.2 &  228 & 0.8 & 10.5 & 1.57 \\ 
 18 & 18013 &  32k & 7& 8.5& circ & 38.7 & 4.48 & 11.2 &  4.87 & 0.58 & 4.22 &  76 & 0.80 &  6.5 & 15.7 &  228 & 0.8 & 10.5 & 1.43 \\
 19 &  8928 &  16k & 7& 8.5& circ & 30.6 & 3.55 &  6.9 &  2.89 & 0.32 & 4.22 &  76 & 0.80 &  6.0 &  8.5 &  228 & 0.8 & 11.0 & 1.21 \\ 
 20 &  4402 &   8k & 7& 8.5& circ & 24.2 & 2.80 &  4.4 &  1.67 & 0.17 & 4.22 &  76 & 0.80 &  5.5 &  4.3 &  228 & 0.8 & 10.1 & 1.14 \\ \hline

 21 & 17981 &  32k & 5& 8.5& e0.2 & 29.5 & 5.51 &  9.0 &  6.34 & 0.80 & 5.76 &  50 & 0.65 & 14.5 &  9.0 &  150 & 0.2 &  8.4 & 1.24: \\ 
 22 & 18300 &  32k & 5& 8.5& e0.3 & 25.7 & 4.81 &  7.8 &  5.21 & 0.65 & 4.65 &  41 & 0.65 & 12.0 &  7.1 &  120 & 0.1 &  7.9 & 1.02: \\
 23 & 17966 &  32k & 5& 8.5& e0.5 & 18.6 & 3.47 &  5.7 &  3.61 & 0.40 & 2.88 &  25 & 0.65 &  8.8 &  5.3 &   75 & 0.0 &  5.0 & 1.23: \\ 
 24 & 17957 &  32k & 5& 8.5& e0.7 & 12.2 & 2.27 &  3.6 &  2.09 & 0.21 & 1.52 &  13 & 0.65 &  5.9 &  3.4 &   39 & 0.0 &  3.0 & 1.29: \\ 
 25 & 18026 &  32k & 5& 8.5& e0.8 &  8.9 & 1.67 &  2.8 &  1.46 & 0.13 & 0.96 &   8 & 0.65 &  4.5 &  2.6 &   24 & 0.0 &  2.3 & 1.28: \\ \hline
\end{tabular}
}

Left section: model parameters; Middle section: cluster time scales; Right section: fit parameters.\\
First three blocks: clusters with $W_0=5$ in circular orbits at $R_{\rm Gal}$= 15, 8.5 and 2.8 kpc.
Fourth block: clusters with $W_0=7$ in circular orbits at $R_{\rm Gal}$=8.5 kpc.
Fifth block: clusters with $W_0=5$ in elliptical orbits with apogalactic distance 
of $R_A$= 8.5 kpc and a perigalactic distance of $R_A(1-\epsilon)/(1+\epsilon)$. 
The values of $r_{\rm h}$ and $r_{\rm t}$ apply to the perigalacticon. 
Clusters with $W_0=5$ or 7 have $\gamma=0.65$ and 0.80 respectively, before core collapse.
The number of stars is given in units of $1k=1024$.

\label{tbl:BM03models}
\end{table*}

In this paper we study the results of $N$-body simulations,
in order to understand the way clusters evolve due to dynamical and evolutionary effects.
We use the models of Roche-lobe filling clusters from BM03 for comparison 
with our predictions, supplemented with a few models of initially Roche-lobe underfilling clusters.
Out of the initial 33 models we have selected 25 representative BM03 cluster models.
These are chosen because they allow the study of the effects of the
different parameters of the models, i.e. initial mass and initial concentration, and of the cluster 
orbits, i.e. Galactocentric distance and eccentricity. (We found that the information
derived from these models also applies to the other 8 models.)
The models  span a range of
dissolution times between 1.5 and 50 Gyr. The selected models are listed in 
Table \ref{tbl:BM03models}.

\subsection{Parameters of the cluster models}
\label{sec:3.1}

The models are divided into five blocks, separated by horizontal lines in Table 
\ref{tbl:BM03models}.
The first three blocks contain models of Roche-lobe filling clusters of different masses with an initial
concentration factor $W_0=5$, in circular orbits at galactocentric distances of 
$\Rgal=15$, 8.5 and 2.83 kpc.
The fourth block contains Roche-lobe filling cluster models in circular orbits and
with different initial masses, 
but with a more concentrated initial density distribution, with a King profile
of $W_0=7$. These will be used to study the effect of the initial concentration
on the cluster evolution, by comparing them with the results of the $W_0=5$ models.
The fifth block contains Roche-lobe filling cluster models with $W_0=5$  
in elliptical orbits with various eccentricities.
All models have a Kroupa stellar initial mass function (IMF), with a mass range of 0.15 to  15 \Msun,
an initial mean stellar mass of 0.547 \Msun\ and a metallicity of $Z=0.001$.
The clusters have no initial binaries, but binaries do form during the dynamical evolution, mainly
in the high density central region during core collapse.
Neutron stars and white dwarfs are retained in the cluster when they are formed (no kick velocities)
but may be lost later by dynamical effects. Black holes are not considered in the BM03
models, because of the adopted upper mass limit of 15 \Msun.  

The clusters in elliptical orbits (nrs 21 to 25)  have about the same initial mass and the same 
apogalactic distance of $R_A=8.5$ kpc, but different elliptical
orbits with eccentricities  $0.2 \le \epsilon \le 0.8$.
This implies perigalactic distances of
$R_P=R_A (1-\epsilon)/(1+\epsilon)$ between $0.667\,R_A$ and $0.111\,R_A$.
For these models the mass loss rates are strongly variable with time: at
perigalacticon the rates are much higher than at apogalacticon. 
The initial values of the tidal radius, half-mass radius etc. in Table \ref{tbl:BM03models} refer to
the values at perigalacticon.

The data of each model in Table \ref{tbl:BM03models} are given in three groups, separated by a vertical line. 
The left group gives the initial model data:
model nr, initial mass, initial nr of stars,  
$W_0$, apogalactic distance, 
type of orbit with eccentricity, tidal radius \rt\ and half mass radius \rh .
The middle group gives the various time scales of the models:
the time $\tone$ when $M(t)=0.01 \Mi$,
the core-collapse time $t_{\rm cc}$, the initial half mass
relaxation time \trh, the half mass crossing time \tcrh\ and the 
initial crossing time at the tidal radius \tcrt.

The right hand group gives the data that describe the mass loss rates of the models:
 \tzero, $\tzero \Mig$ (which is a proxy of the 
expected total life time).
The values of delay-time \tdel\ and \findmax\
together describe how the clusters react dynamically to mass loss by stellar evolution (see Sect. \ref{sec:6}).
The last two columns give the values of $\tzeropostcc$, which describes the mass loss rate after core collapse,
and \jumpcc\, which describes the increase in mass loss due to core collapse.
The determination of these parameters is described below.

\subsection{The mass loss rates of the cluster models} 
\label{sec:3.2}

BM03 define a star to be lost from a cluster if it is outside the Jacobi radius $\rj$
of the cluster. Stars with a velocity $v>v_{\rm esc}$ but $r<\rj$ are still considered
 cluster members. On the other hand, the mass lost by stellar evolution (i.e. by 
winds and supernovae) is assumed to leave the cluster immediately.

We have derived the mass loss rates of the N-body models of BM03.
From the output of these model calculations we can separate the mass loss that is due to 
stellar evolution from the contribution by dynamical effects.
{\it There is an important difference between these two mass loss rates. Mass loss by 
stellar evolution is instantaneous and independent of the structure and orbit of the 
cluster. On the other hand, mass loss by dynamical effects always proceeds on a slow time scale
and needs time to build up.} We will see this in the results.

\subsection{Three phases of mass loss}
\label{sec:3.3}

\begin{figure*}
\centerline{\psfig{figure=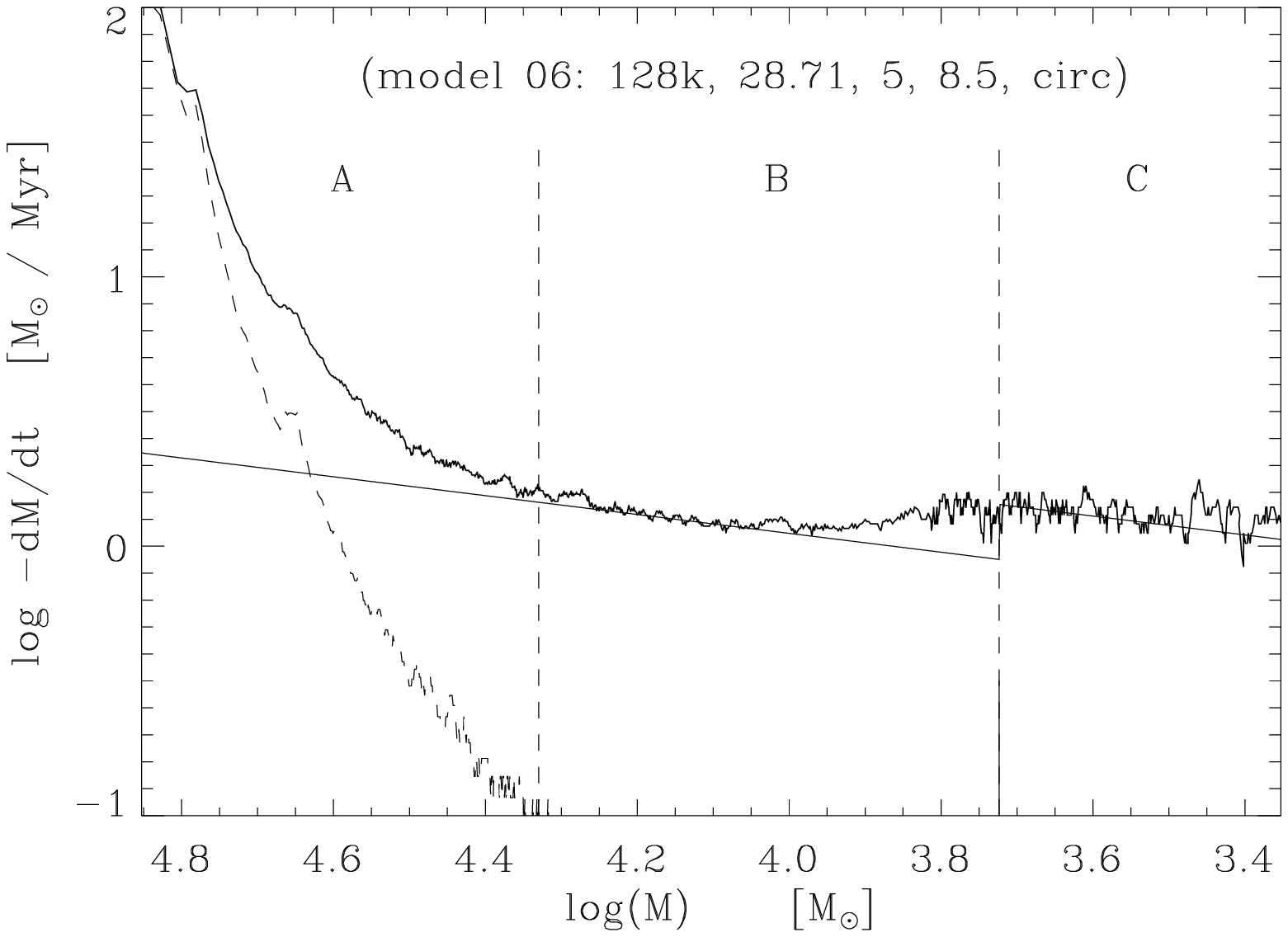,width=6.0cm}\hspace{-0.3cm}
            \psfig{figure=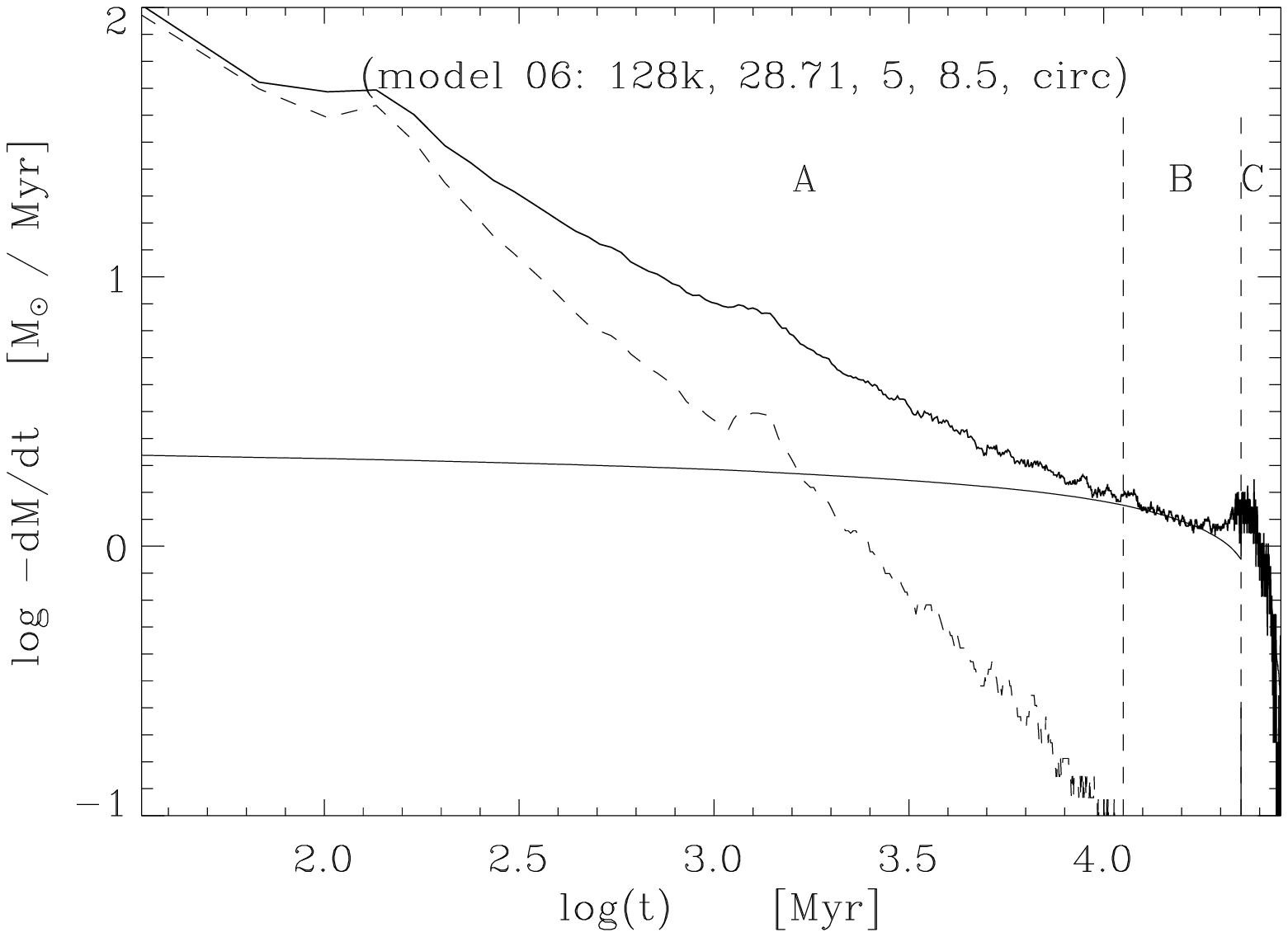,width=6.0cm}}
\vspace{-0.3cm}
\centerline{\psfig{figure=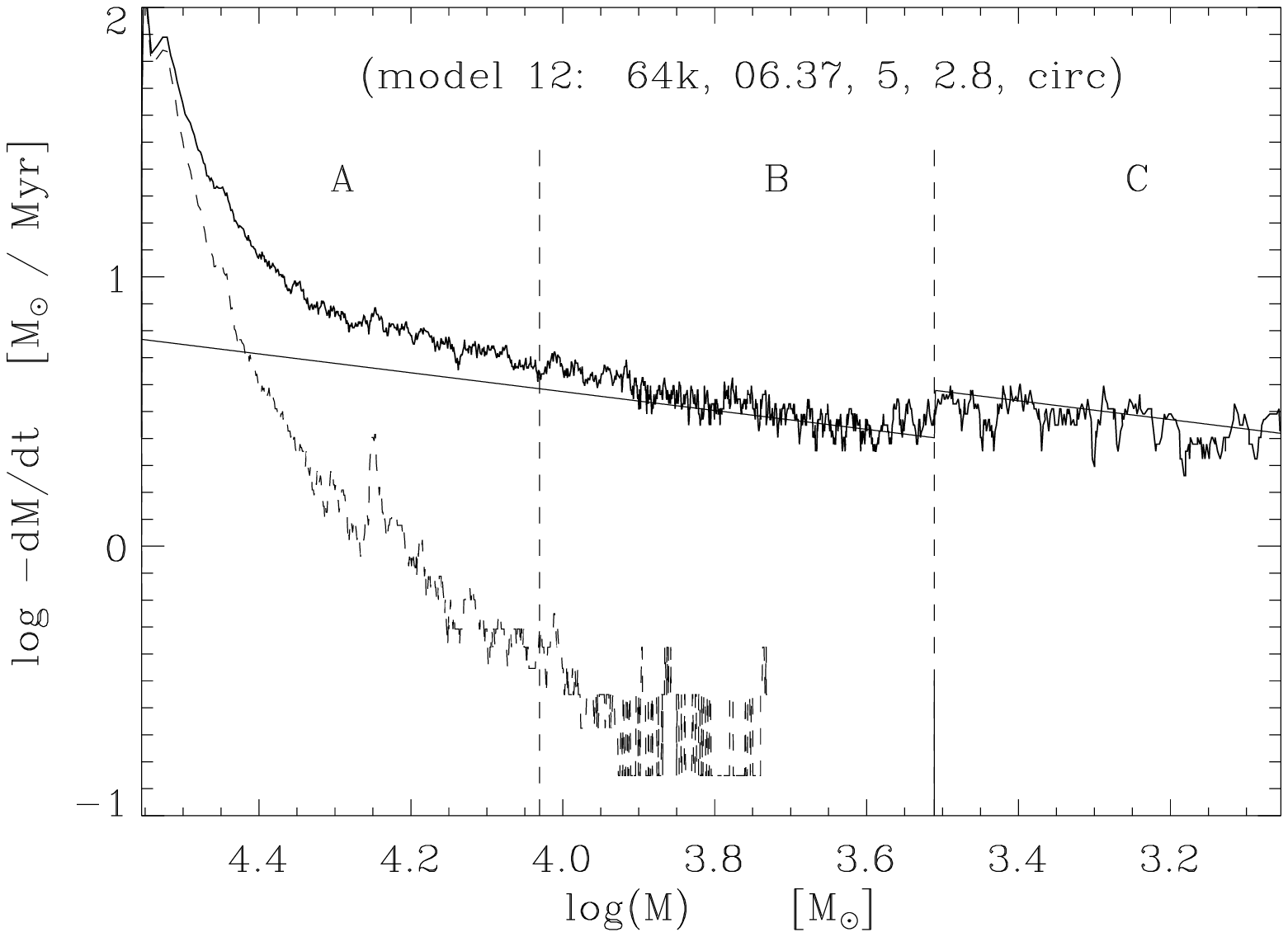,width=6.0cm}\hspace{-0.3cm}
            \psfig{figure=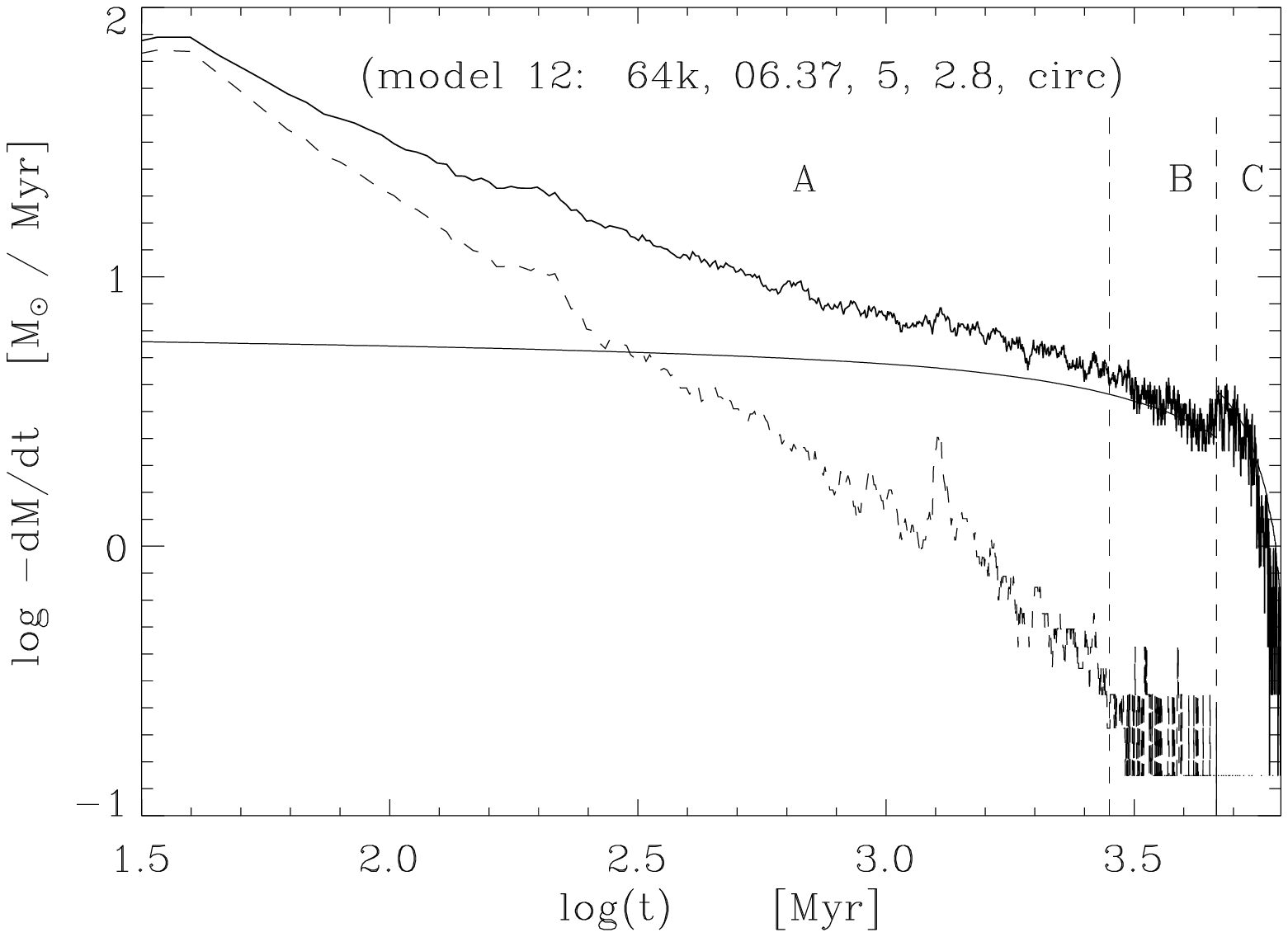,width=6.0cm}}
\caption[] {The three phases of mass loss in two \nbody\ models of BM03:
 A = dominated by stellar evolution, B = dominated by dissolution, C = dominated by dissolution after core collapse.
The line separating phases A and B is when the evolutionary mass loss has dropped to  
10\% of the total mass loss.
The line separating phases B and C indicates the core collapse time.
The figure shows $dM/dt$ versus
$M$ (left) and $t$ (right) in logarithmic units. 
The full upper line is the total mass loss rate; the dashed line is the
mass loss by stellar evolution; the full smooth lines are the mass loss by dynamical effects, assumed to 
scale as a power law of $M$, before and after core collapse.  
The models are defined by a vector containing: model nr, nr of stars, total lifetime (in Gyrs), concentration
parameter $W_0$, $R_G$ (in kpc), and orbit.}
\label{fig:phases}
\end{figure*}

A study of the mass loss rates of the BM03 models shows that three mass loss phases
can be recognized. This is depicted in
Fig. \ref{fig:phases} for two models (nrs 6, 12), which shows the variation of 
\dmdt\ as function of $M$ and $t$. In the first phase (A)  mass loss is dominated by
stellar evolution and so the mass loss rate drops steeply with time. 
In the second phase (B) mass loss is dominated by dynamical effects and 
the mass loss rate behaves approximately as a power law of $\dmdt \propto M^{1-\gamma}$
(see Sects. \ref{sec:2.2.1} and \ref{sec:2.2.2}). The third phase (C) is after core collapse.
Mass loss is also dominated by dynamical effects, but the mass loss rate is higher than
before core collapse. 
This has been noticed before in the mass evolution of cluster models, e.g. 
Baumgardt (2001).
 The separation between the three regions is not as strict as
suggested in Fig. \ref{fig:phases} because the different effects overlap near the boundaries (see below).

In the next three sections we discuss the three mass loss phases and
how they depend on the cluster parameters and the environment.

\section{Direct and induced mass loss by stellar evolution} 
\label{sec:4}

Mass loss during the early history of clusters (phase A in Fig. \ref{fig:phases}) 
is dominated stellar evolution.

At very early stages only stellar evolution contributes 
to the mass loss. But very soon thereafter, the mass loss rate \dmdt\ of all models 
is higher than \dmdtev. 
This is seen best in the right hand panels of Fig. \ref{fig:phases}, where the mass loss rates
are plotted versus time: the difference between the total mass loss rate (top lines) and the
mass loss by evolution (dotted line) is much larger than the mass loss by dissolution 
(almost straight line), which will be discussed below.
{\it This shows that during the early phases,
when the mass loss is dominated by stellar evolution, Roche-lobe filling clusters 
lose an extra amount of mass (in the form of stars)
by dynamical effects, induced by the mass loss due to stellar evolution.} 
This evolution-induced mass loss is due to the fact that the cluster radius expands and the
tidal radius shrinks due to evolutionary mass loss.

For adiabatic models, we expect that 
the evolution-induced mass loss rate, \dmdtind\ will be 
about equal to the mass loss rate by stellar evolution, \dmdtev. 
This can be understood as follows.

If a cluster loses a fraction $ \delta << 1$ of its mass $M_0$ on a time scale longer than 
its crossing time, the cluster will adiabatically expand such that its radius, relative to its initial radius,
 is $r/r_0=M_0/M\approx1+\delta$. At the same time, the mass loss causes the Jacobi radius to 
shrink: $r_J/r_{J0}=(M/M_0)^{1/3}\approx1-\delta/3$. 
The mass in the shell between $1-\delta/3<r/r_{J0}<1+\delta$ is consequently unbound since 
it is outside the new Jacobi radius. For a logarithmic potential, the density at 
$r_{J0}$, $\rho(r_{J0})$, is 6 times lower than the mean density within 
$r_{J0}$: $\bar{\rho_{J0}}=3M_0/(4\pi r_{J0}^3)$. 
The mass $\Delta M$ that is in the unbound shell is thus
 $\Delta M=4\pi r_{J0}^2\rho_{J0}\Delta r$, with $\Delta r=4\delta r_{J0}/3$, such that
 $\Delta M=(2/3)\delta M_0$. In fact, $\Delta M$ will be slightly 
higher because we have adopted the lower limit for the density in the outer layers of the cluster. 
So we may expect that evolutionary mass loss rate induces about the same rate 
of evolution-induced mass loss, $\dmdtind = \find \times \dmdtev$
with $\find \simeq 1$.

A study of all models shows two deviations from this simple expectation.\\
(i) The value of \find\ is only about unity for models for which 
$\dmdtev\ >> \dmdtdis$. This is the case for models with a very long lifetime of $\ttot > 25$ Gyr. 
If the mass loss by dissolution in the early lifetime of the cluster cannot be ignored,
the evolution-induced mass loss rate is smaller. 
Therefore $\find$ will be smaller than unity.\\
(ii) The evolution-induced mass loss does not start at $t=0$ but it needs time to build up.
We can expect that the time scale for this build-up will be of the order of the
crossing time at the tidal radius, because this is the time scale on which stars
can leave the cluster by passing the tidal radius due to the reduction of the depth 
of the potential well.

%
\begin{figure}
\centerline{\epsfig{figure=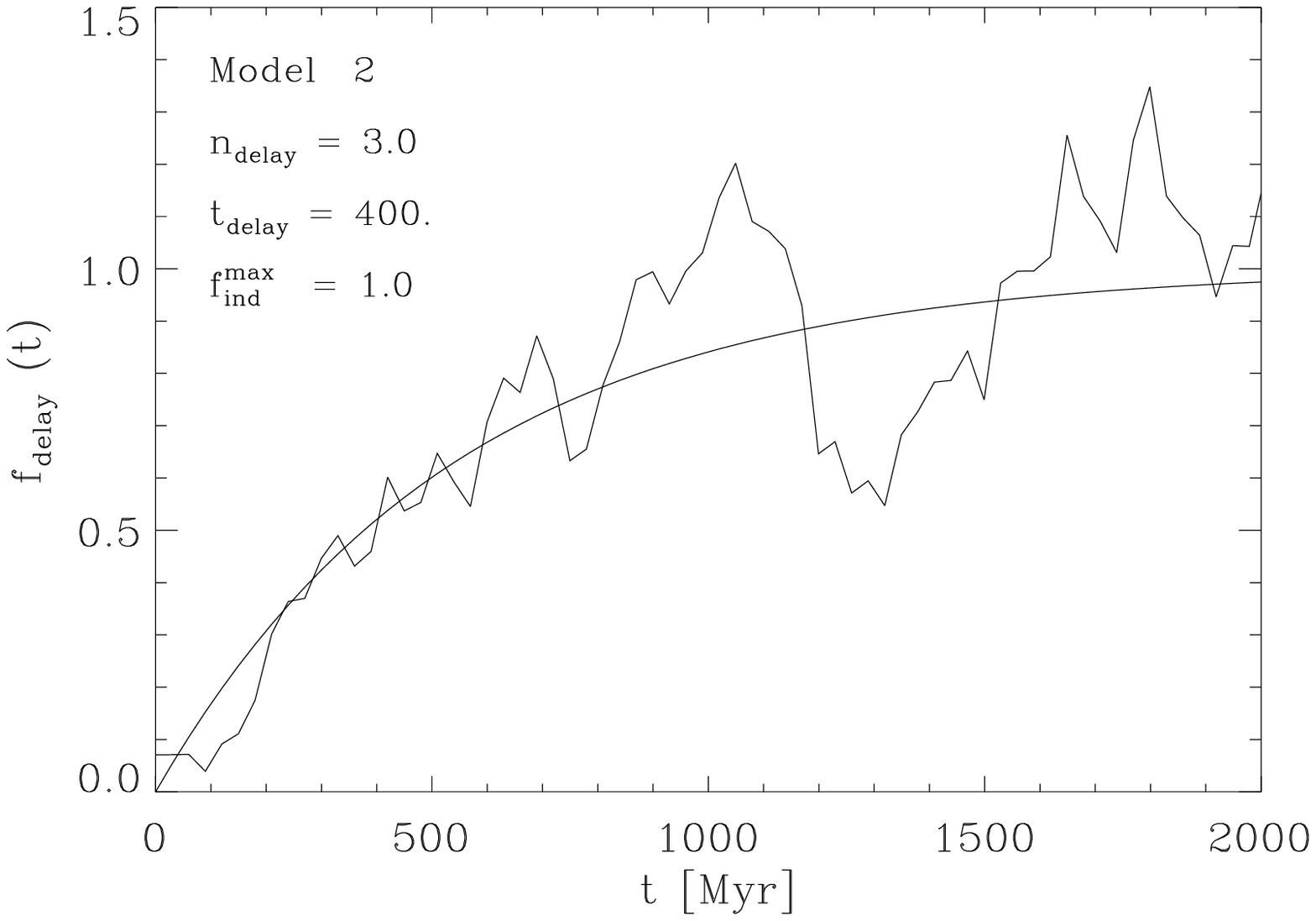,width=8.0cm}}
\centerline{\epsfig{figure=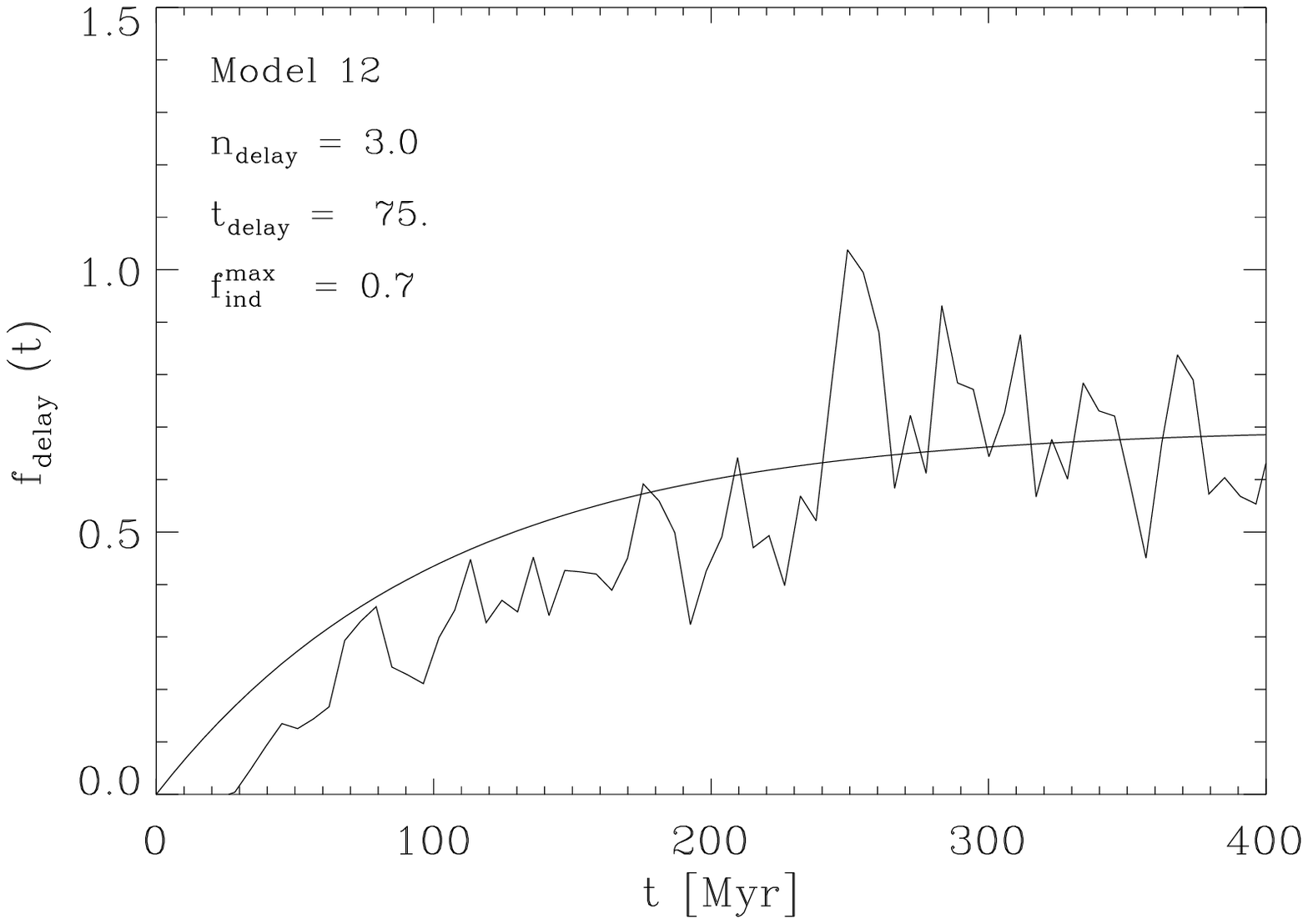,width=8.0cm}}
\caption[]{The function \fdel\ for two N-body models with very different
delay time scales.  
The smooth line is the fit by formula \ref{eq:fdelay}. The parameters are given in the figure. 
}
\label{fig:fdelay}
\end{figure}
%

A study of all the cluster models shows that we can describe the evolution-induced
mass loss rate by

\begin{equation}
\dmdtindeq = \find (t) \times \dmdteveq.
\label{eq:dmdtdynev}
\end{equation}
with 

\begin{equation}
\find (t)= \findmax \times \fdelay (t)
\label{eq:find}
\end{equation}
We assume that the growth of the evolution-induced mass loss 
approaches its maximum value in an exponential function of the
delay time scale

\begin{equation}
\fdelay (t)= 1 - \exp \{ -(t/t_{\rm delay}) \}  
\label{eq:fdelay}
\end{equation}
The delay time scale is expected to depend on the crossing time at the tidal radius,
so 

\begin{equation}
t_{\rm delay} = n_{\rm delay} \times t_{\rm cr}(r_{\rm t})
\label{eq:tdelay}
\end{equation}
Eq. \ref{eq:find} describes the increase of $\find (t)$ from 0 to \findmax\ 
with a time scale of $t_{\rm delay}$. 
Figure \ref{fig:fdelay} shows the function \find\ for two models.
The figure shows that the exponential expression describes the delay function quite well.

The values of \ndelay\ turn out to be about 3.0 for all Roche-lobe filling models.
Therefore we have adopted this value for all models.
(see Table \ref{tbl:BM03models}).

\begin{figure}
\centerline{\epsfig{figure=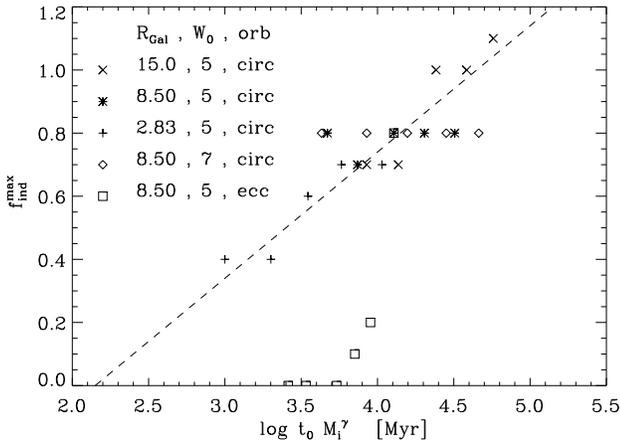, width=09.0cm}}
\caption[]{The values of \findmax\ versus $\log \tmig$
for all Roche-lobe filling cluster models. For models in circular orbits 
the values of \findmax\ decrease with decreasing \tmig, i.e. with increasing 
mass loss by dissolution, as expected. The dashed line shows the approximate
relation which is given by Eq. \ref{eq:findmax}. 
Clusters in highly eccentric 
orbits have smaller values of \findmax (see Sect. \ref{sec:7}).}
\label{fig:findmax}
\end{figure}

The values of \findmax\ listed in Table \ref{tbl:BM03models} show that $\findmax =1$ only for the
clusters with very long lifetimes larger than about 25 Gyr. For clusters with shorter lifetime
\findmax\ does not reach this value, because the contribution by dissolution helps
to restore the equilibrium that was destroyed by the fast mass loss due to stellar evolution.
Therefore we expect the value of \findmax\ to depend on the ratio between the evolutionary
mass loss, $\dmdtev = M \dmuevdt$ and the mass loss by dissolution, $-M/\tdis$ with 
$\tdis=\tzero M^{\gamma}$. 
The smaller the ratio $\dmdtdis\ / \dmdtev$  
the larger \findmax\ with a maximum of $\findmax \simeq 1$. 
Fig. \ref{fig:findmax} shows the values of \findmax\ as a function of \tmig. (The determination of
the values of \tzero\ and $\gamma$ is described in Sect. 5).  
The results can be represented by the relation

\begin{equation}
\findmax = -0.86+ 0.40 \times \log (\tmig)~~{\rm for}~\log(\tmig) \ge 2.15 
\label{eq:findmax}
\end{equation}
and $\findmax=0$ if $\log(\tmig)<2.15$,  with the ages in Myrs.

%

\section{Dissolution before core collapse}
\label{sec:5}

We first consider the dynamical mass loss in phase B, i.e. before core collapse,
which is expected to vary as $\dmdtdis \sim M^{1-\gamma}$ (Sect. \ref{sec:2.2}).
This mass loss is due to two-body relaxation (dissolution)  plus the induced loss of stars due to 
the expansion of the cluster and the shrinking of the tidal radius.
In order to derive the value of $\gamma$  we study the dependence of $\dmdtdis$
on $M(t)$
for all models of clusters in circular orbits. For the determination of the mass loss by dissolution
we have corrected the total mass loss rate for both the evolutionary and the evolution-induced mass loss rate.

\subsection{The empirical value of  $\gamma$ before core collapse}
\label{sec:5.1}

\begin{figure*}
\centerline{\hspace{+3.0cm}\epsfig{figure=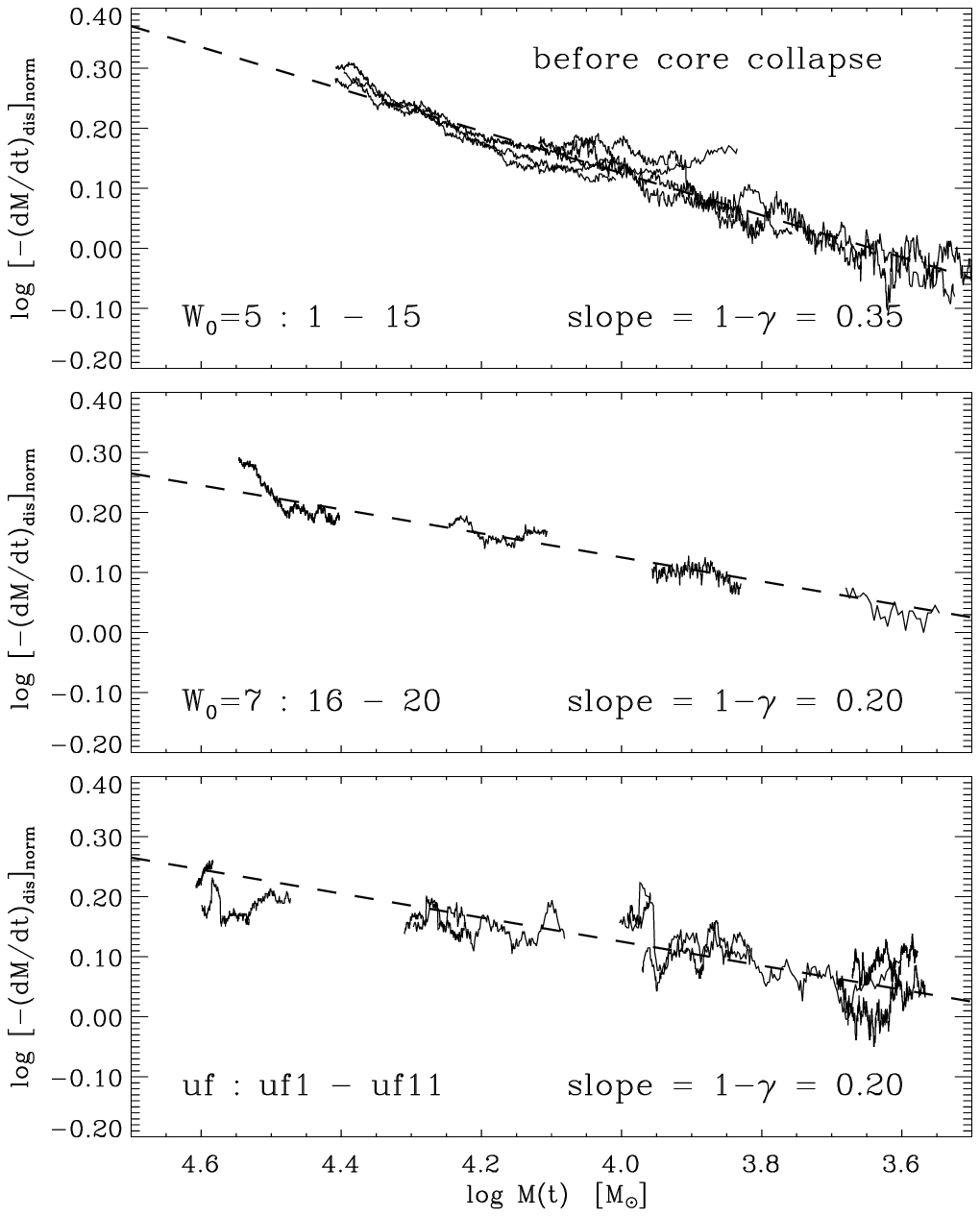, width=12cm}\hspace{-4.0cm}
\epsfig{figure=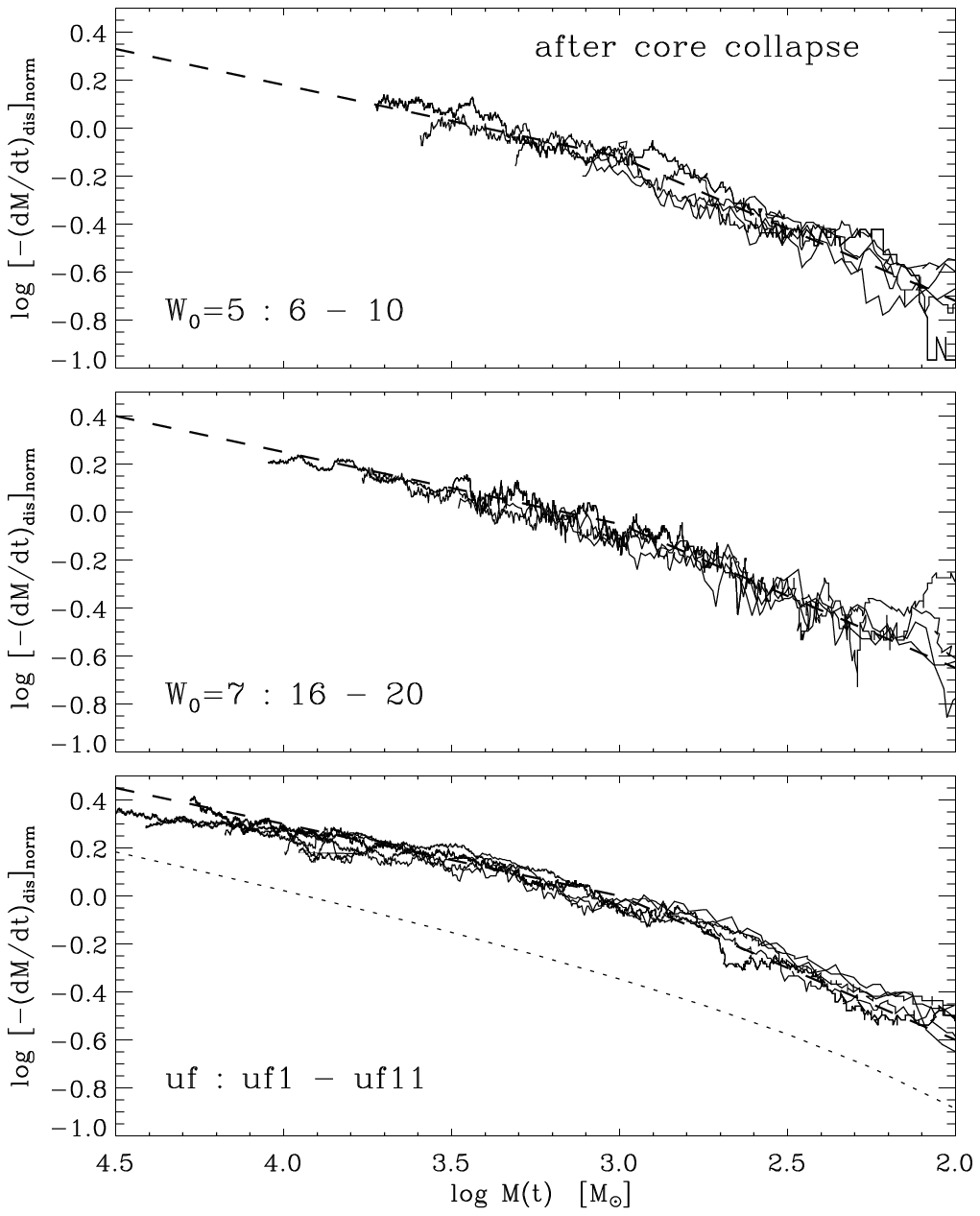,width=12.0cm}}
\caption[]{The dependence of the mass loss rate by dissolution on $M(t)$. The figure shows
the values of $\log -\dmdtdisnorm $, defined by Eq. \ref{eq:dmdtnorm},   
versus $\log(M)$, before (left) and after (right) core collapse for the models with 
$W_0=5$ (top), $W_0=7$ (middle) and Roche-lobe underfilling models (bottom). 
Notice the 
power law dependence of \dmdtdisnorm\ on $M$. 
The dashed lines show the best-fit linear relation. Before core collapse these relations have 
a slope of $1-\gamma=0.35$, 0.20 and 0.20 respectively for $W_0=5$, $W_0=7$ and Roche-lobe underfilling models.
After core collapse the relation is expressed by a double power law with
$1-\gammacc=0.30$ if $M>10^3$ \Msun\ and $1-\gammacc=0.60$ if $M<10^3$ \Msun. The dotted line
in the lower right hand panel shows the shape of the expected relation with the Coulomb logarithm
taken into account if the mean stellar mass is $\mmean=0.5$ \Msun\ (see Sect. \ref{sec:6.2}).
}
\label{fig:gamma}
\end{figure*}

The upper left panel of Fig. \ref{fig:gamma} shows the mass loss rates as a function of mass 
of models 1 to 15, i.e. for clusters with $W_0=5$. For each model the mass loss rates are 
measured in the mass interval where the evolutionary mass loss is small, 
$\dmdtev < 0.1 \dmdt$, up to core collapse.
So each model occupies a limited region in the $M(t)$-range.
The mass loss rates are corrected for evolutionary and evolution-induced mass loss.

The mass loss rates are normalized to $\rgal = 8.5$ kpc and to the mean stellar mass,
 to make the curves overlap. This is possible  
because we expect from Eqs. \ref{eq:dndtdmdt} and \ref{eq:tnrefdef} that

\begin{equation}
\left(\frac{{\rm d}M}{{\rm d}t}\right)\left( \frac{\Rgal}{8.5 {\rm kpc}}\right) 
\left( \frac{\mmean}{\mmean_i}\right)^{-\gamma}
\equiv \left(\frac{{\rm d}M}{{\rm d}t}\right)^{\rm norm} \propto M^{1-\gamma}
\label{eq:dmdtnorm}
\end{equation}
where $\mmean$ is the mean stellar mass and $\mmean_i$ is the initial mean stellar mass.
We see that the curves nicely overlap. 
We have fitted a straight line through the data 
with a slope $1-\gamma = 0.35\pm 0.02$. This implies that $\gamma = 0.65 \pm 0.02$.
(The appearance of $\gamma$ in the term $(\mmean/\mmean_i)^{\gamma} \simeq 1$ implies that we had to 
derive the value of $\gamma$ in two iterations.)
We will adopt $\gamma=0.65$ for the $W_0=5$ models in the rest of the paper.
This agrees with observations of star clusters in M51 (Gieles 2009). 
and with the N-body simulations of clusters without stellar evolution by Gieles \& Baumgardt
(2008).

The middle left panel shows the relation for the normalized mass loss rates for the models
with $W_0=7$. Each model occupies only a small part in this diagram because
the pre-core collapse time of these models is short: only about half as long as that of the
$W_0=5$ models. Nevertheless, we see that there is a clear power law dependence
on $M$ with a slope of $1-\gamma = 0.20 \pm 0.04$ which implies $\gamma=0.80 \pm 0.04$. 
We adopt $\gamma=0.80$ for the pre core collapse phase of the $W_0=7$ models. 
This value is higher than for the $W_0=5$
models, as expected (see Sect. \ref{sec:2.2.1}). 

\subsection{The dissolution parameters  \tzero\ and \tnref }  
\label{sec:5.2}

We have fitted the mass loss rates of the BM03 models in phase B, i.e. the pre-core collapse phase,
with a function $\dmdtdis =- M^{1-\gamma}/\tzero$ with $\gamma=0.65$ and 0.80 for the models with
$W_0=5$ and 7 respectively. The values of the mean stellar mass in that phase is also derived from 
the details of the BM03 models.
The values of $t_0$ are listed in Table \ref{tbl:BM03models}.
We see that they are approximately constant for the $W_0=5$ cluster models at the same value of
$\Rgal$ and that for clusters in different orbits $\tzero \propto \Rgal$, as expected (Eq.
\ref{eq:tnrefdef}).

\begin{figure*}
\centerline{\hspace{+3.0cm}\epsfig{figure=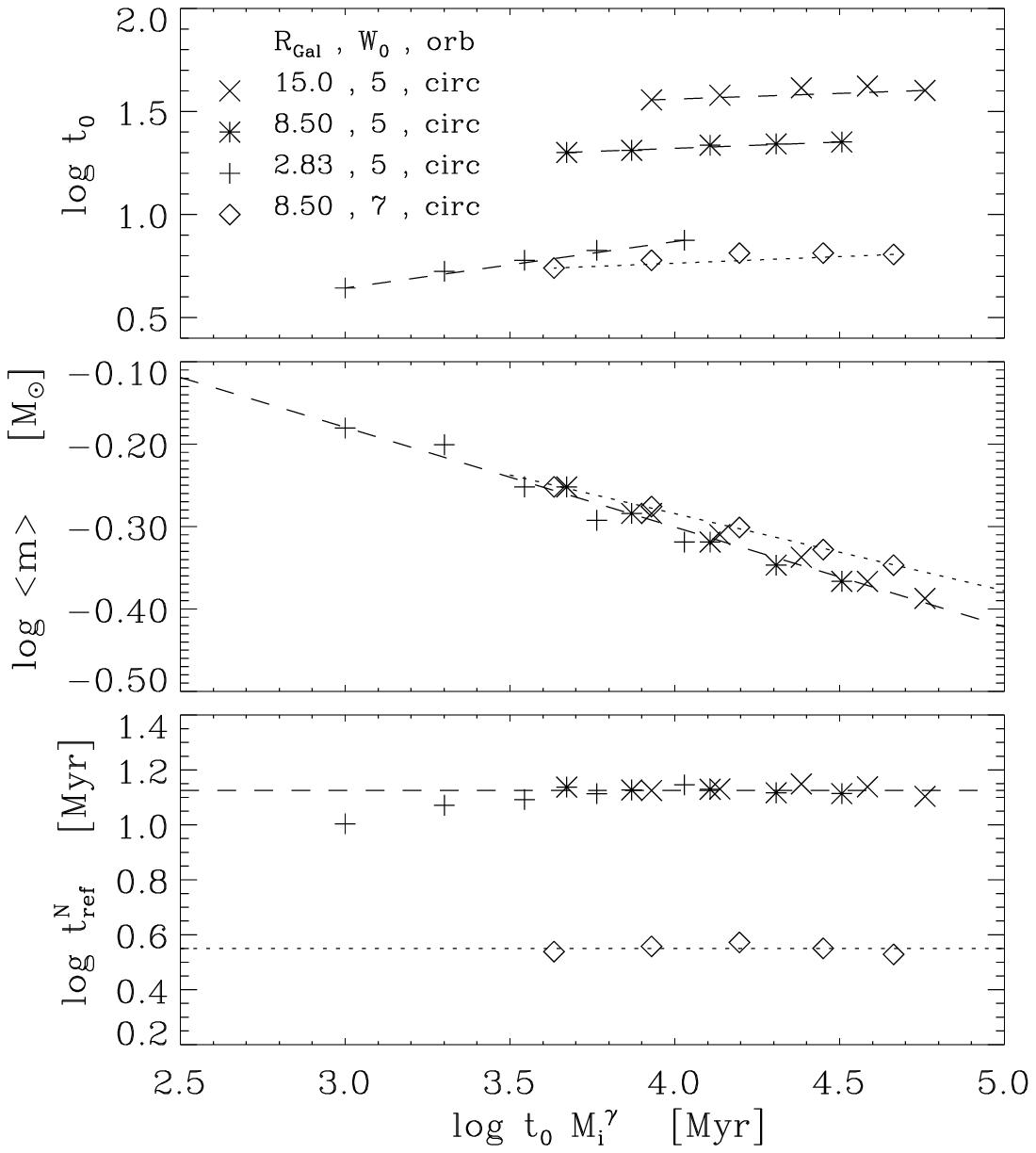,width=12.0cm}\hspace{-4.0cm}
\epsfig{figure=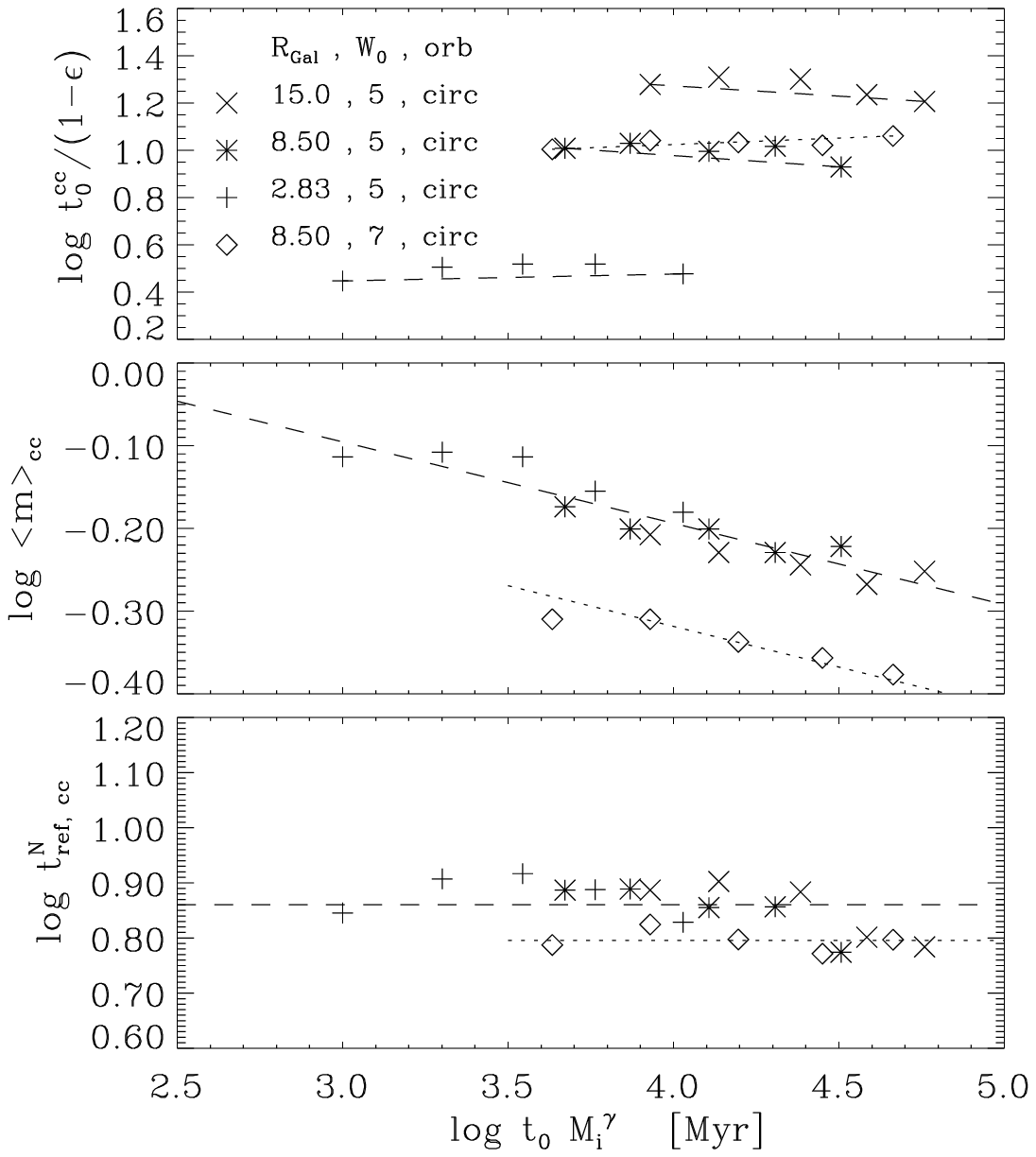,width=12.0cm}}

\caption[]{
The parameters of the models 1 to 20 of clusters in circular orbits before core collapse (left)
and after core collapse (right).
Top: The relation between $\tzero M_i^\gamma$, which is a proxy for the total lifetime of a cluster,
 and \tzero\ (left) or \tzerocc (right).
Dashed and dotted lines show the relations for clusters with $W_0=5$ or 7 respectively 
for the same value of \Rgal.
Middle: The mean stellar mass after the evolution dominated phase and before core collapse,
i.e. at $\mu \simeq 0.25$ (left) or at core collapse (right).
Bottom: The resulting values of $\tnref$ (left) or $\tnrefcc$ (right)  as a function of   
$\tzero M_i^\gamma$. The mean values of $\tnref = 13.3$ Myr and $\tnrefcc= 7.2$ Myr
for $W_0=5$ models and $\tnref=3.5$ Myr and $\tnrefcc = 6.2$ Myr for
$W_0=7$ models are indicated.
}
\label{fig:tnref}
\end{figure*}

Fig. \ref{fig:tnref} (top panel) shows the relation between $\tzero$ and $\tmig$, 
which is a proxy for the total lifetime of the cluster, for all models. 
For each value of \Rgal\ the values of \tzero\ decrease slightly with decreasing $\tmig$.
This is because \tzero\ is expected to depend on $\mmean^{-\gamma}$ (Eq. \ref{eq:tnrefdef})
and the mean mass in the pre-core collapse phase depends on \tmig.
The vertical offset between the relations for different values of \Rgal\ is because \tzero\ is
also expected to be proportional to \Rgal\ (Eq. \ref{eq:tnrefdef}).  
The difference in the values of \tzero\ between models of $W_0=5$ and 7 at the same \Rgal\
is due to the difference in the value of $\gamma$. 
In fact, 
for clusters with $\Mi =10^4 \Msun$
the dissolution time scale $\tdis=\tzero 10^{4\gamma}$ of a $W_0=5$ cluster is 
about the same as that for  
a $W_0=7$ cluster.

The values of \tzero\ are measured after the stellar evolution-dominated 
phase but before core collapse. This is when $M(t)/\Mi \equiv \mu$ is 
approximately between 0.4 and 0.15, with a mean value of $<\mu> \simeq 0.25$.
The middle panel of Fig. \ref{fig:tnref} shows the values of \mmean\ in the pre-collapse phase at $\mu=0.25$
derived from the BM03 models. These values of \mmean\ are approximately 

\begin{eqnarray}
\log \mmean & = & +0.184 - 0.121 \times \log ~\tzero \Mi ^{0.65}~~{\rm for}~W_0=5 \nonumber \\
\log \mmean & = & +0.090 - 0.094 \times \log ~\tzero \Mi ^{0.80}~~{\rm for}~W_0=7 
\label{eq:mmeanprcc}
\end{eqnarray}
These two relations are shown in the figure. The increase of \mmean\ towards shorter lifetimes 
is because at a given fraction of its lifetime the maximum mass of a star 
in a cluster with a short lifetime is higher than in case of a longer lifetime due to stellar
evolution (see also Sect. \ref{sec:5.3}).

The lower panel of Fig. \ref{fig:tnref} shows the value of 
$\tzero \mmean^{\gamma} (\Rgal /8.5)^{-1} (\vgal/220)(1-\epsilon)^{-1}$, which is expected to be
constant, \tnref\ (see Eq. \ref{eq:tnrefdef}). We see that the values are indeed about constant 
with $\log (\tnref/ {\rm Myr}) = 1.125 \pm 0.016$ for $W_0=5$ models and $0.550\pm 0.015$
for $W_0=7$ models.
For clusters with total ages less than about 3 Gyrs, the values are slightly smaller. This is due to the
fact that these clusters still contain massive stars over most of their lifetime and massive stars are
effective in kicking out lower mass stars (BM03).

\subsection{The evolution of the mean stellar mass}
\label{sec:5.3}

\begin{figure}
\centerline{\epsfig{figure=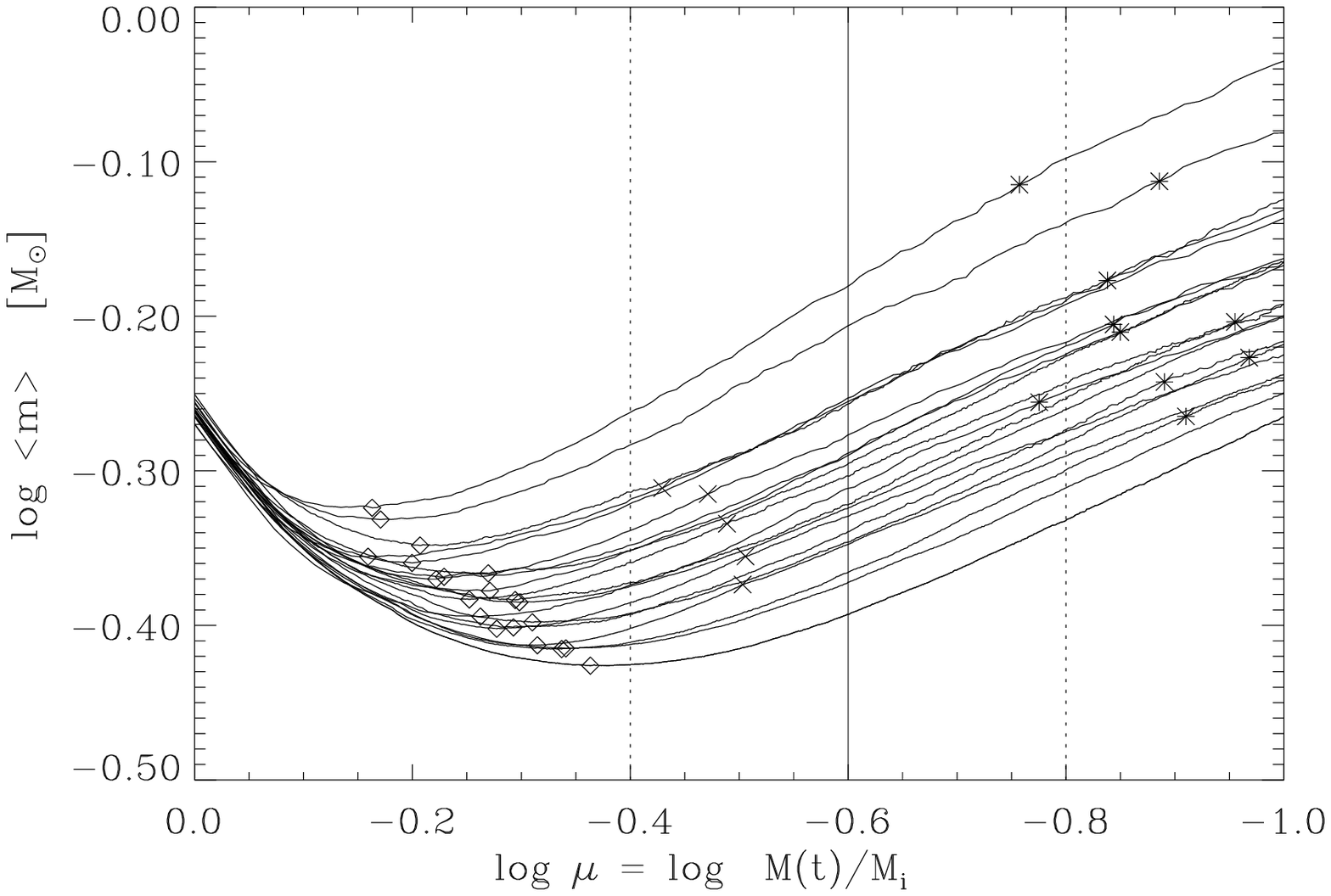,width=7.8cm}}
\centerline{\epsfig{figure=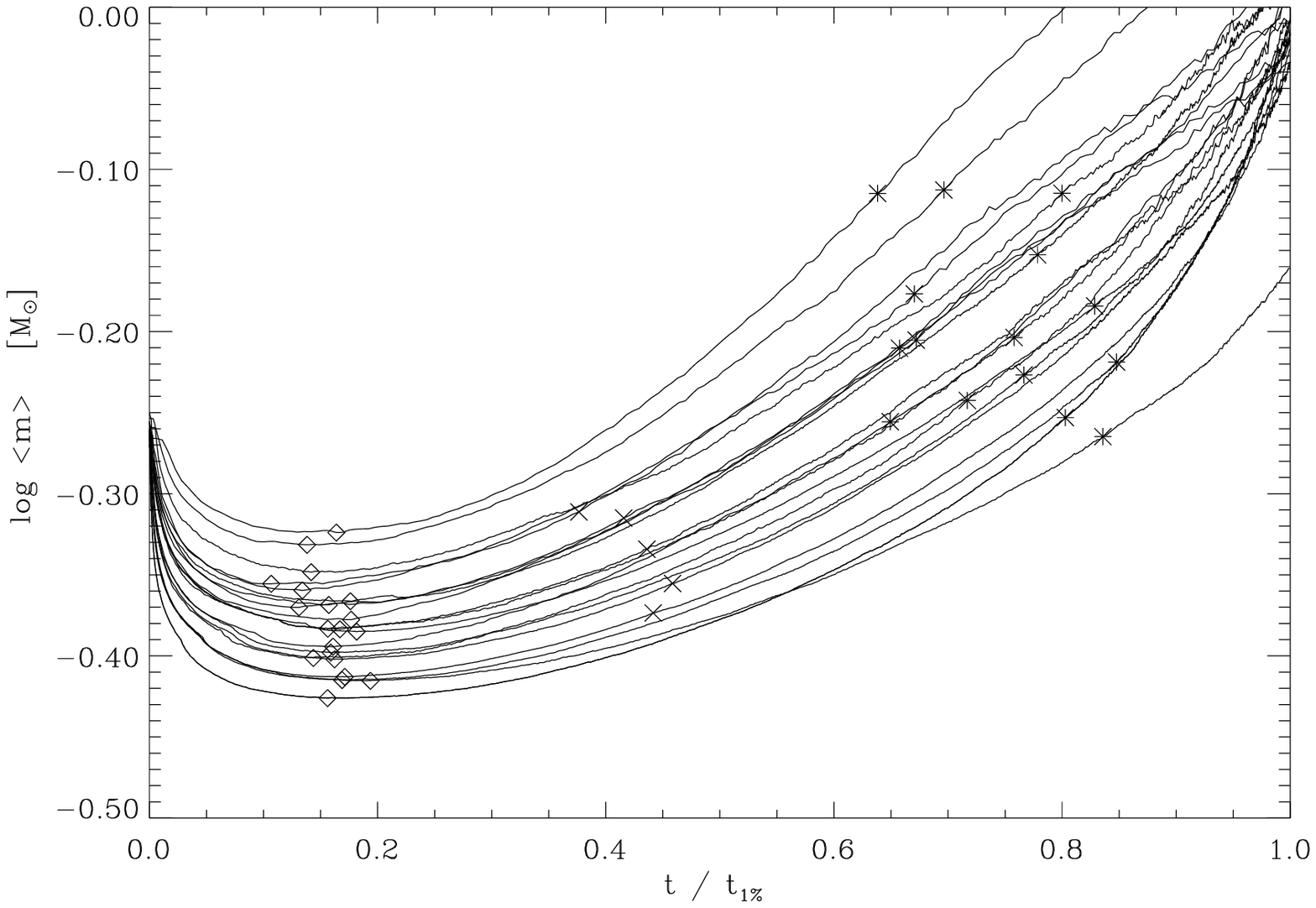,width=7.8cm}}
\caption[]{
Top: The mean stellar mass of $W_0=5$ and $W_0=7$ clusters as a function of the 
remaining mass fraction for models 1 to 20. 
The shorter the lifetime of a cluster,
the higher the line in this figure.
The vertical dotted lines indicate approximately the range where \tzero\ was determined 
in the pre-core collapse phase, with the
central value indicated by a full line. The diamonds indicate the minimum values.
The mean mass at core collapse is indicated by an asterisk or a cross for models with $W_0=5$ and
$W_0=7$ respectively.
Bottom: The mean stellar mass as a function of the fraction of the total cluster lifetime.
For all models the minimum is reached at $t \simeq 0.15 \tone$.
}
\label{fig:mmean}
\end{figure}

The time variation of \mmean\ for the different models is shown in  Fig. \ref{fig:mmean}.  
The top  panel shows 
the time evolution of \mmean\ (plotted in terms of $\mu$) for all models with $W_0=5$ and 7 
in circular orbits.
The lower panel shows the mean mass as function of time in terms of $t/\tone$. 
The mean stellar mass initially decreases
due to the loss of high mass stars by stellar evolution, but then increases again with 
age due to the 
preferential loss of low mass stars by dynamical effects. The minimum value is reached
for all models at $t \simeq 0.2\tone $. 
This is the time when
the cluster is fully mass segregated after which the low mass stars in the outer part of the 
cluster are lost preferentially (BM03).
For models with a short total dissolution time  the minimum
mass is higher than for models with a long lifetime. 
Notice that after mass segregation \mmean\ increases as a power-law of $\mu$
with $\mmean  \propto \mu^{-\delta}$ with $\delta \simeq 0.30$ for all BM03 models, including the 
ones not shown here.

The middle panel of Fig. \ref{fig:tnref} shows that the mean stellar mass of the $W_0=7$ 
models is slightly larger
than for the less concentrated $W_0=5$ models. 
This is due to the fact that core collapse
occurs earlier in the more concentrated $W_0=7$ models and so the value of \tzero\ in the
pre-core collapse phase refers to an earlier time than in the $W_0=5$ models. Fig. \ref{fig:mmean}
shows that \mmean\ increases with age, so it is smaller in the
pre-core collapse phase of the $W_0=7$ models.

\subsection{The start of the dissolution process}
\label{sec:5.4}

We have derived the mass loss by dissolution for the BM03 models from the output files of these models.
The data show that the dissolution does not start at $t=0$ but needs time to develop, just like the 
evolution-induced mass loss needs time to get going. 
This can be seen in Fig. \ref{fig:phases} which shows that at very early times  
the total mass loss rate is equal to the evolutionary mass loss, without any contribution
by dynamical effects.
The time needed to develop the dynamical dissolution depends on the 
crossing time at the tidal radius and is expected to behave in the same way as the 
growth of the evolution-induced mass loss rate.
Therefore we can describe the dissolution before core collapse as

\begin{equation}
\left(\frac{{\rm d}M}{{\rm d}t}\right) = - \fdelay \times \frac{M^{1 - \gamma} }{\tzero} 
\label{eq:dmdtdis+fdelay}
\end{equation}
with \fdelay\ given by Eq. \ref{eq:fdelay}.

The delay-time of the dissolution is a result of the initial conditions of the cluster 
models which start without stars with escape velocities. This might not be realistic. If clusters 
form in a collapsing cloud and go through a phase of violent relaxation, the tail of the
Maxwellian velocity distribution is initially filled and dissolution will start immediately.

\section{Dissolution after core collapse}
\label{sec:6}

We now consider the dissolution in phase C,  i.e. after core collapse.
Fig. \ref{fig:phases} shows that the mass loss rate increases by a small factor at about \tcc\
and that the slope $1-\gamma$ of the $\log -\dmdtdis$ vs $\log M$ relation is different from
that in phase B.

\subsection{The time of  core collapse, \tcc\ }
\label{sec:6.1}

\begin{figure}
\epsfig{figure=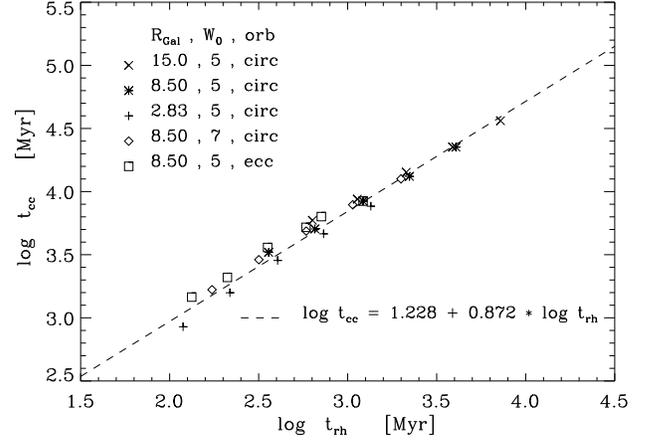,width=9.0cm} 
\caption[]{
The relation between the initial half mass relaxation time, $\trh$ and the core collapse
time $\tcc$ for Roche-lobe filling clusters, including those in elliptical orbits.
}
\label{fig:tcc}
\end{figure}

The core collapse times, listed in Table \ref{tbl:BM03models}, are derived
empirically from the \nbody\ calculations. Theory predicts that core collapse occurs after
about a fixed number of central relaxation times, $t_{\rm rc}$, (e.g. Spitzer 1987, p.95). 
However as the 
cluster evolves, it loses mass and expands, so $t_{\rm rc}$ changes  continuously. 
Therefore we search for
an empirical expression of \tcc\ in terms of the {\it initial} values of 
\trh\ and \Rgal\  (see Fig. \ref{fig:tcc}).  A linear regression 
analysis shows that we can approximate \tcc\ of all Roche-lobe filling cluster models,
including those with $W_0=7$ and elliptical orbits,  to an accuracy better than about 0.05 dex 
by the following relations

\begin{equation}
\log~(\tcc) =   1.228 + 0.872 \times \log(\trh) 
\label{eq:tcctidal}
\end{equation}
%

\subsection{The value of $\gamma$  after core collapse}
\label{sec:6.2}

 Core collapse changes the density distribution of the stars in the clusters, so that it becomes
independent of the initial distribution. Therefore we expect the values of $\gamma$ in phase C
to be the same for all the models.
The right hand panels of Fig. \ref{fig:gamma} show the normalized dissolution rate,
given by Eq. \ref{eq:dmdtnorm},
after core collapse for cluster models with $W_0=5$ (top) and  $W_0=7$ (middle).
 Each model occupies
a certain mass range, starting at core collapse down to about $M(t)=10^2~ \Msun$.

Notice that, apart from a small vertical offset, the sets of models show very similar 
lines with a slight curvature, in the sense that the slope gets steeper towards lower mass. 
The empirical relations can be fitted very well with a broken power-law relation 

\begin{equation}
\left(\frac{{\rm d}M}{{\rm d}t}\right)^{\rm postcc} =  -\frac{M(t)^{1-\gammacc}}{ t_0^{\rm cc}}
\label{eq:dmdtdispostcc}
\end{equation}
with $\gammacc=0.70$ for $M(t)>10^3$ and $\gammacctwo=0.40$,
where the subscript 2 refers to the mass loss at $M(t)<10^3 \Msun$.
These values are the same for $W_0=5$ and 7 clusters, because core-collapse
results in a redistribution of the density profile which becomes nearly independent
of that in the pre-collapse phase.
Continuity of the mass loss rate at $M(t)=10^3 \Msun $ requires that 
$\tzerocctwo = \tzerocc \times 10^{3(\gammacc - \gammacctwo)} =10^{0.90} ~ \tzerocc$.

The steepening of the slopes in Fig. \ref{fig:gamma}  is the result of the changes in the 
Coulomb logarithm towards smaller numbers of stars (Eq. \ref{eq:tdisN}).
This can be shown as follows.
 Adopting for simplicity a mean stellar mass
of $\mmean \simeq 0.5$ \Msun\ we predict (Eq. \ref{eq:tdisN}) that 
$\dmdtdis \simeq  \mmean \dNdt \simeq \mmean~N/ \tdis \sim
M^{1-x} (\ln 0.02M/\mmean)^x$ (Giersz \& Heggie 1996) 
with $x \simeq 0.75$ (see Sect. \ref{sec:2.2.1}). The  
variation of $\log \dmdtdis$ with $\log M$, predicted with this aproximation is 
shown in the lower part of Fig.
\ref{fig:gamma} by a dotted line with an arbitrary vertical offset. The shape of this
predicted line is very similar to the one derived empirically.

\subsection{The values of \tzerocc\ and \tnrefcc\  after core collapse}
\label{sec:6.3}

The top right part of Fig. \ref{fig:tnref} shows the values of \tzeropostcc\ as a function of
\tmig, which is a proxy for the lifetime of the cluster. The pattern is the same as on the left side
of this figure, i.e. before core collapse. The mean stellar mass at core collapse is shown
in the middle right part of the figure. We find that we can approximate

\begin{eqnarray}
\log \mmean_{\rm cc} & = & +0.200 - 0.0984 \times \log ~\tzero \Mi^{0.65}~~~~~~{\rm for}~~ W_0=5 \nonumber \\
\log \mmean_{\rm cc} & = & +0.075 - 0.0984 \times \log ~\tzero \Mi^{0.80}~~~~~~{\rm for}~~ W_0=7 
\label{eq:mmeanpostcc}
\end{eqnarray}
The smaller mean mass of the $W_0=7$ models at core collapse is due to the fact that core collapse
occurs much earlier than for the less concentrated $W_0=5$ models, so that the 
preferential loss of low mass stars had a smaller effect (see Fig. \ref{fig:mmean}).

The lower right part of Fig. \ref{fig:tnref} shows the derived values of $\tnref$. 
The mean values are
$\log (\tnrefcc)= 0.864 \pm 0.044$ and $0.796 \pm 0.017$ for the models with $W_0=5$ and 7 
respectively. The scatter in \tnrefcc\ is larger than that of \tnref\ in Fig. \ref{fig:tnref}
because of the larger noise in the \dmdt\ - vs - $M$ relation due to the smaller numbers 
of stars (see Fig. \ref{fig:phases}).
We see that \tnrefcc\ is slightly different for $W_0=5$ and 7 models.
This can be understood because inside a cluster the relaxation time increases with radius,
so the outer cluster parts still keep some memory of the initial distribution by the time the 
center has gone into collapse.

\section{Clusters in elliptical orbits}
\label{sec:7}

\begin{figure}
\centerline{\epsfig{figure=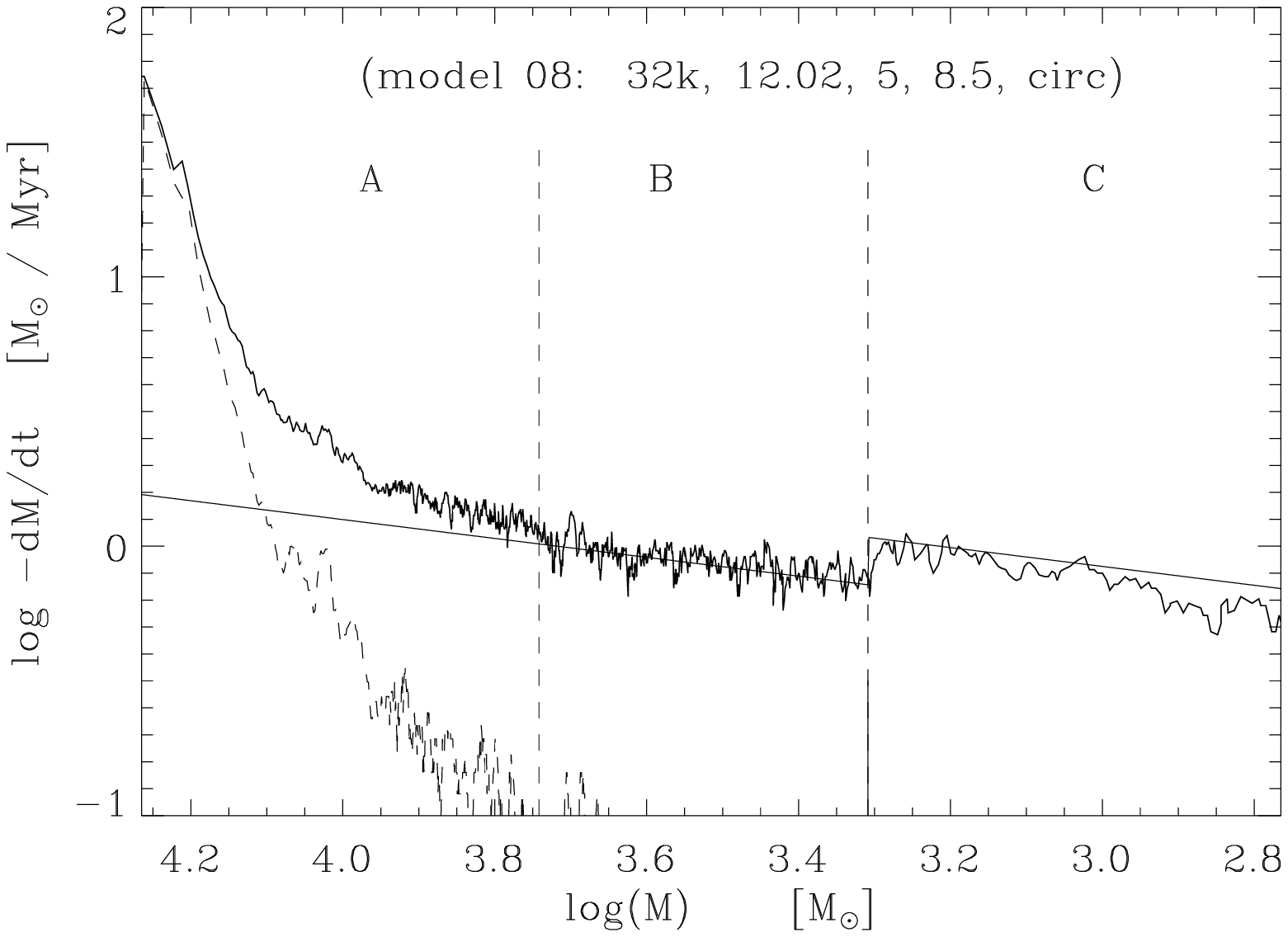,width=7.0cm}} \vspace{-0.3cm}
\vspace{-0.3cm}
\centerline{\epsfig{figure=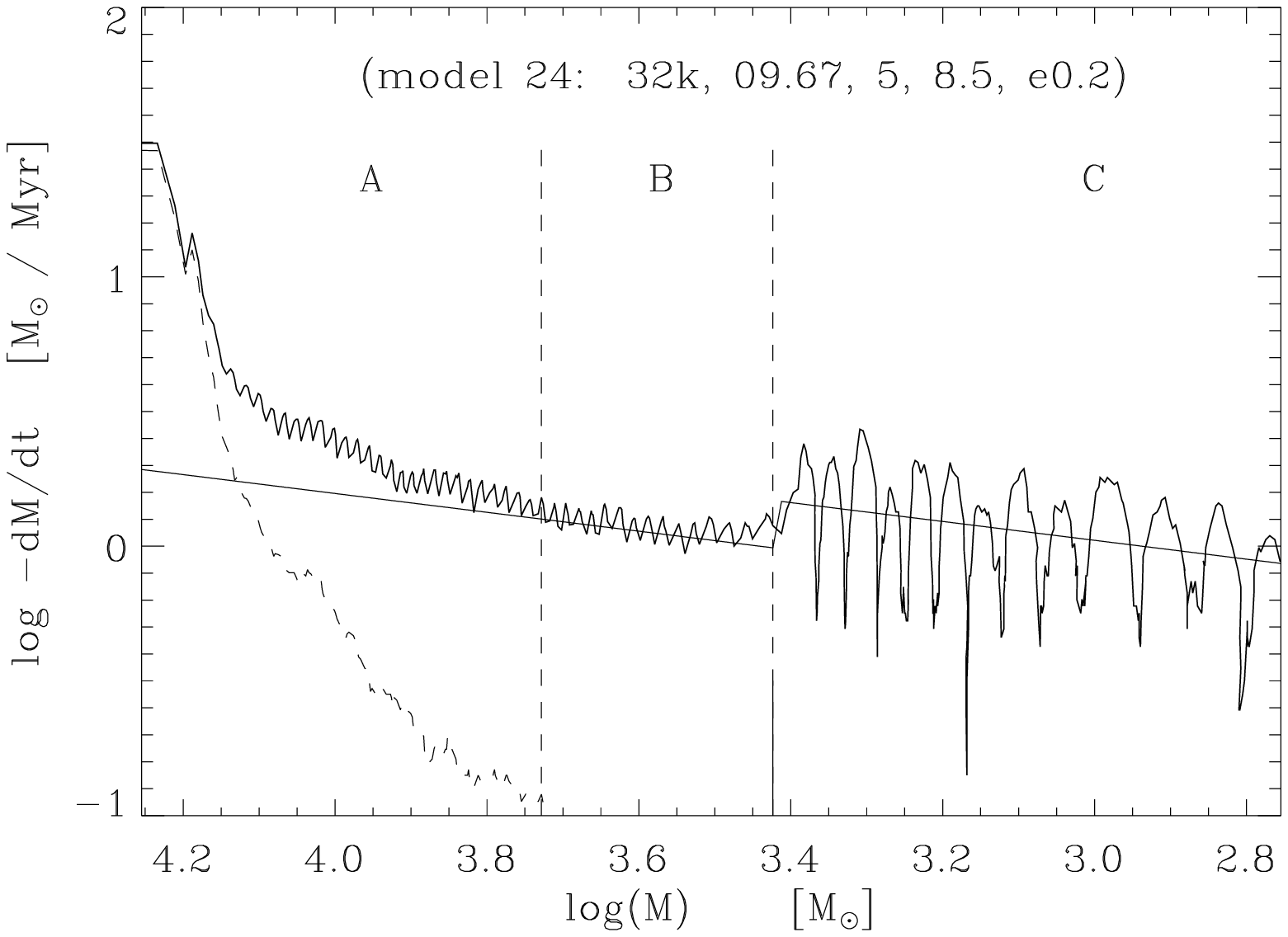,width=7.0cm}} \vspace{-0.3cm}
\vspace{-0.3cm}
\centerline{\epsfig{figure=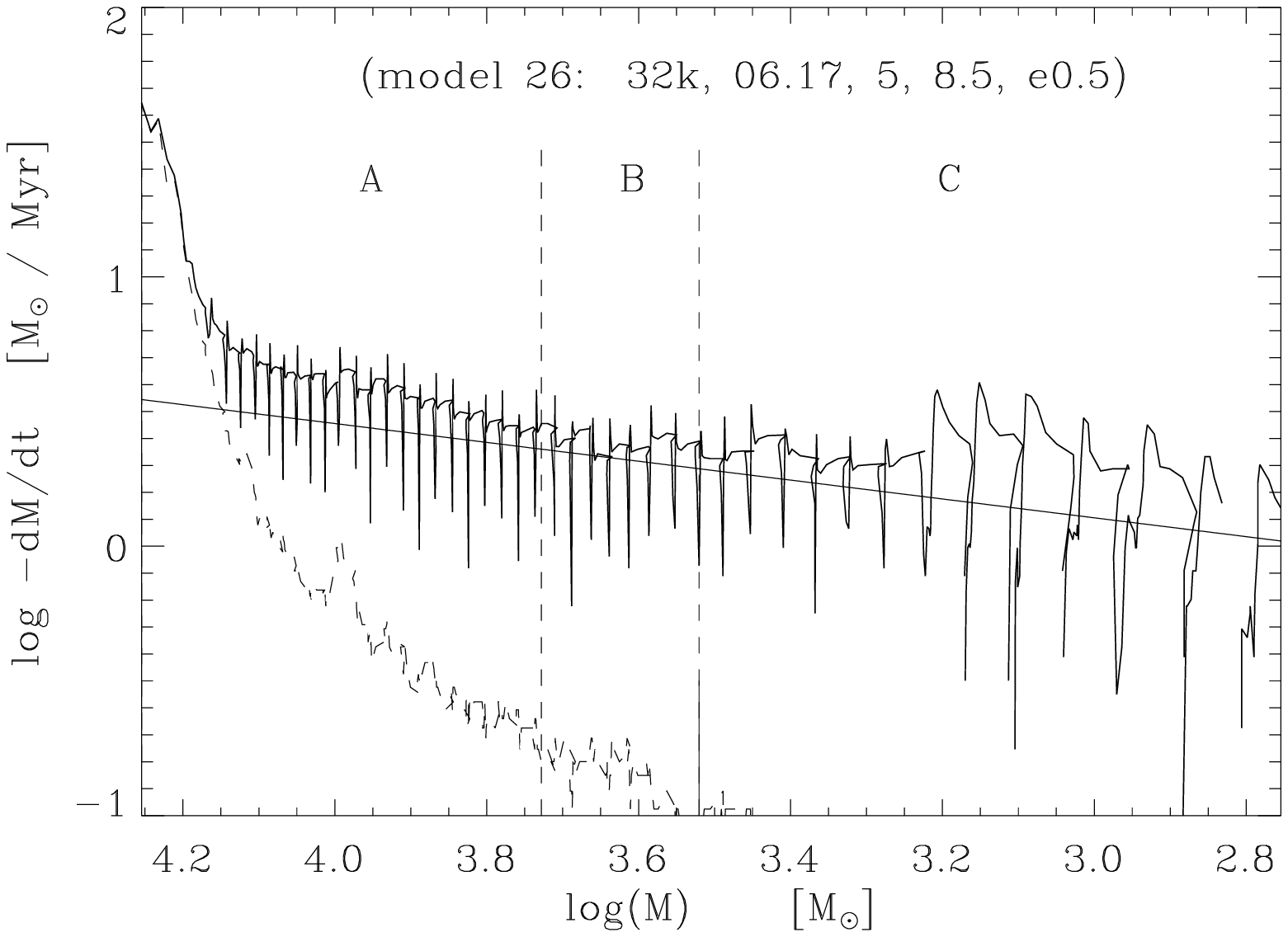,width=7.0cm}} 
\caption[] {The mass loss rates of clusters with elliptical orbits with eccentricities of 0 (upper), 0.20 (middle)
and 0.5 (bottom).
 A = dominated by stellar evolution, B = dominated by dissolution, C = dominated by dissolution after core collapse.
The line separating phases B and C indicates the core collapse time.
The figure shows $\dmdt$ versus $M$ in logarithmic units. 
The full upper line is the total mass loss rate; the dashed line is the
mass loss by stellar evolution; the full smooth lines are the mass loss by dynamical effects, assumed to 
scale as a power law of $M$, before and after core collapse.  
The models are defined by a vector containing: model nr, nr of stars, total lifetime (in Gyrs), concentration
parameter $W_0$, $R_G$ (in kpc), and orbit.
}
\label{fig:elliptical}
\end{figure}

Models 21 to 25 are for clusters in elliptical orbits with a apogalactic distance of 8.5 kpc
and various eccentricities. These can be compared with an otherwise similar model nr 8 in a circular
orbit. Fig. \ref{fig:elliptical} shows the mass loss of models 8, 21 and 23, with 
$\epsilon=0.0$, 0.2 and 0.5 
respectively.  
Since the total lifetime of the clusters scales approximately as $(1-\epsilon)$, 
the lifetimes
of the clusters are very different (see Table \ref{tbl:BM03models}).
The evolution can be described by the same three phases that we found for all other models:
a stellar evolution dominated phase (A), a dissolution dominated phase before core collapse (B)
and the phase after core collapse (C). The mass loss rates in all phases is variable 
with a periodicity of the orbital period. The mass loss rate is highest at perigalacticon and the
amplitude of the variations increases with increasing ellipticity. Especially after core collapse
the amplitude increases drastically. This is due to the expansion of the outer layers of the cluster
as a reaction to the core collapse. The stars in the outer layers are then more susceptible
to the periodically changing tidal field.
Only the models with $\epsilon=0$ and $0.2$ show a jump in the mass loss rate at the time of core collapse.
For clusters in more eccentric orbits no clear jump is observed at core collapse, but the 
mass loss rate does increase compared to the simple power-law extrapolation of phase B.

The straight full lines in Fig. \ref{fig:elliptical} show the values of $\gamma$ defined by $-\dmdtdis = M^{1-\gamma}/ \tzero$. 
Both the values of $\gamma$ and of \tzero\ describe the ``time averaged'' mass loss rate
as a function of the mass. These values were derived from a study of the $M(t)$ history
for these models. The values of \tzero, \ndelay, \findmax\  and \tzeropostcc\ are 
listed in Table \ref{tbl:BM03models}.

\begin{figure*}
\centerline{\hspace{+3.0cm}\epsfig{figure=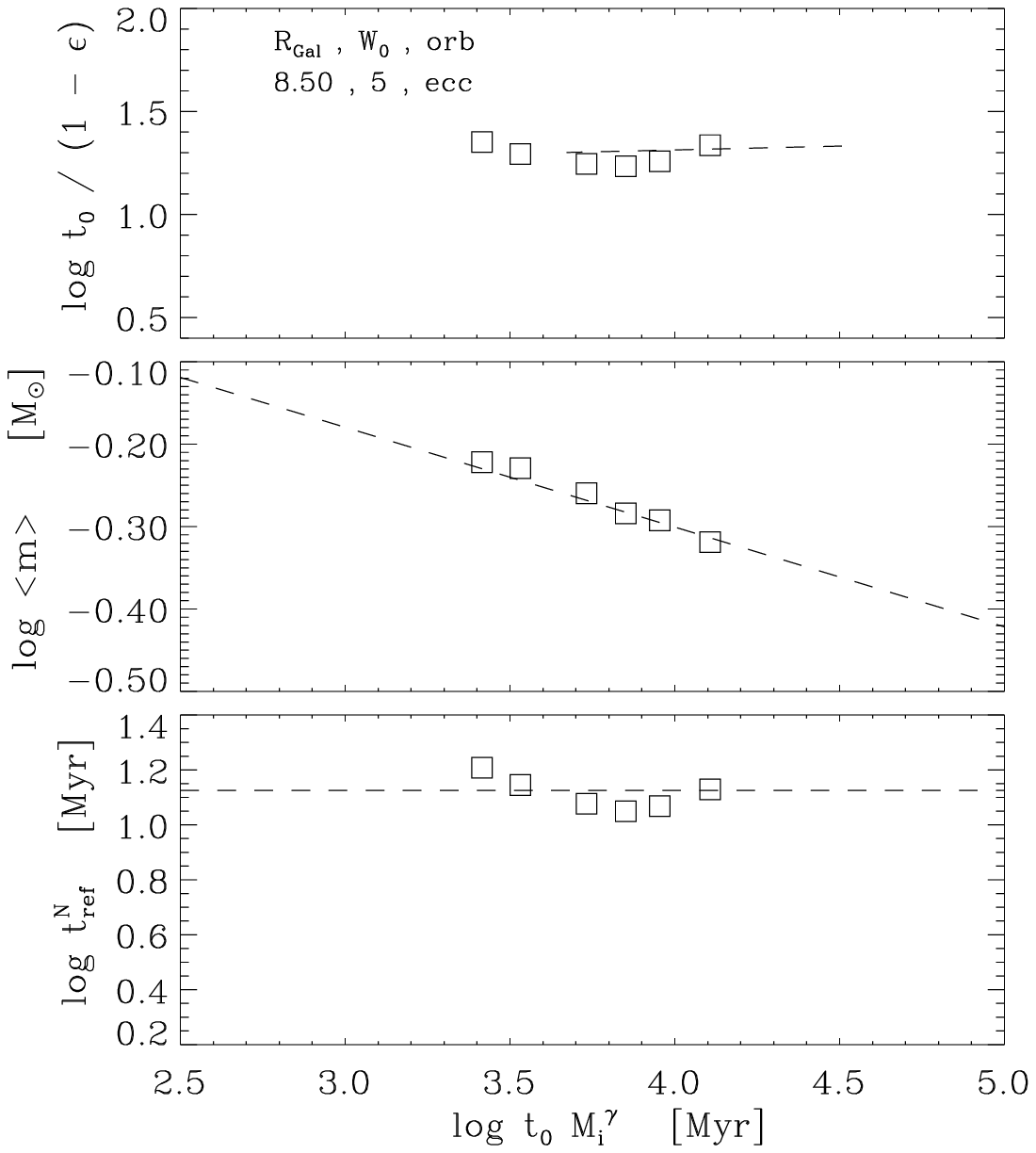,width=12.0cm} \hspace{-4.0cm}
\epsfig{figure=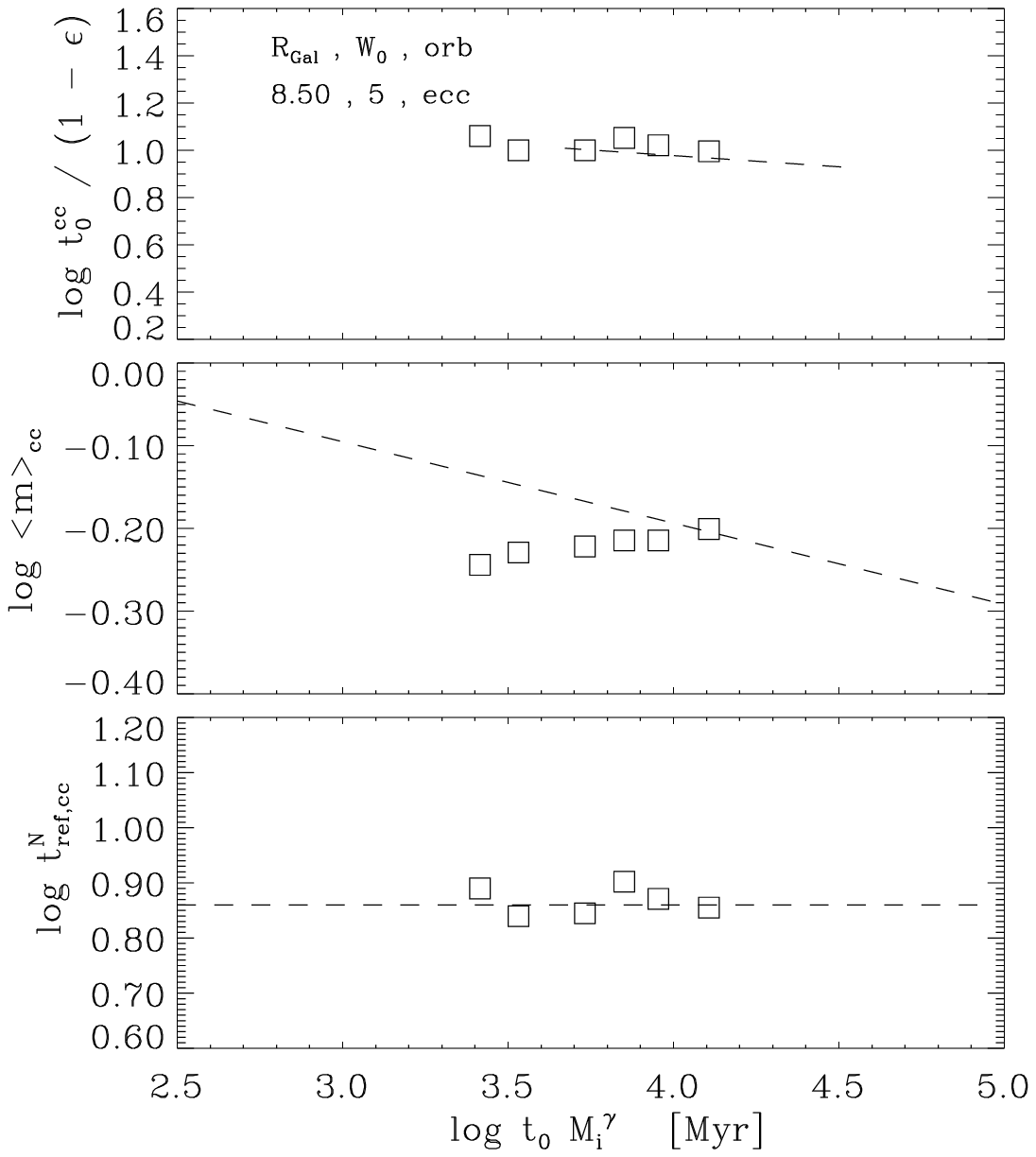,width=12.0cm}} 

\caption[]{The parameters of the models of clusters in elliptical orbits, nrs 20 to 25, 
combined with model 8 of the same mass but with $\epsilon=0$ before (left)
and after (right) core collapse. In all panels the right data point is for $\epsilon=0$
and the left one is for $\epsilon=0.8$.
Top: The relation between $\tzero M_i^\gamma$  and \tzero\ before (left) or \tzerocc\ (right) 
after core collapse.
 The dashed lines show the relation for clusters in
circular orbits at $\rgal = 8.5$ kpc (models 6 to 10) for comparison (Fig. \ref{fig:tnref}).
Middle: The mean stellar mass before (left) and at core collapse (right), 
again compared to the that of models 6 to 10.
Bottom: The resulting values of $\tnref$ (left) and $\tnrefcc$ (right) as a function of   
$\tzero M_i^\gamma$. The mean values of $\tnref = 13.3 $ Myr and $\tnrefcc = 7.2$ Myr are 
the same as for models 6 to 10 with circular orbits.
}
\label{fig:tnref-ecc}
\end{figure*}

Fig. \ref{fig:tnref-ecc} shows the values of \tzero, \mmean\ and \tnref\
of clusters in elliptical orbits (models 20 to 25) before and after core collapse respectively.
We have added the data of model 8 which has the same initial mass, $W_0$ and the same \Rgal\ but  
a circular orbit. The values are compared with those of clusters with $W_0=5$ at $\Rgal=8.5$ kpc
but with different masses (dashed lines), taken from Figs. \ref{fig:tnref}. 
Notice that the models in eccentric orbits have very similar characteristics
compared to those in circular orbits, if we correct for the effect of eccentricity by
a factor $(1-\epsilon)^{-1}$ in \tzero\ and \tzerocc. 
BM03 already concluded that the total life time 
of clusters is proportional to $(1-\epsilon)$. 
We find here that the 
``orbital averaged'' mass loss rates before and after core collapse both scale with
$(1-\epsilon)^{-1}$.

\begin{figure}
\epsfig{figure=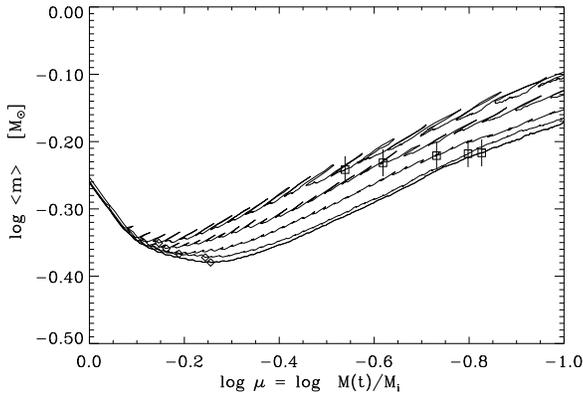,width=8.0cm} 
\caption[]{
The variation of the mean stellar mass as function of $\mu$ for clusters in
elliptical orbits. Lowest curve: $\epsilon=0.2$, upper curve: $\epsilon=0.8$.
 The wiggles are the result of the variation in 
the mass of the cluster within the variable tidal radius. 
Diamonds indicate the minimum values and squares indicate
the moment of core collapse. Notice that the mean mass at core collapse is about 
the same for all models.
}
\label{fig:mmeancc-ecc}
\end{figure}

The main difference between the cluster models in circular and elliptical orbits is in the
mean stellar mass at core collapse (compare the middle panels of Figs.  
\ref{fig:tnref} and \ref{fig:tnref-ecc}). For clusters in circular orbits \mmeancc\ increases
towards shorter lifetimes, whereas \mmeancc\ is about constant for clusters with the same
apogalactic radius but different ellipticities. This is because the BM03 models are 
Roche-lobe filling at perigalacticon so both  \trh\ and \tcc\ decrease steeply with
increasing eccentricity. The combination of a short lifetime (i.e. a higher maximum star mass 
at \tcc) and a smaller ratio of $\tcc / \tmig$ (i.e. fewer low mass stars are lost after
mass segregation) results in \mmeancc\ being almost independent of $\epsilon$.
This is illustrated  in Fig. \ref{fig:mmeancc-ecc}, which shows the variation of $\mmean$ with
$\mu$ and the mean mass at core collapse, which is almost constant.
 
The core collapse time of the clusters in elliptical orbits, shown in Fig. \ref{fig:tcc},
is slightly longer than predicted by Eq. \ref{eq:tcctidal} by about 20 percent. This is because the
values of \trh, listed in Table \ref{tbl:BM03models}, refer to perigalacticon, which is smaller than  the
orbital-averaged values of \trh.


\section{Initially Roche-lobe underfilling cluster models}
\label{sec:8}

\subsection{The parameters of the initially Roche-lobe underfilling models }
\label{sec:uf-table}
 
Because we are also interested in the mass history of initially highly concentrated clusters,
i.e. with a half-mass radius much smaller than their tidal radius,
we have supplemented the set of Roche-lobe filling cluster models with \nbody\
simulations of a series of clusters that start severely Roche-lobe underfilling.
These are for clusters with an initial mass in the range of 10400 to 84000 \Msun, 
in circular orbits at $\rgal=8.5$ kpc and with an initial density distribution described by 
a King parameter of $W_0=5$. The metallicity is $Z=0.001$. 
These cluster models have an initial half mass radius of $\rh=1$, 2
or 4 pc. The stellar IMF of these clusters is different from those of the Roche-lobe filling model.
They have a Kroupa mass function in the range of 0.10 to 100 \Msun, with an initial mean 
 stellar mass of 0.623 \Msun. 
In these models 10\% of the formed neutron stars and 
black holes are retained in the cluster. The models were calculated for this study.
The parameters of the models are listed in Table \ref{tbl:models-uf09}.

For the study of the mass loss by dissolution we define an ``underfilling factor'' $\mathfrak{F}$,
which is defined as the ratio of the initial half-mass radius of the cluster model, \rh, and the half-mass radius
$r_{\rm h}^{\rm rf}$ of a Roche-lobe filling cluster with the same King parameter $W_0$. 

\begin{equation}
\mathfrak{F}_{\rm W_0} \equiv \frac{\rh}{r_{\rm h}^{\rm rf}} = \frac{(\rh/\rt)_{\rm W_0}}{(\rh/\rt)^{\rm rf}_{\rm W_0}}
 = \frac{r_{\rm t}^{\rm W_0}}{r_J}
\label{eq:uffactor}
\end{equation}  
where $r_J= r_{\rm t}^{\rm rf}$ is the Jacoby radius, i.e the tidal radius of a Roche-lobe filling cluster, 
and $r_{\rm t}^{\rm W_0}$ is the end of the density profile
of a cluster with fixed values of $\rh$ and $W_0$. Note that $\mathfrak{F}=1$ for Roche-lobe 
filling clusters and $<1$ for Roche-lobe underfilling clusters.
In this expression
$(\rh / \rt)^{\rm rf}_5=0.187$ and $(\rh / \rt)^{\rm rf}_7=0.116$ are the ratios for a Roche-lobe filling cluster with 
an initial density distribution of $W_0=5$ and 7 respectively. 
(The same definition of $\mathfrak{F}$ was used by Gieles an Baumgardt (2008) in their theoretical study of the mass loss of
Roche-lobe underfilling clusters.)
The values of $\mathfrak{F}_5$ are listed in 
Table \ref{tbl:models-uf09}. We will show below that for the description of the dissolution
of the Roche-lobe underfilling models the parameter $\mathfrak{F}_7 = 1.612 \mathfrak{F}_5$ is more important than $\mathfrak{F}_5$.

\begin{table*}
\caption{The N-body models of initially Roche-lobe underfilling clusters. The left block of contains the model parameters, the
middle section the cluster time scales and the right block contains our fit parameters.}
\centering

\resizebox{\textwidth}{!}{

\begin{tabular}{r r r r r r r r | r  r  r r r | r r r r r r r}
\multicolumn{8}{l}{Input parameters} & \multicolumn{5}{l}{Timescales} & \multicolumn{7}{l}{Output parameters} \\
 $\#$  & Mass      & nr & $W_0$ & $R_{\rm Gal}$ & $r_t$ & $r_{\rm h}$ & $\mathfrak{F}_5$& $t_{\rm 1\%}$ & $t_{\rm cc}$ & $t_{\rm rh}$ & $t_{\rm cr}(r_{\rm h})$ & $t_{\rm cr}(r_{\rm t})$ &
 $\gamma$ &  $t_0$ & $\tmig$ & $t_{\rm del}$ & $f_{\rm ind}^{\rm max}$  & $\tzeropostcc$ & $j_{\rm cc}$     \\ 
     & $M_{\odot}$ & stars &    & kpc        &  pc    & pc &       & Gyr       & Gyr     & Gyr        & Myr
     & Myr  &     &  Myr  & Gyr     &  Myr       &    &  Myr  &    \\ \hline

  uf1 & 10831 &  16k & 5& 8.5& 32.6 & 1.00 & 0.164 & 7.22 & 3.00: & 0.045 & 0.57 & 75.5 & 0.80 & 5.1 &  8.6 &   50 & 0.0 & 10.5 & 1.09 \\ 
  uf2 & 10426 &  16k & 5& 8.5& 32.2 & 2.00 & 0.332 & 7.59 & 3.50: & 0.127 & 1.65 & 75.5 & 0.80 & 6.2 & 10.2 &  100 & 0.0 & 10.4 & 1.34 \\ 
  uf3 & 10589 &  16k & 5& 8.5& 32.4 & 4.00 & 0.660 & 5.89 & 5.30: & 0.361 & 4.64 & 75.5 & 0.80 & 5.0 &  8.3 &  100 & 0.2 &  8.4: & 1.15\\ 
  uf4 & 21193 &  32k & 5& 8.5& 40.8 & 1.00 & 0.131 &11.42 & 3.36  & 0.058 & 0.41 & 75.5 & 0.80 & 5.0 & 14.5 &  200 & 0.0 & 11.1 & 1.16 \\
  uf5 & 21095 &  32k & 5& 8.5& 40.7 & 2.00 & 0.263 &13.40 & 7.00  & 0.165 & 1.16 & 75.5 & 0.80 & 6.0 & 17.3 &  350 & 0.0 & 12.8 & 1.11 \\ 
  uf6 & 20973 &  32k & 5& 8.5& 40.7 & 4.00 & 0.526 &12.75 & 9.30  & 0.466 & 3.29 & 75.5 & 0.80 & 6.0 & 17.2 &  200 & 0.1 & 10.6 & 1.24 \\ 
  uf7 & 41465 &  64k & 5& 8.5& 51.0 & 1.00 & 0.105 &17.79 & 3.73  & 0.076 & 0.29 & 75.5 & 0.80 & 5.0 & 24.7 &  300 & 0.0 & 10.8 & 1.18 \\ 
  uf8 & 40816 &  64k & 5& 8.5& 50.8 & 2.00 & 0.211 &20.76 & 8.32  & 0.212 & 0.83 & 75.5 & 0.80 & 6.5 & 31.7 &  800 & 0.0 & 11.8 & 1.44 \\
  uf9 & 42114 &  64k & 5& 8.5& 51.3 & 4.00 & 0.417 &21.18 &12.65  & 0.608 & 2.32 & 75.5 & 0.80 & 6.0 & 30.0 &  800 & 0.0 & 11.5 & 1.31 \\ 
 uf10 & 83853 & 128k & 5& 8.5& 64.5 & 2.00 & 0.116 &34.77 &10.07  & 0.282 & 0.58 & 75.5 & 0.80 & 7.0 & 60.8 & 1200 & 0.0 & 13.7 & 1.51 \\ 
 uf11 & 83700 & 128k & 5& 8.5& 64.5 & 4.00 & 0.332 &36.58 &18.40  & 0.796 & 1.65 & 75.5 & 0.80 & 7.2 & 62.4 & 1500 & 0.0 & 11.9 & 1.67 \\ 
  \end{tabular}
}
Number of stars: $1k=1024$.
\label{tbl:models-uf09}
\end{table*}

\subsection{Dissolution before core collapse}
\label{sec:8.2}

The mass loss of the Roche-lobe underfilling models can be described in the same way as for the Roche-lobe filling models.
The resulting fitting parameters are listed in the right hand block of Table \ref{tbl:models-uf09}.

The first phase is dominated by stellar evolution. However in this case there is no evolution-induced mass loss.
This is because the clusters are initially well within their tidal limit. So the mass loss by stellar evolution
does produce an expansion of the radius, but this expansion does not immediately 
reach the tidal radius. This is reflected in
the values of $\findmax =0.0$ for all models, except uf3 and uf6. These two are the least Roche-lobe underfilling models
with $\mathfrak{F}_5=0.66$ and 0.53 respectively.
The value of \findmax\ can then be described in a similar way as for Roche-lobe filling clusters, 
Eq. \ref{eq:findmax}, but 
with a correction term that depends on the underfilling factor,

\begin{equation}
\findmax =  -0.86+ 0.40 \times \log (\tmig)+2.75 \times \log \mathfrak{F}_5 
\label{eq:findmax-uf}
\end{equation}
with a maximum of 1 and a minimum of 0 and $\gamma=0.80$ (see below).

The dissolution before core collapse can be expressed by a power law approximation of 
$\dmdtdisnorm$ versus $M$, with $\dmdtdisnorm$ defined by Eq. \ref{eq:dmdtnorm}. 
The lower left panel of Fig. \ref{fig:gamma} shows the relation between $\log(\dmdtdisnorm)$ and
log $(M)$. Because the core collapse time of the Roche-lobe underfilling models is short, each model
contributes only to a small part of the mass range. 
We found that $1- \gammauf=0.20 \pm 0.05$ gives a good fit to these plots, so we adopted $\gammauf=0.80$.
We note that this value is the same 
as for the Roche-lobe filling clusters with $W_0=7$, whereas the initially Roche-lobe underfilling clusters have an initial
concentration of $W_0=5$. However, a study of the expansion of severely Roche-lobe underfilling clusters
by means of \nbody - simulations has shown that the initial expansion due to mass loss redistributes the
density close to that of a $W_0=7$ model, in agreement with our 
derived value of $\gammauf=0.80$.

The tendency of clusters to evolve to a $W_0 \simeq 7$ model was 
noticed by Portegies Zwart et al. (1998).
In initially strongly concentrated clusters, $W_0>7$, dynamical friction quickly drives the massive stars
to the center where they will lose mass due to stellar evolution. This results in an expansion of the core
and a less steep density profile, so $W_0$ decreases. On the other hand, in clusters with a less steep
initial concentration, $W_0 \le 5$, dynamical friction is less efficient and the massive stars lose
mass by stellar evolution before they reach the center. 
So the cluster expands more homogeneously due to evolutionary mass loss.
After the massive stars have undergone stellar evolution, dynamical effects take over and the core
shrinks due to the approaching core collapse,  
so the density distribution becomes more concentrated and $W_0$ increases.

Once the cluster has expanded to about the tidal radius, the dissolution is very similar to that of
a Roche-lobe filling cluster with $W_0=7$. This is reflected in the values of \tzerouf\ which range from
5.4 to 7.5 Myr, whereas the values of \tzero\ for the comparable Roche-lobe filling models, nrs 16 to 19,
range from 6.0 to 6.5 Myr. Part of the difference is due to differences in the mean stellar mass,
because the two sets of models have different IMFs.  

\begin{figure}
\centerline{\epsfig{figure=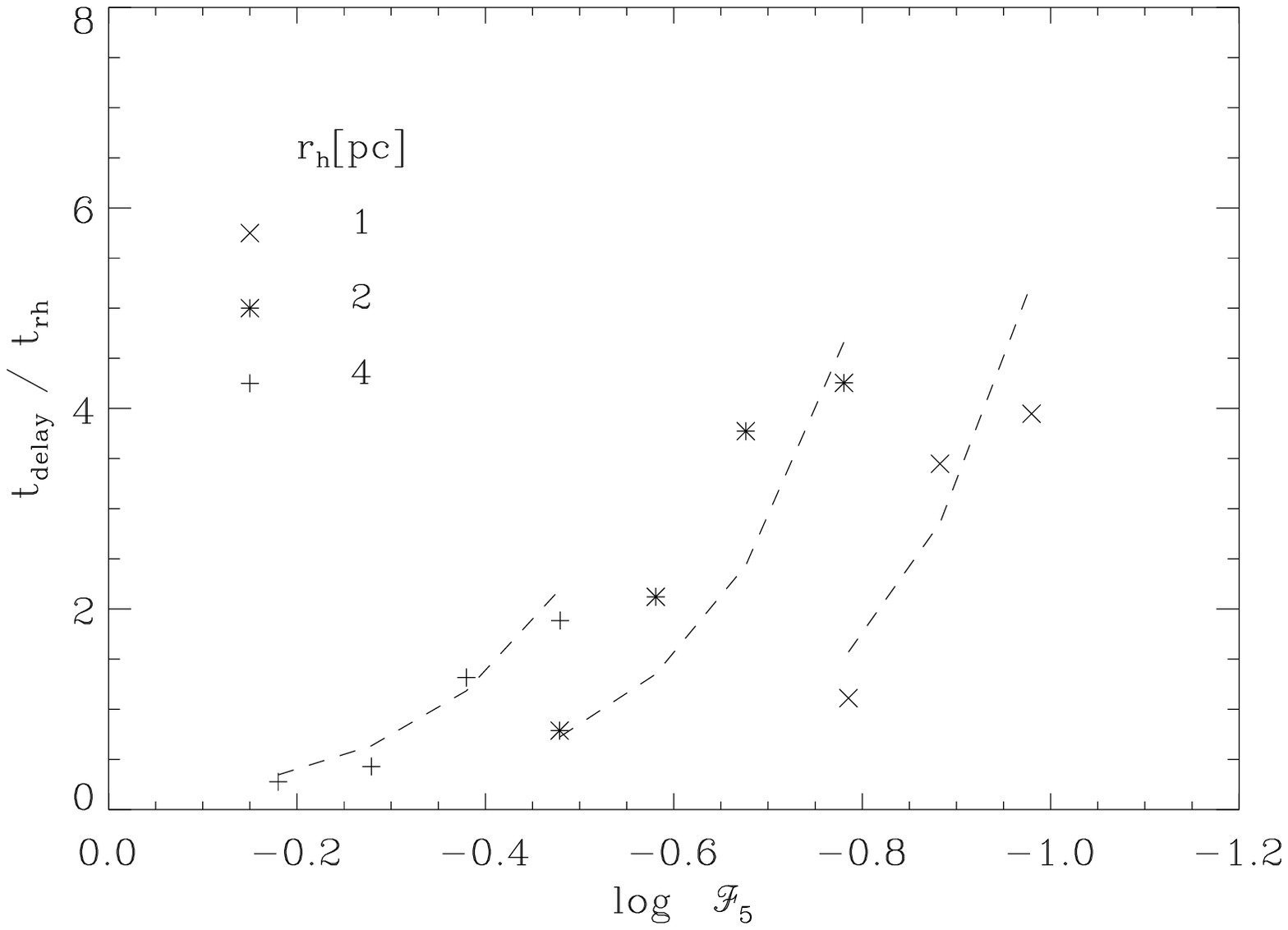, width=8.0cm}}
\caption[]
{The ratio between the delay time $\tdelay$ for the establishment of dissolution and
the half mass relaxation time \trh\ for initially Roche-lobe underfulling clusters. The dashed lines show the fit
of Equ. \ref{eq:tdelaytrh}.}
\label{fig:tdelay-uf}
\end{figure}

Dissolution of the Roche-lobe underfilling models needs more time to get started than Roche-lobe filling models, 
because the clusters
first have to expand to the tidal limit. This is a slow process that occurs on the relaxation time scale.
This is shown in Fig. \ref{fig:tdelay-uf} which shows the ratio $\tdelay / \trh$ as a function of the
Roche-lobe underfilling factor $\mathfrak{F}_5$, for different values of \rh.
The values of $\tdelay /\trh$ range from about 0 to 4 for the models considered here.
The figure shows that we can approximate 

\begin{equation}
\tdelay  ~ \simeq~ 4.31~10^{-3}\times  (\mathfrak{F}_5)^{-1.989} ~ \trh^{1.605}
\label{eq:tdelaytrh}
\end{equation}
with all ages in Myr.
This equation is valid for $ -1.0 < \log( \mathfrak{F}_5) < -0.20$. For Roche-lobe filling clusters 
with  $ \log( \mathfrak{F}_5) > -0.20$ the delay time does not scale with 
\trh\ but with the crossing time at the tidal radius and $\tdelay \simeq 3.0~ \tcrt$ (Sect. \ref{sec:4}).
So, for strongly Roche-lobe underfilling models the delay time scales with the initial value of $\trh$, 
but if the underfilling factor approaches $\mathfrak{F}=1$ the delay time is much shorter and scales with $\tcrt$.

The dissolution before core collapse can now be expressed by Eq. \ref{eq:dmdtdis+fdelay}
with \fdelay\ given by Eq. \ref{eq:fdelay} and \tdelay\ approximated by Eq. \ref{eq:tdelaytrh}.

\begin{figure*}
\centerline{\hspace{+3.0cm}\epsfig{figure=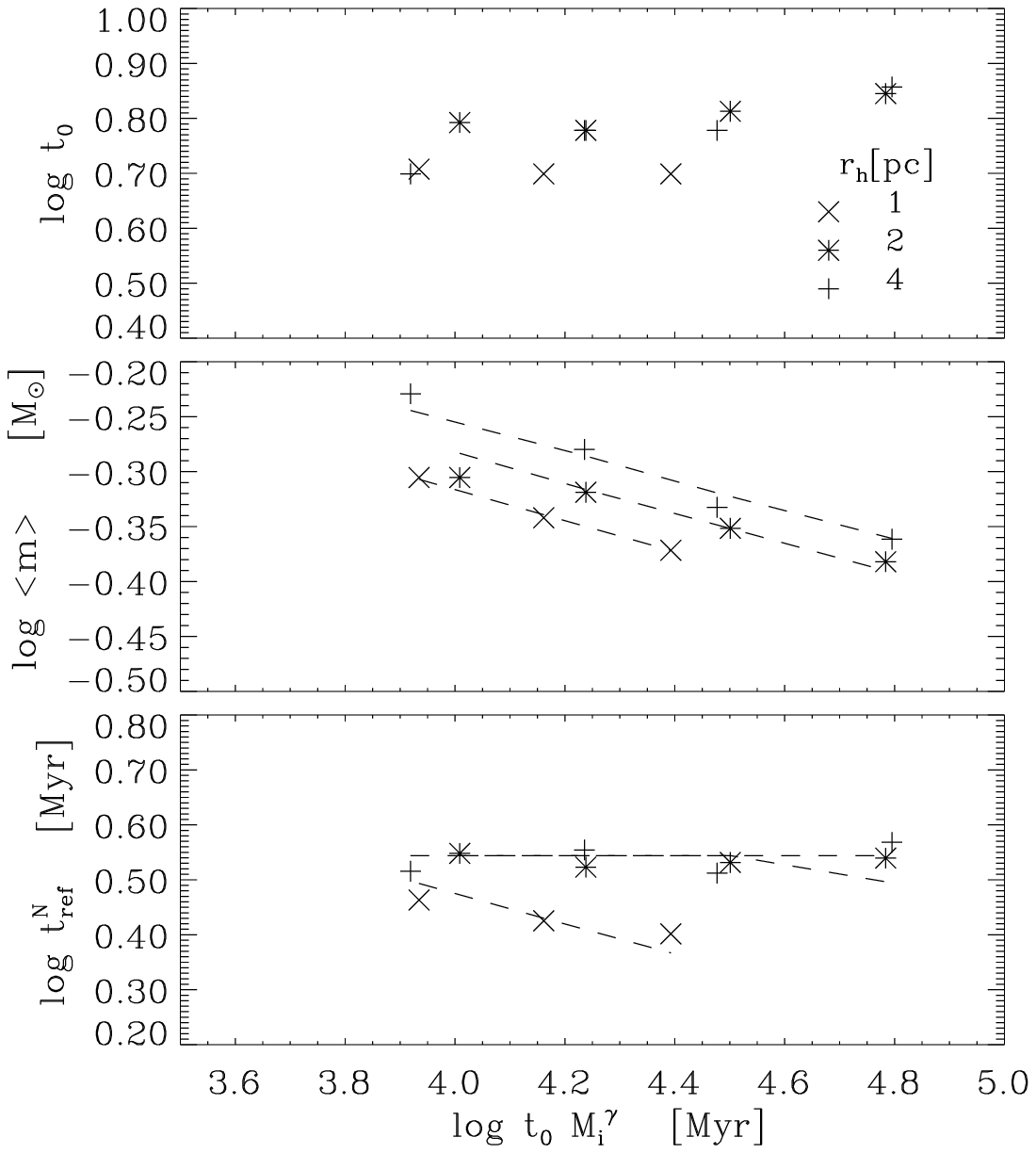,width=12.0cm} \hspace{-4.0cm}
\epsfig{figure=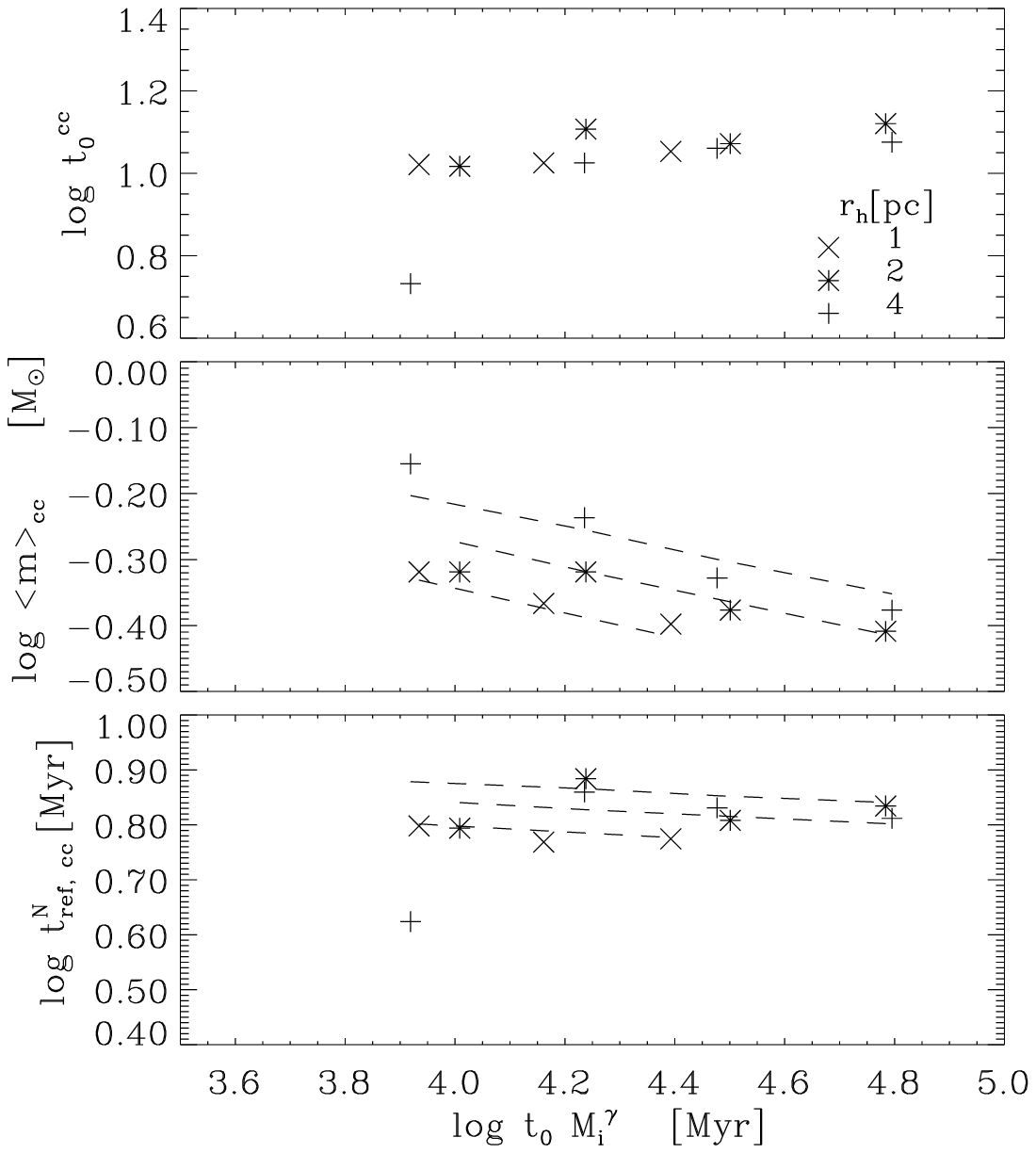,width=12.0cm}} 

\caption[]{The parameters of the models of initially Roche-lobe underfilling clusters, nrs uf1 to uf11 
before (left) and after core collapse (right).
Top: The relation between $\tmig$  and \tzerouf\ (left) or \tzeroccuf\ (right), 
i.e. after core collapse. 
Middle: The mean stellar mass before (left) and at core collapse (right).
The dashed lines indicate mean relations of the different values of \rh.
Bottom: The resulting values of $\tnrefuf$ (left) and $\tnrefccuf$ (right) as a function of   
$\tzero M_i^\gamma$. The mean relations are given in Sect. \ref{sec:8.2} and 
\ref{sec:8.3}.
}
\label{fig:tnref-uf}
\end{figure*}

Fig. \ref{fig:tnref-uf} (left) shows the values of \tzerouf, the mean mass before core collapse \mmean\
and the resulting values of \tnrefuf\ as a function of \tmig.
 We used different symbols for different initial half mass radii.

The mean stellar mass before core collapse depends both on $\tmig$ and on $\mathfrak{F}$ and not only on 
$\tmig$ as is the case for the Roche-lobe filling clusters. This is because the core collapse time $\tcc$
depends on $\mathfrak{F}$  and so the amount of mass that the cluster has lost before core collapse 
also depends 
on both \tmig\ and \trh.
We find that we can express the mean stellar mass before core collapse, i.e. between $t(\mu=0.25)$ and
\tcc, as 

\begin{equation}
\log \mmean ~=~ 0.139 -0.0984 \log (\tmig) + 0.101 \log (\mathfrak{F}_7)
\label{eq:mmean-precc-uf}
\end{equation}
We have expressed \mmean\ in terms of $\mathfrak{F}_7$ (with $\mathfrak{F}_7=1.612 \mathfrak{F}_5$, Eq. \ref{eq:uffactor}) 
instead of the initial value of $\mathfrak{F}_5$ 
because the initial expansion of the clusters redistributes the density to about
a $W_0=7$ model (see above). 
The relation has about the same dependence on \tmig\ as in the case of 
Roche-lobe filling clusters (Eq. \ref{eq:mmeanprcc}).

The lower left part of Fig. \ref{fig:tnref-uf} shows the resulting values of \tnrefuf.
For Roche-lobe filling clusters we found one value of $\tnref=3.5$ Myr ($\log \tnref =0.544$) 
for all models of $W_0=7$ (see Fig. \ref{fig:tnref}). The Roche-lobe underfilling models show a large range in \tnrefuf,
indicating that the underfilling factor plays is a role for small values of $\mathfrak{F}_7$. 
We found that we can approximate

\begin{eqnarray}
\log (\tnrefuf)& =& 0.544 ~~~{\rm if}~ \log(\mathfrak{F}_7)>-0.50 \nonumber \\
               &  & 0.544 + 0.65 \times \{\log(\mathfrak{F}_7) +0.50\} \nonumber \\
               &  &  ~~{\rm if}~ \log(\mathfrak{F}_7)<-0.50
\end{eqnarray}
This shows that there is a smooth transition between the dissolution parameter for Roche-lobe filling and 
underfilling clusters.
The fits are shown as three partially overlapping dashed lines for models of \rh=1, 2 and 4 pc 
in the lower part of Fig. \ref{fig:tnref-uf}.

\subsection{Dissolution after core collapse}
\label{sec:8.3}

\begin{figure}
\epsfig{figure=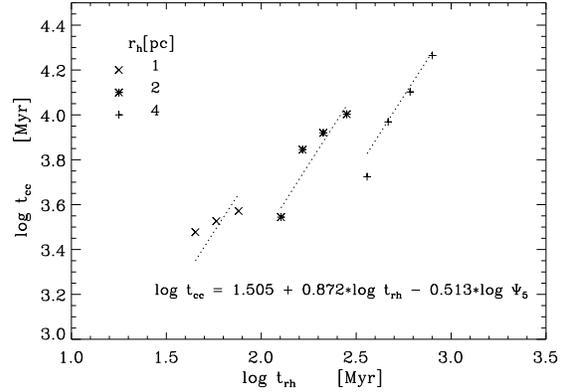, width=8.0cm}
\caption[]
{The core collapse time of the Roche-lobe  underfilling models as a function of 
the initial half mass relaxation time and the underfilling factor $\mathfrak{F}_5$.
}
\label{fig:tcc-uf}
\end{figure}

The time of core collapse is expected to scale with the half mass relaxation time.
Fig. \ref{fig:tcc-uf} shows $\tcc$ as a function of \trh\ and the 
initial Roche-lobe underfilling factor $\rh / \rt$. In Sect. \ref{sec:6.1} we found that 
$\tcc \propto \trh^{0.872}$ for Roche-lobe filling clusters.
 We find that for Roche-lobe underfilling clusters 
the dependence of $\tcc$ on $\trh$ can be described by the same power law
dependence, but there is an additional dependence on the underfilling factor

\begin{equation}
\log (\tcc) ~\simeq~ 1.505 ~+~ 0.872 \log(\trh) ~-~ 0.513 \log (\mathfrak{F}_5)
\label{eq:tcc-uf}
\end{equation}
(In this case $\mathfrak{F}_5$ is the crucial parameter, rather than $\mathfrak{F}_7$, because the initial 
relaxation time \trh\ is defined for the initial density distribution with $W_0=5$.) 
The smaller the underfilling factor, the larger \tcc\ for a given value of \trh. 
This is because clusters that are initially strongly Roche-lobe underfilling expand
more strongly and so $\trh (t)$ increases more strongly with time, which results in a larger
ratio of $\tcc / \trh$. In the limit of $\mathfrak{F}_5=1$, i.e. for Roche-lobe filling clusters 
Eq. \ref{eq:tcc-uf} predicts that
$ \log (\tcc) \simeq 1.50 + 0.872 \log(\trh)$. This can be compared with the value for 
Roche-lobe filling clusters of $ \log (\tcc) \simeq 1.23 + 0.872 \log(\trh)$ (Eq. \ref{eq:tcctidal}). 
So Roche-lobe underfilling clusters need about twice as many ``initial'' relaxation times to go into
core collapse as Roche-lobe filling clusters. 
This is because of their stronger expansion
and the fact that they do not push stars over the tidal boundary while evolving towards
core collapse.

A study of the relation between $\dmdtdisnorm$ and $M(t)$ of the various Roche-lobe underfilling models shows that 
the dissolution after core collapse can be  described by the same values of $\gammacc$
as for the Roche-lobe filling cluster models, i.e $\gammacc=0.70$ if $M(t)>10^3$ \Msun\ and
$\gammacctwo=0.40$ if $M(t)<10^3$ \Msun\ (see lower right panel of Fig. \ref{fig:gamma}).
This is due to the fact that core collapse results in a redistribution of the density
in the cluster and erases the memory of the pre-collapse phase.

The values of $\tzerocc$ of the Roche-lobe underfilling models are listed in Table \ref{tbl:models-uf09}.
They are plotted versus \tmig\ in the right hand panel of Fig. \ref{fig:tnref-uf}
together with the mean stellar mass at core collapse and the resulting values
of $\tnrefcc$.
The dissolution parameter after core collapse when $M(t)<10^3 \Msun$ is
$\tzerocctwo = \tzerocc \times 10^{3(\gammacc - \gammacctwo)} =10^{0.90} ~ \tzerocc$.

The mean mass at core collapse can be approximated by

\begin{equation}
\log(\mmeancc) = 0.178 -0.0984 \log(\tmig)+0.207 \log(\mathfrak{F}_7)
\label{eq:mmeancc-uf}
\end{equation}
The relation has the same slope as that of the Roche-lobe filling clusters after core collapse
(Eq. \ref{eq:mmeanpostcc}) but the constant 0.178 for clusters with $\mathfrak{F}_7=1$ is different
from the 0.075 of Roche-lobe filling clusters with $W_0=7$, which indicates
a higher mean mass at core collapse.
This is mainly due to differences in the IMF of the two sets of models.

Fig. \ref{fig:tnref-uf}b shows that $\tnrefcc= \tzerocc \times \mmeancc^{\gamma}$ with $\gamma=0.70$
for these models 
is about constant but with a significant scatter. This shows that there is a small residual
effect of the initial underfilling factor in the dissolution after core collapse. We found that we 
can approximate

\begin{equation}
\log(\tnrefcc) = 0.875 + 0.127 \times \log(\mathfrak{F}_7)
\label{eq:tnrefccuf}
\end{equation}
The constant $0.875$ is slightly larger than the value of $0.796$ for Roche-lobe filling
models of $W_0=7$. 

\begin{figure}
\epsfig{figure=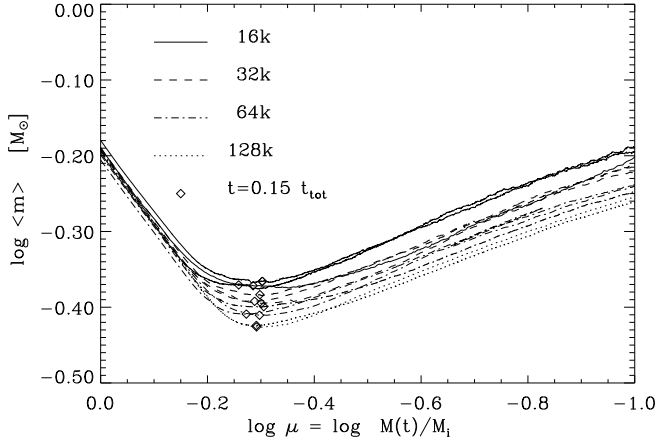,width=9.0cm}
\caption[]
{The evolution of the mean stellar mass in the Roche-lobe underfilling clusters as a function
of $\mu$. The diamonds indicate the moment when $t=0.15 \ttot$, at which time the mean mass reaches
its minimum value due to mass segregation.}
\label{fig:mmeanuf}
\end{figure}

Fig. \ref{fig:mmeanuf} shows the history of the mean stellar mass in the Roche-lobe underfilling cluster 
models as a function of $\mu$. The trends are approximately the same as in Fig. \ref{fig:mmean},
i.e. an initial decrease due to the loss of massive stars by stellar evolution followed by an 
increase of \mmean\ after mass segregation has been established and low mass stars are lost 
preferentially. 
The difference is due to the different initial stellar IMF, which have an initial 
$\mmean=0.623$ \Msun\ instead of 0.547 for the Roche-lobe filling models. 
Moreover, in the Roche-lobe underfilling models 90$\%$ of the black holes and neutron stars is ejected
whereas they were all retained in the Roche-lobe filling models. 
Both effects influence the evolution of $\mmean$ (Kruijssen 2009).
The decrease of \mmean\ is 
stronger than for the Roche-lobe filling models because the IMF of the Roche-lobe underfilling
models reaches up to 100 \Msun , whereas the IMF of the other models reaches to 15 \Msun.
We see that \mmean\ of the Roche-lobe underfilling models reaches its minimum at the same value of 
$t \simeq 0.15 t_{\rm tot}$ as the other models.
This shows that full mass segregation is reached at the same fraction of the total lifetime,
independent of the initial radius. 

\section{The relation between the cluster  lifetime and \tmig }
\label{sec:9}

The lifetime of clusters depends on 
time scales for mass loss by stellar evolution, and the dissolution constants \tzero\ and \tzeropostcc\
before and after core collapse. Stellar evolution dominates the mass loss only during the first part
of the cluster lifetime and typically removes about 20 to $40\%$ of the cluster mass,
after which dissolution takes over.
Since the dissolution timescales before and after core collapse both depend
on the strength of the tidal field in which the cluster moves, 
we may expect that the cluster lifetime depends largely on $\tzero \Mi ^\gamma$.

\begin{figure}
\centerline{\epsfig{figure=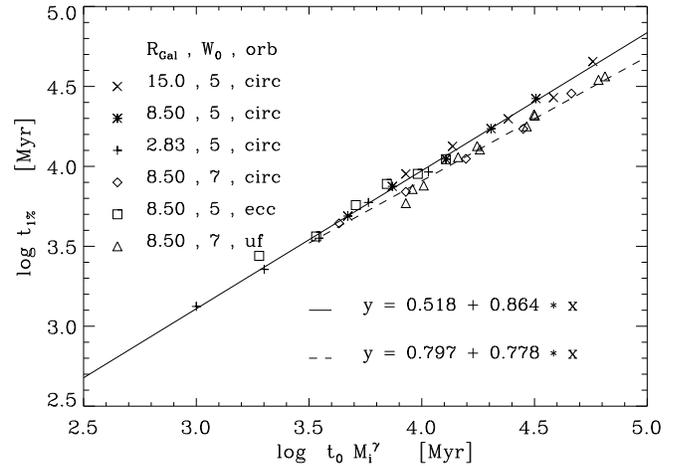,width=9.5cm}}
\caption[] {The relation between $\tzero \times \Mi^\gamma$ and \tone\ for all Roche-lobe filling 
and underfilling cluster models in circular and elliptical orbits.
The tight relation can be described by two linear equations  (\ref{eq:tone-tmig}) 
for clusters with $W_0=5$ (full line)
and $W_0=7$ (dashed line).}
\label{fig:tone-tmig}
\end{figure}

This is confirmed in Fig. \ref{fig:tone-tmig}, which shows a very tight relations between 
$\tzero \times \Mi^\gamma$ and \tone. For clusters with an initial density concentration described by
$W_0=5$ and $W_0=7$ in circular and elliptical orbits we find
 
\begin{eqnarray}
\log ( \tone)& = & 0.518 + 0.864 \times \log ( \tzero \Mi^{0.65})~~{\rm if}~W_0=5 \nonumber \\
             & = & 0.797 + 0.778 \times \log ( \tzero \Mi^{0.80})~~{\rm if}~ W_0=7
\label{eq:tone-tmig}
\end{eqnarray}
The relation between $\tmig$ (with $\gamma=0.80$) and \tone\ for the Roche-lobe underfilling clusters
is indistinguishable from that of the Roche-lobe filling clusters of $W_0=7$.
The tight correlations show that \tmig\ 
can be used as accurate 
indicator of the lifetime of a cluster.

Equation \ref{eq:tone-tmig} might suggest that the lifetime of a cluster is proportional
to $\Mi^{0.56}$ for $W_0=5$ clusters and $\Mi^{0.62}$ for $W_0=7$ cluster. However, we remind that
$\tzero$ is proportional to $\mmean^{- \gamma}$ (Eq. \ref{eq:tnrefdef}) and 
$\mmean \propto (\tmig)^{-0.121}$ and $(\tmig)^{-0.094}$ (Eq. \ref{eq:mmeanprcc}) for $W_0=5$ and 
7 respectively. This implies that $\tone \propto \Mi^{0.61}$ and $\Mi^{0.67}$ for $W_0=5$ and 7,
in agreement with the values of the indices 0.62 and 0.67 derived by BM03.

\section{The predicted mass history of star clusters}
\label{sec:10}

In the previous sections we have described the interplay between the different mass loss processes
of star clusters. Based on these results we derived a recipe for calculating the mass evolution
of star clusters in different environments. The recipe is described in Appendix A.

\subsection{The contribution of different effects to the mass loss}
\label{sec:10.1}

\begin{figure}
\centerline{\hspace{+3.5cm}\epsfig{figure=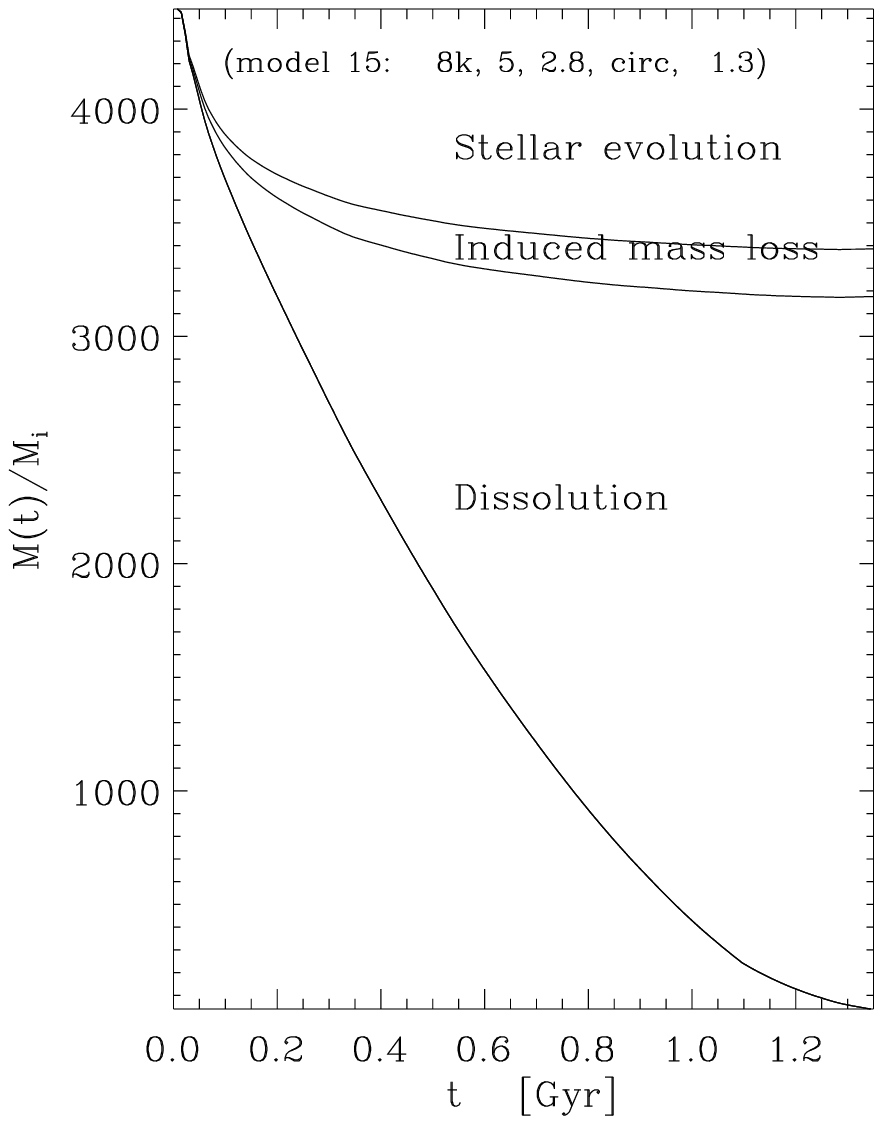, width=8.0cm}\hspace{-3.5cm}
            \epsfig{figure=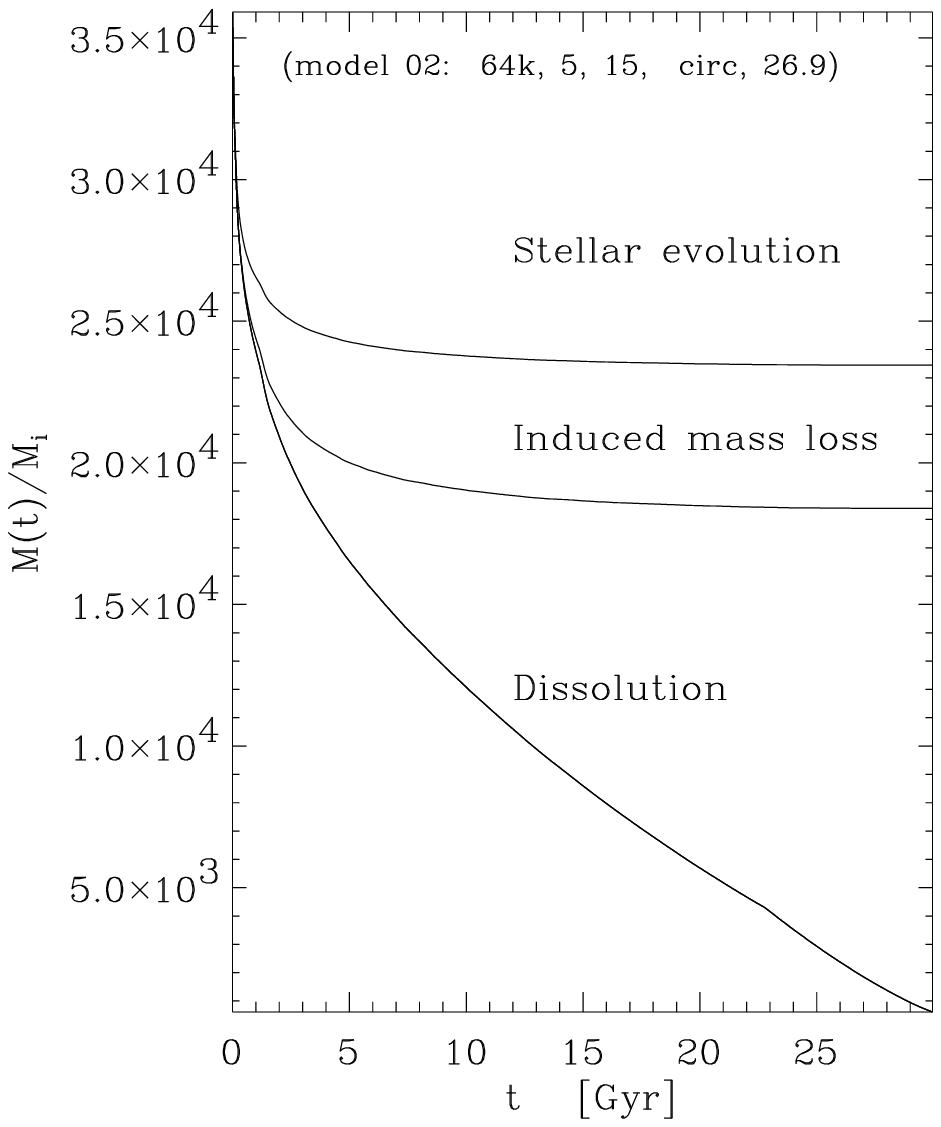, width=8.0cm}}
\caption[]{The contribution of different mechanisms to the mass loss of two cluster
models with a short (model 15, left) and a long (model 2, right) lifetime. 
The induced mass loss is small for clusters with a short lifetime of $\tone=1.3$ and 26.9 
Gyr respectively.}
\label{fig:contributions}
\end{figure}

The mass loss from star clusters is due to several effects: stellar evolution, evolution-induced
loss of stars, and dissolution (relaxation-driven mass loss) before and after core collapse.
Fig. \ref{fig:contributions} shows the contributions of these different effects 
for two characteristic models, $\#$ 15 which has a lifetime of $\tone=1.3$ Gyr and $\#$ 2 with $\tone=26.9$ Gyr.
The two models show that clusters with a long lifetime ($>20$ Gyr) lose about 35\% of their mass 
by stellar evolution,
15\% by induced mass loss and the remaining 50\% by dissolution. Clusters with a short lifetime
($<5$ Gyr) lose more than 60\% by dissolution, less than about 30\% by stellar evolution, and less than 10\%
by induced mass loss. This is because the short lifetime is the ``result'' of a strong mass loss
by dissolution, which does not leave much time for the cluster to lose a large fraction of its mass
by stellar evolution.

\subsection{Predicted M(t) of the total mass of clusters}
\label{sec:10.2}

We have calculated the $M(t)$ history of all cluster models listed in Table \ref{tbl:BM03models} 
with the recipe described in Appendix A.  A subset of the results is
shown in Fig. \ref{fig:M(t)-comparison}. 
The sample shown contains models with $W_0=5$ with a large (64k or 128k) and small (8k) number of stars
at respectively $\Rgal=15$, 8.5 and 2.63 kpc; two models with $W_0=7$ (128k and 8k); two models
in elliptical orbits ($\epsilon=0.2$ and 0.5); and five Roche-lobe underfilling models with different
numbers of stars and initial half mass radii.
The agreement is good for all models of clusters
in circular and elliptical orbits, with initial concentrations $W_0=5$ and 7, and for the
Roche-lobe underfilling clusters, including the ones not shown here. 
For cluster models in the original BM03 sample that are not discussed in this paper, the
agreement is equally good.

The different models have different shapes of $M(t)/\Mi$ versus $t/\tone$. 
Clusters with a long lifetime ($\tone > 20$ Gyr, high \Mi) show a strong drop in mass during the first 5\% 
of their life, due to stellar evolution and induced mass loss, followed by a more gentle decrease.
Clusters with a short lifetime ($\tone < 10$ Gyr, low \Mi) show a more gradual concave shape.
All models show a bump in the $M(t)$-plot near the core collapse time: the mass loss rate is
about twice as high after core collapse than before.

The shapes of the $M(t)$ relations are all convex with various degrees of curvature.
Only those cluster models for which core collapse occurs about halfway through their
lifetime show a more or less linear mass history (e.g. models 16, uf09 and uf10).

\begin{figure*}
\centerline{\epsfig{figure=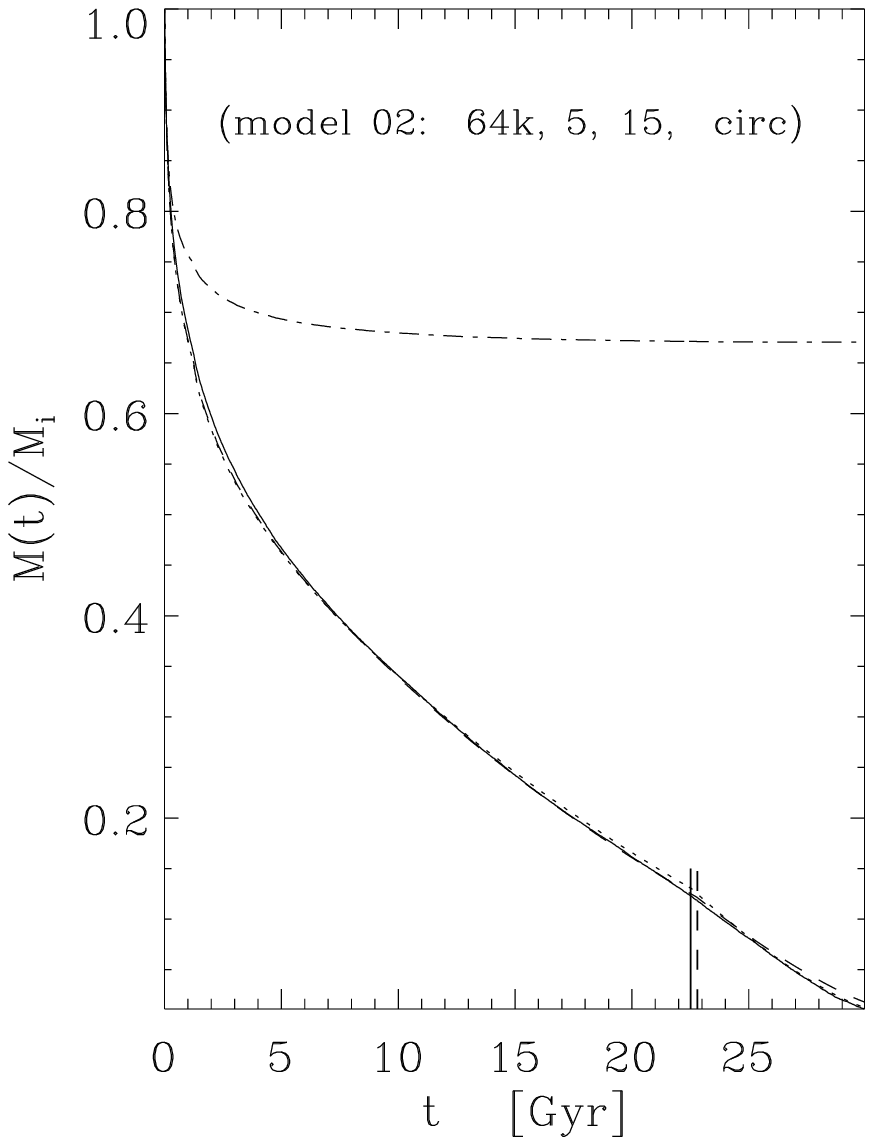,width=7.5cm}\hspace{-4.4cm}
            \epsfig{figure=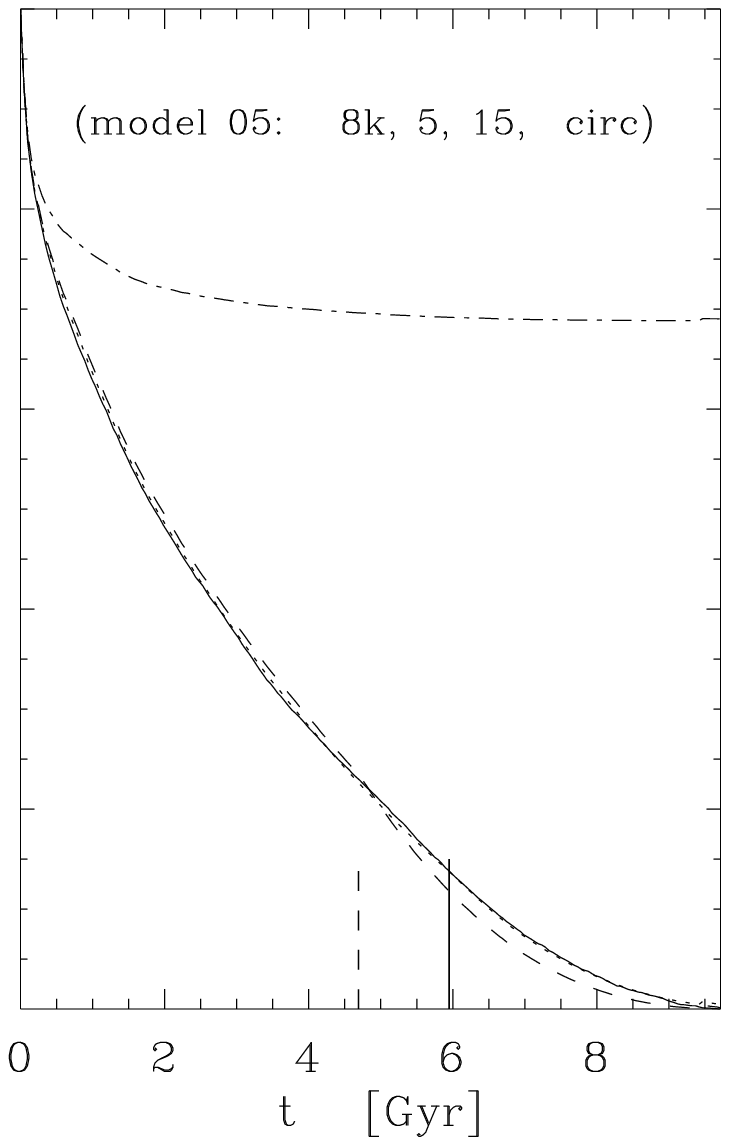,width=7.5cm}\hspace{-4.4cm}
            \epsfig{figure=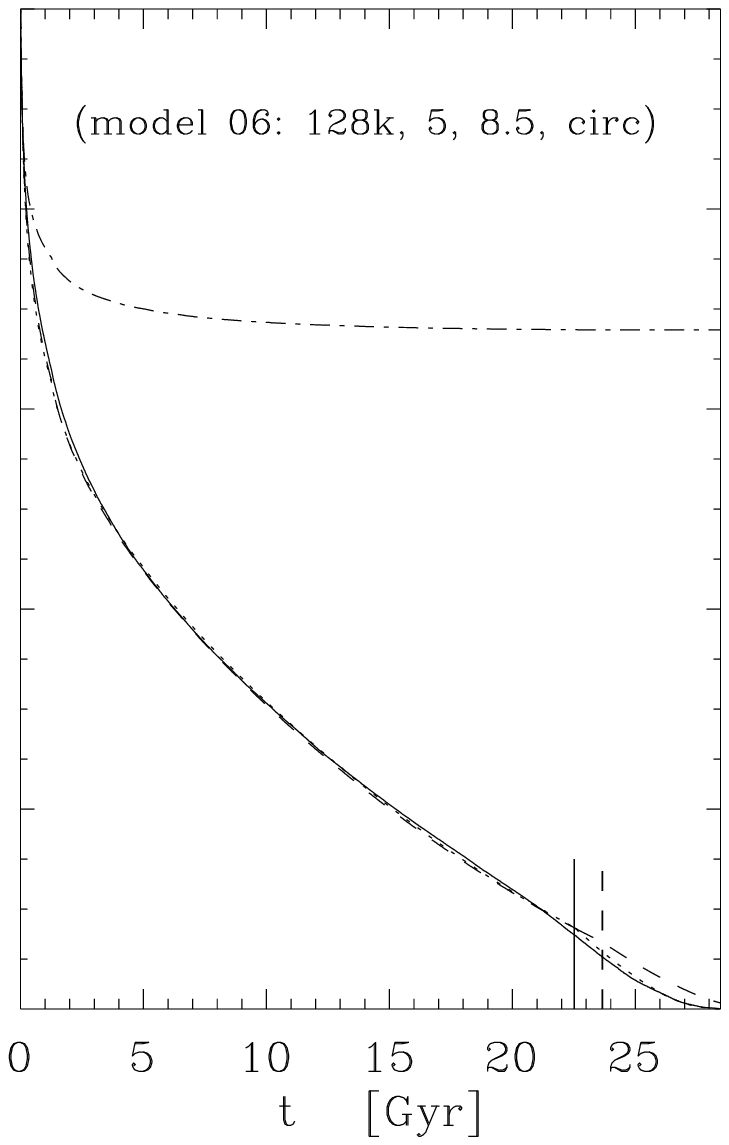,width=7.5cm}\hspace{-4.4cm}
            \epsfig{figure=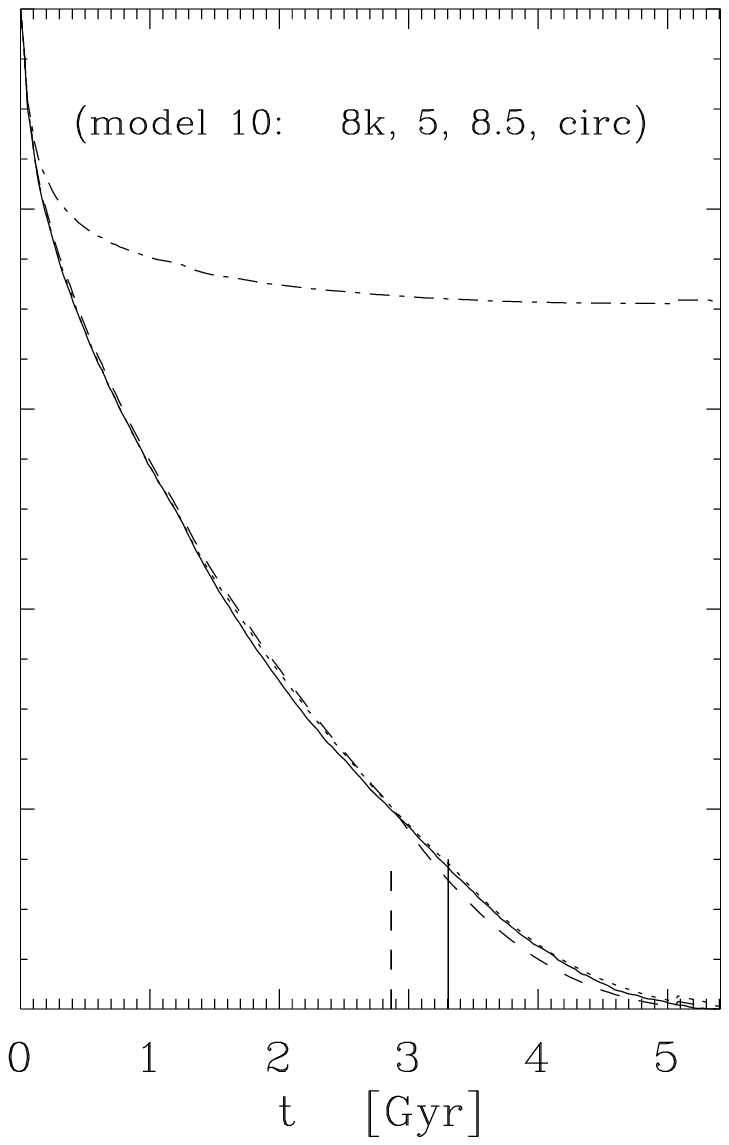,width=7.5cm}\hspace{-4.4cm}
            \epsfig{figure=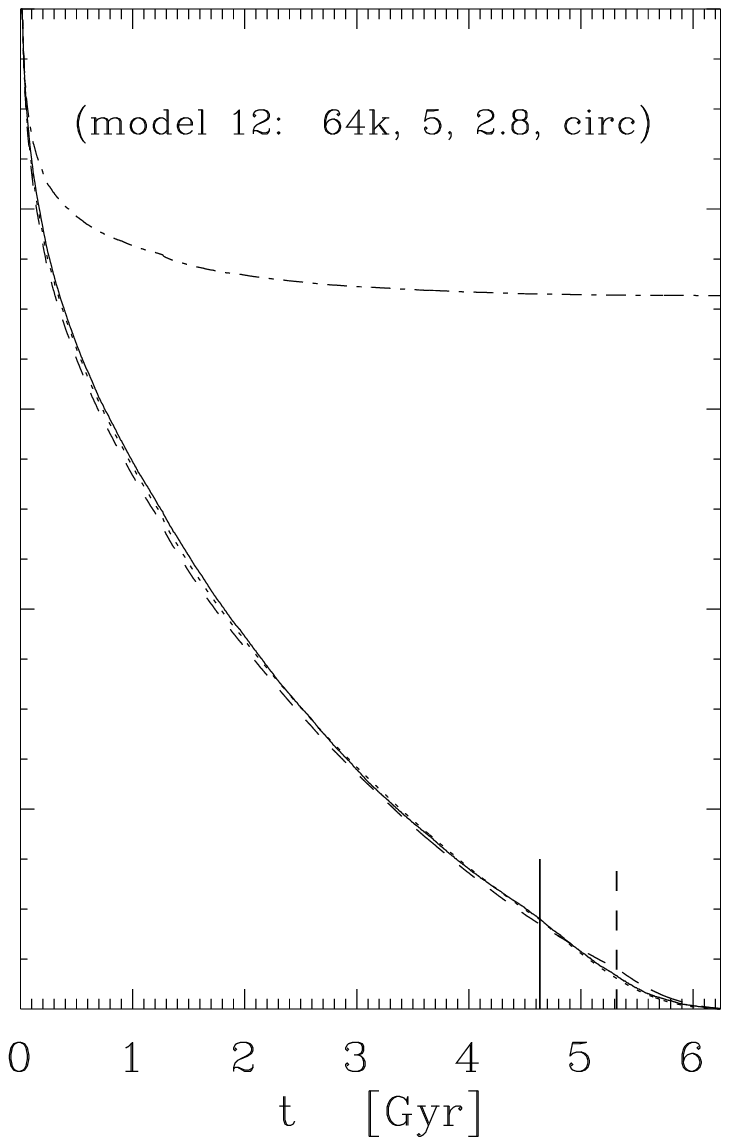,width=7.5cm}\hspace{-4.4cm}}
\vspace{-0.3cm}
\centerline{\epsfig{figure=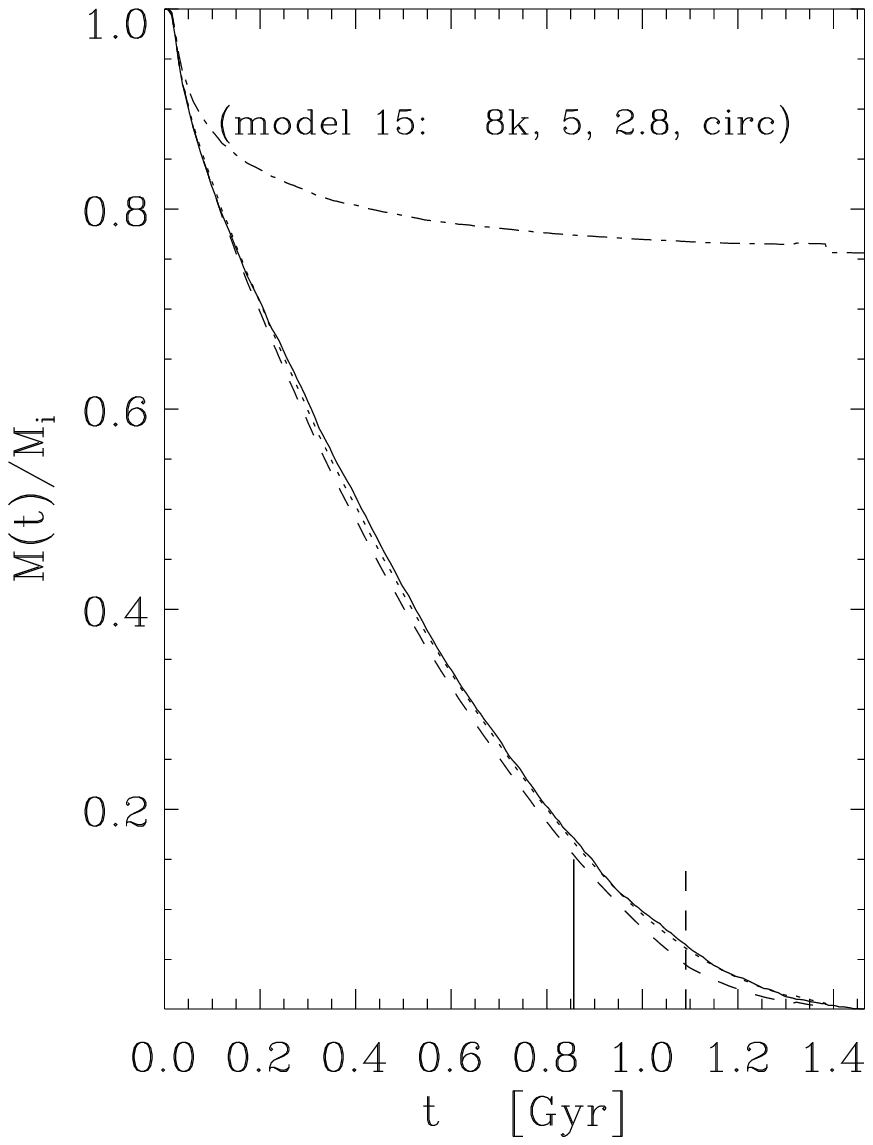,width=7.5cm}\hspace{-4.4cm}
            \epsfig{figure=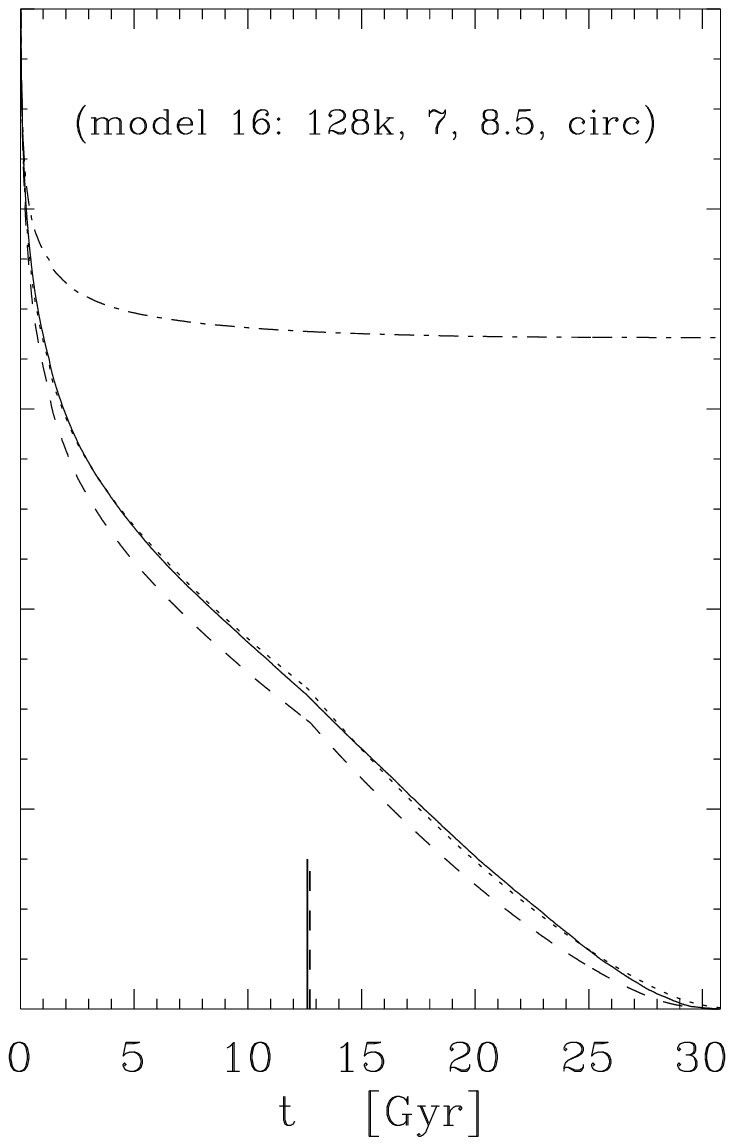,width=7.5cm}\hspace{-4.4cm}
            \epsfig{figure=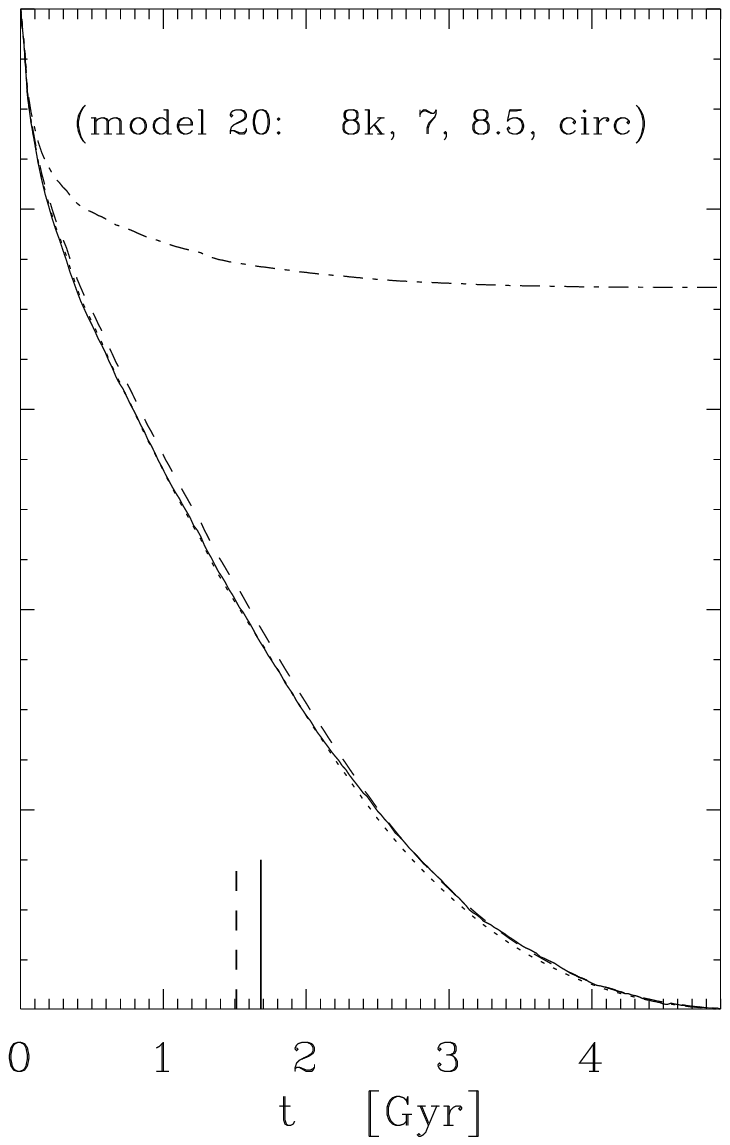,width=7.5cm}\hspace{-4.4cm}
            \epsfig{figure=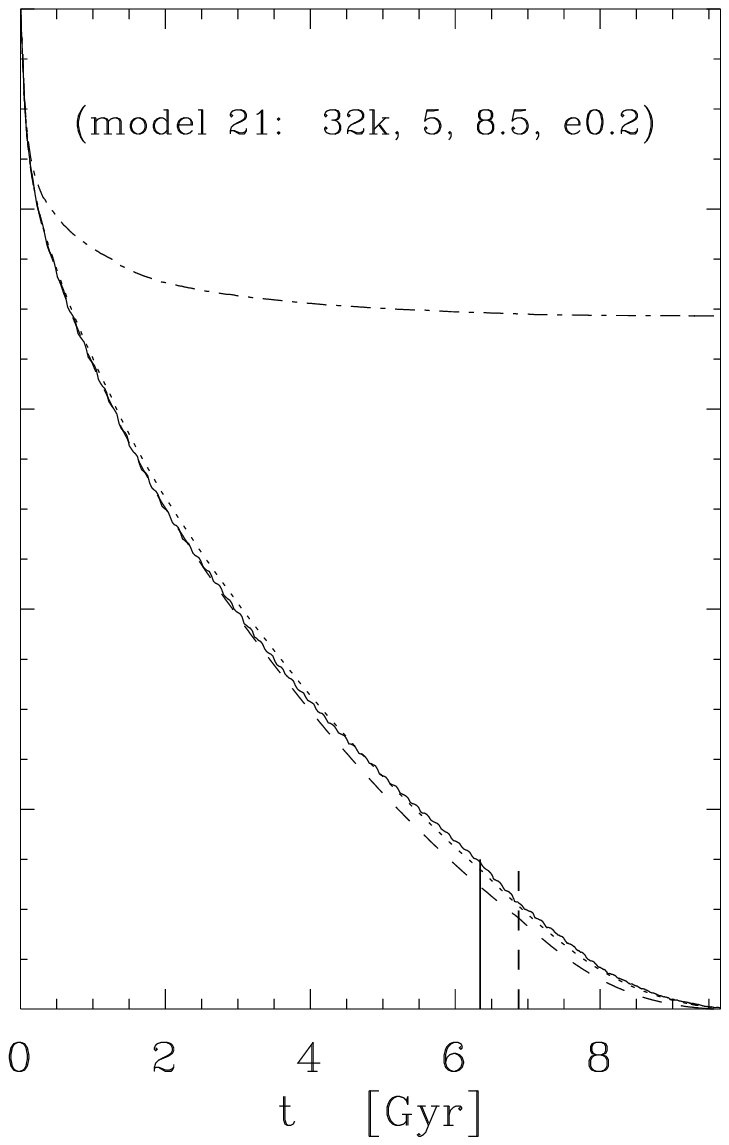,width=7.5cm}\hspace{-4.4cm}
            \epsfig{figure=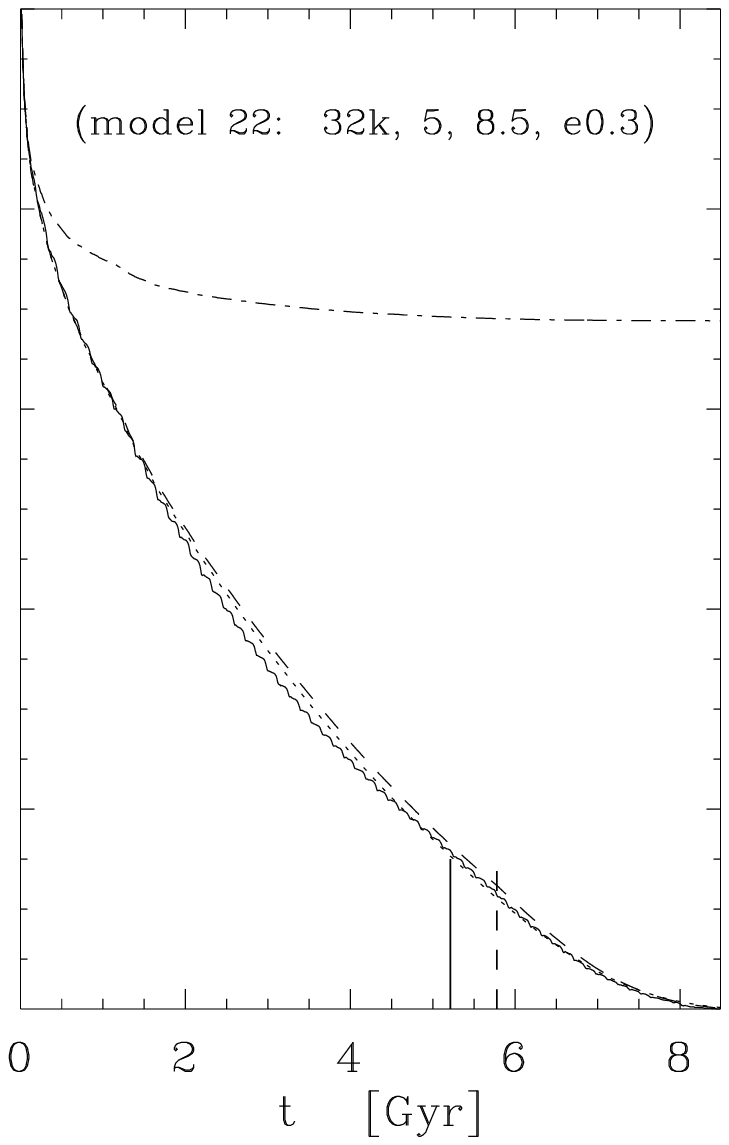,width=7.5cm}\hspace{-4.4cm}}
\vspace{-0.3cm}
\centerline{\epsfig{figure=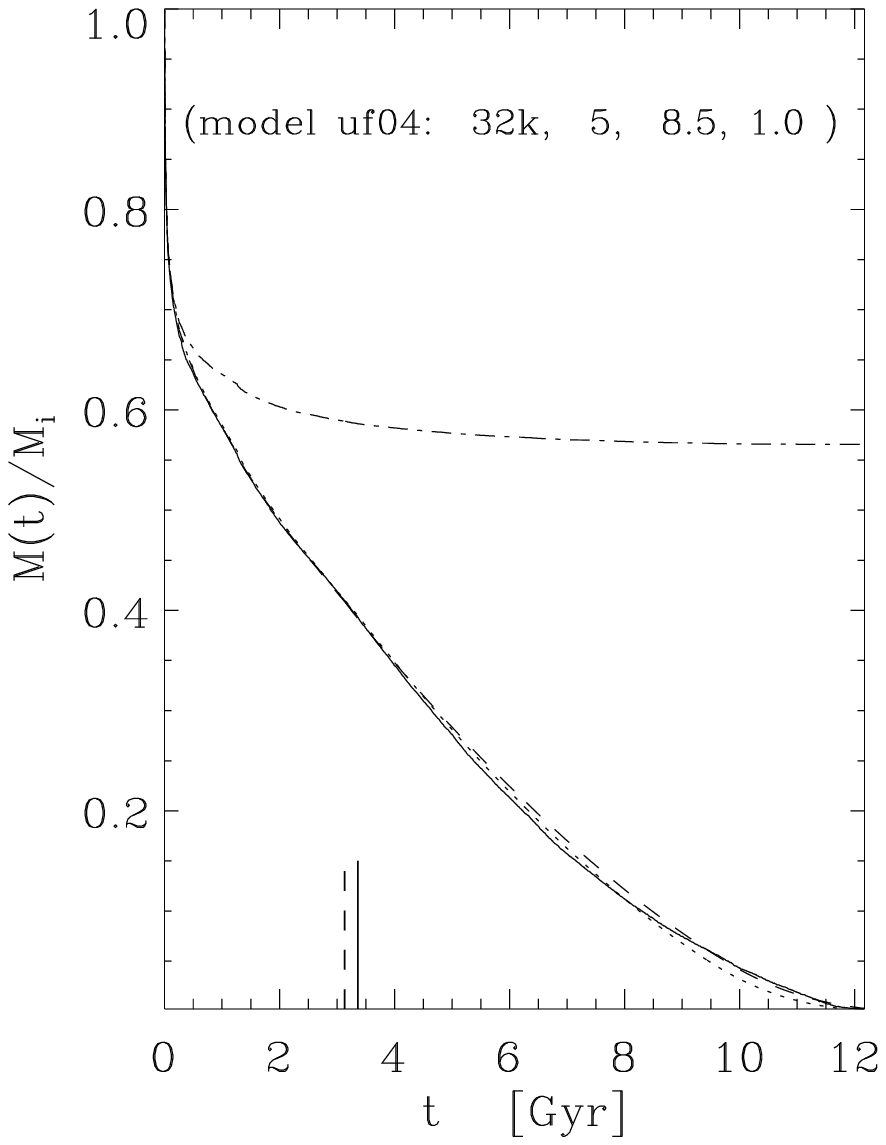,width=7.5cm}\hspace{-4.4cm}
            \epsfig{figure=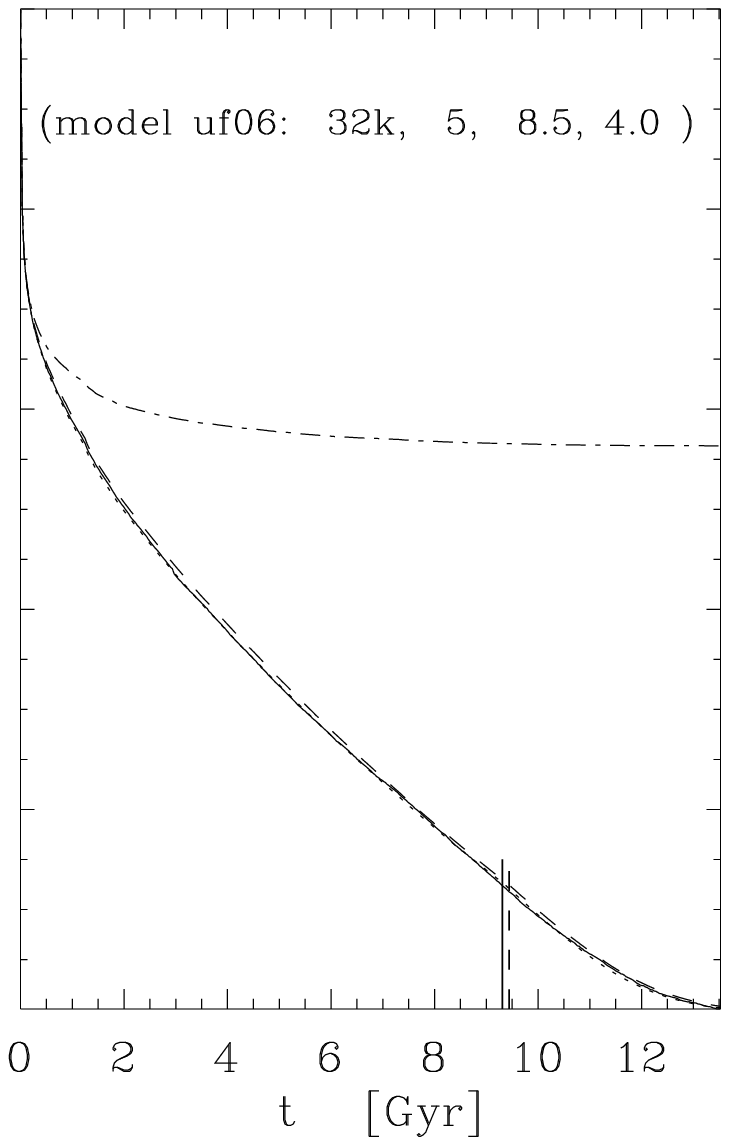,width=7.5cm}\hspace{-4.4cm}
            \epsfig{figure=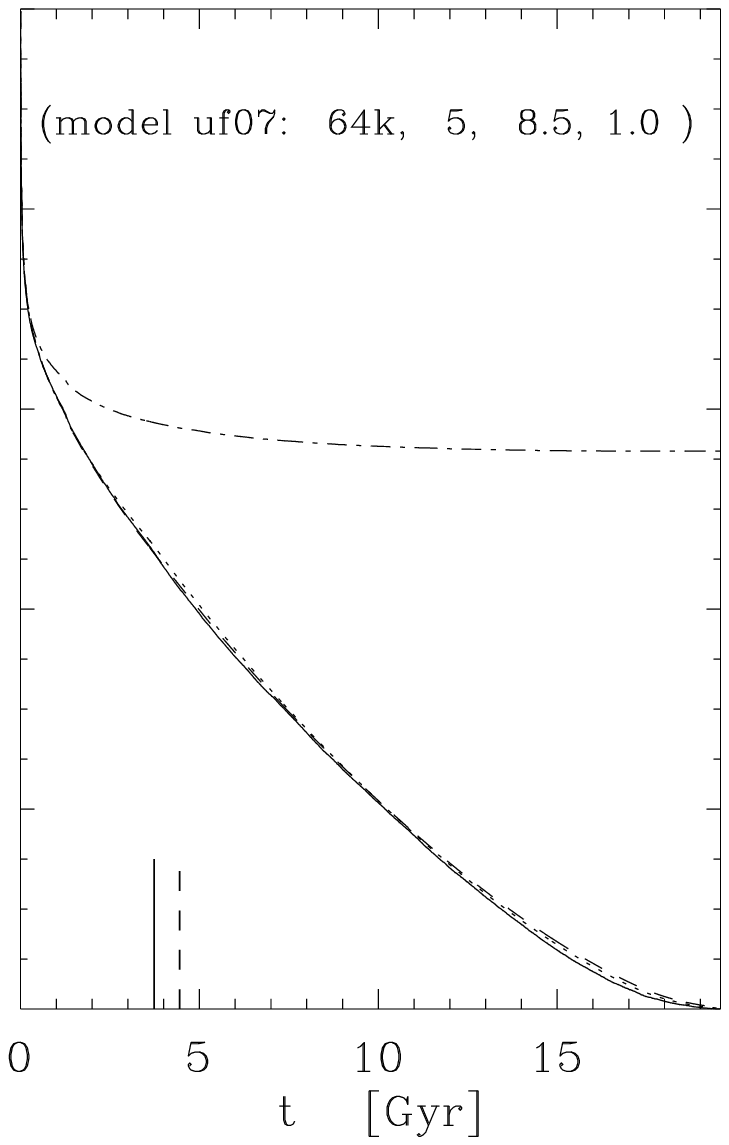,width=7.5cm}\hspace{-4.4cm}
            \epsfig{figure=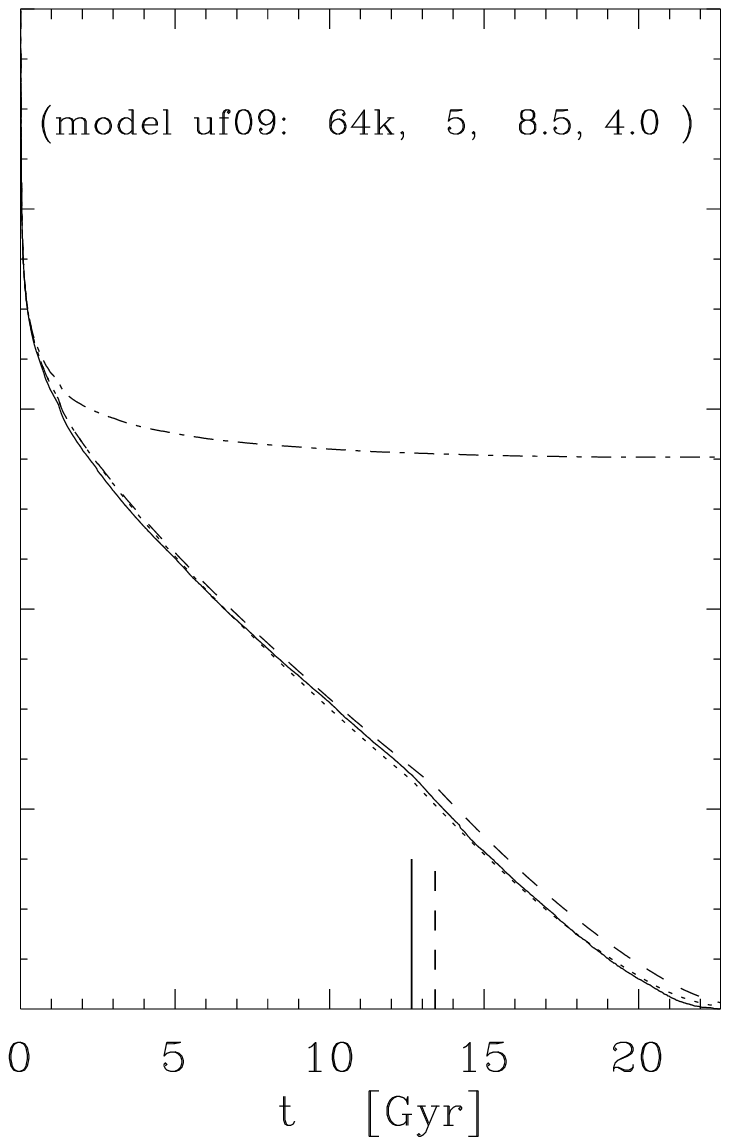,width=7.5cm}\hspace{-4.4cm}
            \epsfig{figure=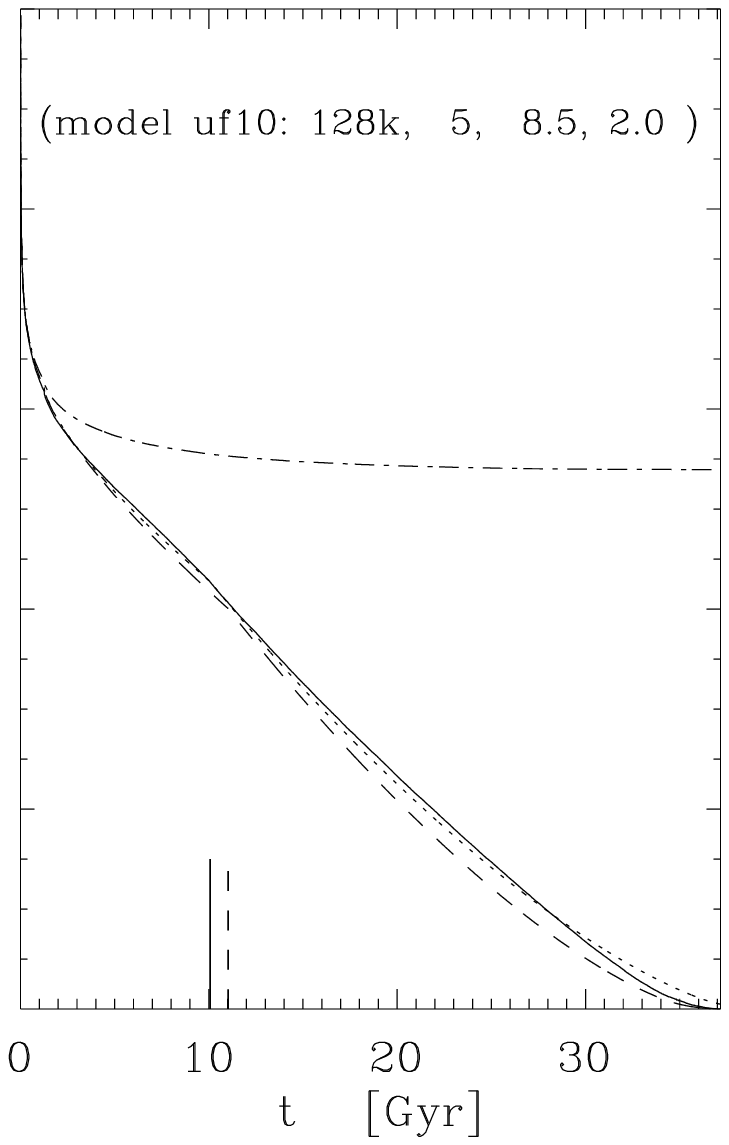,width=7.5cm}\hspace{-4.4cm}}
\caption[]{The $M(t)$ history of a representative set of cluster models:
6 models in circular orbits with $W_0=5$, two with $W_0=7$, two in elliptical orbits,
and 5 Roche-lobe underfilling models.
The models of the Roche-lobe filling clusters are specified by a vector containing: 
model nr, number of stars, $W_0$, \Rgal\ in kpc and orbit.
The models of the Roche-lobe underfilling clusters are specified by: model nr, number of stars, 
$W_0$, \Rgal\ in kpc and $\rh$ in pc.
Full lines: derived from \nbody\ simulations of BM03 or from the sample of Roche-lobe underfilling
models presented here. Dotted line: predicted with the
parameters listed in Tab. \ref{tbl:BM03models}. Dashed lines: predicted
by the method described in Appendix A. The upper dash-dotted line shows the fraction
of the mass that is lost directly by stellar evolution.
The vertical tickmarks indicate
the time of core collapse from the models (full) and calculated with Eqs. \ref{eq:tcctidal} and \ref{eq:tcc-uf} (dashed).}
\label{fig:M(t)-comparison}
\end{figure*} 


\section{Discussion}
\label{sec:discussion}

We have shown how the different mass loss effects of star clusters interact in the 
determination of their mass history. 
We have also derived a recipe for calculating the mass loss history of star clusters
in different environments and with different metallicities and stellar initial mass functions.
 This study is based on the \nbody\ simulations by BM03, supplemented with newer $N$-body
simulations of Roche-lobe underfilling clusters, so the results are 
dependent on the characteristics of these models. 
The \Nbody\ models that we used are relatively simple: the clusters start in virial equiliblium,
without primorial mass segregation, without primordial binaries and with stars in isotropic orbits.
The study of these simple models, which are valid after the gas expulsion phase, 
are a first step to understand the complicated interplay between the 
various dynamical effects in clusters. The models can be refined later
when observational evidence shows which assumptions have to be improved.

We discuss the major assumptions
of the models, how they may have influenced our results, and how they can be
taken into account in the recipe for computing $M(t)$.\\

(a) The BM03 models are Roche-lobe filling. This is a good assumption for open clusters
  and globular clusters which are close to the galactic centre or have very large half-mass radii. 
   However, the majority of globular clusters probably formed with half-mass radii around
   1 pc and therefore started strongly 
   Roche-lobe underfilling (Baumgardt, Kroupa \& Parmentier 2008b). 
   Some of these are still underfilling at present (Baumgardt et al. 2010) 

(b) The models that we used do not have initial mass segregation.
The question of the initial mass segregation is still open. Baumgardt et al. (2008a) found 
that the present overall mass function of most globular clusters can be explained without invoking 
initial mass segregation, but some clusters require initial mass segregation to explain
their present mass function. Observations of young clusters, age $<$50 Myr, show evidence for
mass segregation, e.g. Brandl et al. (1996) for R136; Hillenbrand and Hartmann (1998) for the Orion Nebula Cluster;
McGrady et al. (2005) for M82-F. However,
de Grijs et al. (2002) argued that this does not necesaarily imply {\it initial} mass segregation
because the timescale for the dynamical segregation of high mass stars in young massive clusters 
may be very short.
Dynamical mass segregation has been taken into account in the models we used in this study.

(c) The models have no primordial binaries, but only dynamically formed binaries.
It is expected that real clusters contain a large fraction of initial binaries
(Elson et al. 1998, Hut et al. 1992, Hu et al. 2006, Sommariva et al. 2009). 
However only hard binaries influence the cluster
dynamics as they can heat the cluster and prevent core collapse (Hut et al. 1992).
K\"upper et al (2008) have shown that the escape rate is hardly affected
by binaries. 

(d) Stellar remnants are initially retained in the BM03 cluster models, i.e. they are not 
ejected by a kick-velocity. The new Roche-lobe underfilling models have a 10\% retention factor of 
black holes and neutron stars.
The retention of the remnants implies that
the model clusters at older ages contain a large fraction of neutron stars which are more
massive than the average stellar mass. Clusters with a large fraction of massive remnants
will dissolve faster due to the higher average stellar mass. Also the depletion rate of low 
mass stars could be different (Kruijssen 2009).

(e) The effects of bulge shocks are included in the models of clusters in
eccentric orbits. However,
disk-shocking is not included in our models. Vespirini and Heggie (1997) have studied
the effect of disk-shocking. Based on their results (see their Fig. 21) we conclude that 
the effect of disk-shocks on decreasing the lifetime of clusters is small, especially for
clusters beyond the solar circle and for massive clusters.
Our models also do not include the effects of shocks by spiral density waves or passing GMCs. 
The latter effect is thought to be the main destruction mechanism
for clusters in the Galactic plane (e.g. Lamers \& Gieles 2006)
and probably also in GMC-rich interacting galaxies (Gieles et al. 2008).
These effects will increase the dissolution compared to that of the 
BM03 models. However they can easily be accounted for in the recipe that we derived for 
calculating $M(t)$, by simply adopting a smaller value of the dissolution  
parameter \tzero. This correction is justified because shocks will remove stars
from the outer regions of the cluster in approximately the same way as the tidal
field. 

(f) The models are calculated for a given stellar initial mass function, (a Kroupa
mass function of $0.15<m<15$ \Msun\ for the BM03 models and a Kroupa mass function 
of $0.10 < m < 100$\Msun\ for the new Roche-lobe underfilling models)
and for a given metallicity of $Z=0.001$. The mass loss by stellar evolution
depends on these assumptions. However, the recipe that we derived allows the
choice of different metallicities and different IMFs, by applying the approximate
formulae that describe the mass loss by stellar evolution and the formation of
remnants for a grid of metallicities listed in Appendix B.

(g) We assumed that the cluster move in a spherical logarithmic potential with a 
constant rotation speed.
This implies that we may have underestimated the effect of disk-shocking, which 
is important for clusters in disk galaxies. 
A study of the effects of a non-spherical halo and 
the resulting non-circular orbits with disk-shocking has to be postponed to future studies. 
(The use of GPUs
for the computations of cluster dynamics will allow a significant expansion of the
parameter space of cluster models.)

\section{Summary and Conclusions}
\label{sec:conclusions}

Based on \nbody\ simulations by BM03 of the evolution of Roche-lobe filling star clusters 
of different initial concentrations and in different orbits in the 
Galaxy, and on a new sample of Roche-lobe underfilling clusters in circular orbits, 
we have studied the interplay between the different mass loss effects:
mass loss by stellar evolution, loss of stars induced by stellar evolution,
and dynamical mass loss (referred to as ``dissolution'') before and after core collapse.

At young ages stellar evolution is the dominant effect. The fast (adiabatic) evolutionary 
mass loss results in a simultaneous expansion of the cluster and a shrinking of its tidal (Jacobi) 
radius. So the outer cluster layers become unbound. 
This {\it evolution-induced mass loss} contributes  to the overall mass loss
if the cluster is deeply emersed in the tidal field, i.e. if the cluster is initially
filling its tidal radius. 
The evolution-induced mass loss rate is proportional to the mass loss rate by stellar evolution
but it is smaller for clusters 
with a mass loss rate by dissolution larger than the mass loss rate by stellar evolution.
This is for instance the case for low mass clusters or for clusters in orbits close to the
Galactic center.
The \nbody\ models
show that the induced mass loss does not start immediately, but that it needs time to build up. 
This build-up can be described by an exponential function (Eqs. \ref{eq:find} and 
\ref{eq:fdelay}) with a delay time scale of
a few, typically 3, times the crossing time at the tidal radius.
 For Roche-lobe underfilling clusters the delay time scale is much longer, 
of the order of a few half mass relaxation times, because the cluster first has to
expand to the tidal radius. The actual value depends on the initial Roche-lobe underfilling factor.
As the evolution-induced mass loss rate needs time to get going, 
the {\it total amount of evolution-induced 
mass loss} is considerably smaller, typically 10 to 50\%\ of the total amount of mass 
lost by stellar evolution  (see Fig. \ref{fig:contributions}).

The mass loss of the cluster models by dissolution needs time to build-up, just like the
evolution-induced mass loss. However, this is a consequence of the initial conditions of
the cluster model and the start of the dissolution might be very different in 
real clusters (Sect. 5.3).

We have shown from both theory and the model simulations that the dissolution 
rate depends on the environment of the clusters 
and can be described accurately by a formula of the type $\dmdtdis = -M(t)^{1-\gamma}/\tzero$,
with $M$ and $t$ in units of solar mass and Myr.
The value of the dissolution parameter $\tzero$ depends on the environment, 
e.g. the Galactic potential, the orbit and shocks by spiral arms and passing GMCs.

We have derived expressions for
estimating $t_0$ for clusters in galaxies where tidal
evaporation is the main dissolution effect. This value depends on the Galactic potential
(i.e. the galactic rotation velocity), the orbit of the cluster, the initial concentration 
characterized by $W_0$, and on the evolution of the mean stellar mass. We have derived 
an expression for the mean stellar mass during the pre-core collapse phase
and the post-core collapse phase for various initial mass functions and metallicities.
For clusters in an environment where shock-heating by encounters with spiral arms or GMCs
are important, the value of \tzero\ can be estimated using the descriptions by Gieles et al. 
(2006, 2007). 

The slope of the stellar mass function 
depends mainly on the remaining mass fraction $\mu=M(t)/\Mi$ of the clusters 
and hardly on the initial parameters, such as  mass, mass function,   
concentration factor and strength of the tidal field (e.g. Vesperini \& Heggie, 1997;
BM03; Trenti et al. 2010).
 This effect can also be seen in 
our results in the evolution of the mean stellar mass of a cluster. The data in 
Figs. \ref{fig:mmean}, \ref{fig:mmeancc-ecc} and \ref{fig:mmeanuf} show a very similar evolution of $\mmean$
as function of $\mu$ in almost all models. The main difference is that clusters with  
short lifetimes have an offset of \mmean\ to higher values. This is because the 
mean stellar mass not only depends on the slope of the mass function, but also on the mass of the
most massive stars that have survived stellar evolution. This upper mass depends on 
the age of the cluster and not on its mass fraction.

The details of the 
\nbody\ simulations have shown that $\gamma = 0.65$ for clusters with an initial
density distribution of a King-profile with $W_0=5$ and $\gamma=0.80$ if $W_0=7$.
The difference in $\gamma$ is due to the fact that the dissolution timescale
depends on both the half-mass relaxation time and the crossing time.
Initially Roche-lobe underfilling clusters quickly expand due to mass loss by stellar evolution
and reach a density distribution of approximately $W_0=7$, so their dissolution 
is also described by $\gamma=0.80$.
These values of $\gamma$ apply to the pre-core collapse phase\footnote{
We
point out that this formula describes the time-dependent mass loss rate per cluster. 
It is different 
from the formula that was derived by BM03 (their Eq. 7) to describe the dependence between the
total lifetime of a  cluster and its initial mass.}.

We note that our Roche-lobe underfilling models have half-mass relaxation times between 40 and 800 Myr.
The central relaxation times are about 10 times shorter, i.e. 4 to 80 Myr, but still longer
than the evolution time of the most massive stars. If the initial radius is smaller than those of our
models, e.g. $\le$ 0.5 pc, the core relaxation time may be shorter than the evolution time
and the cluster concentration might decrease rather than increase due to stellar evolution.  

The \nbody -simulations showed that cluster dissolution does not start right away
but that it also needs time to get going. We find that this build-up can be described by the
same exponential function with the same time scale as the evolution-induced mass loss
(Eq. \ref{eq:fdelay}).

The core collapse time \tcc\ of the models can be expressed in terms of the initial 
half mass relaxation time \trh\ by a simple relation that depends on the 
underfilling factor $\mathfrak{F}$ (Eq. \ref{eq:uffactor}).
For a Roche-lobe filling cluster of $\Mi = 10^5~ \Msun$ in a circular orbit
$\tcc \simeq 5~ \trh (t=0)$.

 The mass loss rate by dissolution increases at core collapse by about a 
factor 2 depending on the model, and has a different mass dependence after core collapse
than before with $\dmdtdis = -M(t)^{1-\gammacc}/ t_0^{\rm cc}$ with $\gammacc=0.70$ for all models.
This is independent of the initial density distribution because this is erased
by the core collapse. 
When the mass of the cluster decreases to $M(t) \le 10^3~ \Msun$
the mass loss dependence changes to  $\gammacc=0.40$. This is due to the 
variation of the Coulomb logarithm in the dependence of relaxation time on the number of
stars $N$ in the cluster. We derived an expression for $t_0^{\rm cc}$, i.e. after core collapse
(Sect. \ref{sec:6.3}).

We have derived simple expressions for the parameters that describe the
evolution-induced mass loss and the dissolution, in terms of the initial cluster
parameters (\Mi, $W_0$) and the orbit (\Rgal\ and eccentricity). We also derived
parameters that describe the mass loss by stellar evolution for different 
stellar IMFs and metallicities.
With these parameters we can describe the different mass loss effects throughout
the lifetime of a cluster. By integrating 
$ (dM/dt)_{\rm tot} = \dmdtev + \dmdtind +\dmdtdis $, starting from the initial mass \Mi,
we can calculate the mass loss histories of clusters.  
For this purpose we describe a simple recipe for calculating $M(t)$ in Appendix A, 
that provides a summary of the equations.  The resulting mass histories are compared with those
derived from the \nbody\ simulations. Some of the characteristic results are shown in Fig.
\ref{fig:M(t)-comparison}. The agreement is very good, within a few percent of the 
initial mass. 
The agreement is equally good for the cluster models of BM03 that were not used in this paper.

The method described here provides a description of the variation of the 
total mass, i.e. stars and remnants, of a cluster with age. To derive the luminous mass,
one has to correct the total mass for the contribution by remnants. 
In the calculations of BM03 the newly formed remnants were retained in the cluster 
(no kick velocity was assumed). In the method that we present here the kick fractions
of black holes, neutron stars and white dwarfs can be specified as free parameters.
In later phases part of the remnants can be lost by 
dissolution. At late ages remnant neutron stars and black holes are the most massive
objects in the cluster. They will sink to the center and are not likely to be 
lost by dynamical effects.

The results of this paper and the methods can be used to predict the mass histories
of  star clusters with different stellar IMFs and different metallicities
in different environments. This can then be used to predict the evolution of the mass function 
of cluster systems.

\section*{Acknowledgments}

We thank Onno Pols for providing us with updated evolutionary
calculations and Diederik Kruijssen for his comments on the manuscript.
Simon Portegies Zwart has given important advice.
HJGLM and HB thank ESO for
Visiting Scientist Fellowship during a few moths in 2008 and 2009 in Garching 
and Santiago, when this study was performed. 
This research was supported by the DFG cluster of excellence Origin
and Structure of the Universe (www.universe-cluster.de).

\bibliographystyle{mn2e}

\begin{thebibliography}{}
\bibitem[\protect\citeauthoryear{}{1999}]{}
 {Aarseth} S.J., 1999, PASP, 111, 1333
\bibitem[\protect\citeauthoryear{}{9999}]{}
 Aarseth S.J., Lecar M., 1975, ARAA, 13, 1
\bibitem[\protect\citeauthoryear{}{1999}]{}
 Ambartsumian V.A. 1938, Ann. Leningrad State Univ., No 22, Vol 4, p.19
\bibitem[\protect\citeauthoryear{}{1999}]{} 
 Anders P.,  Fritze-v. Alvensleben U., 2003, A\&A 401, 1063
\bibitem[\protect\citeauthoryear{}{1999}]{} 
  Baumgardt H., 2001, MNRAS, 325, 132
\bibitem[\protect\citeauthoryear{}{1999}]{}
  Baumgardt H., de Marchi G., Kroupa P., 2008a, ApJ, 685, 247 
\bibitem[\protect\citeauthoryear{}{1999}]{}
  Baumgardt H., Kroupa P., Parmentier G., 2008b, MNRAS, 384, 1231
\bibitem[\protect\citeauthoryear{}{1999}]{}
  Baumgardt H., Makino J., 2003, MNRAS, 340, 227 (BM03)
\bibitem[\protect\citeauthoryear{}{1999}]{}
  Baumgardt H., Parmentier G., Gieles M., Vespirini E., 2010, MNRAS, 401, 1832
\bibitem[\protect\citeauthoryear{}{1999}]{} 
  Boutloukos S.G., Lamers, H.J.G.L.M., 2003, MNRAS, 338, 717  
\bibitem[\protect\citeauthoryear{}{1999}]{}
  Brandl et al. 1996, ApJ, 466, 254
\bibitem[\protect\citeauthoryear{}{1999}]{}
 Bruzual G., Charlot, S., 2003, MNRAS, 344, 1000
\bibitem[\protect\citeauthoryear{}{1999}]{}
 Chandrasekhar S., 1943, ApJ, 98, 54
\bibitem[\protect\citeauthoryear{}{1999}]{}
  Chernoff P., Weinberg M., 1990, ApJ, 351, 121
\bibitem[\protect\citeauthoryear{}{1999}]{}
  De Grijs R., Johnson R.A., Gilmore G.F., Frayn C.M., 2002, MNRAS, 331, 228
\bibitem[\protect\citeauthoryear{}{1999}]{} 
Elson R.A.W., Sigurdsson S., Davies M., Hurley J., Gilmore G., 1998, MNRAS, 857, 862
 \bibitem[\protect\citeauthoryear{}{1999}]{}
 Fioc M., Rocca-Volmerange B., 1997, A\&A, 326, 950
\bibitem[\protect\citeauthoryear{}{1999}]{}
 Fukushige T.,  Heggie, D.C., 2000, MNRAS, 318, 753
\bibitem[\protect\citeauthoryear{}{1999}]{} 
Gieles M., 2009, MNRAS, 394, 2113 
\bibitem[\protect\citeauthoryear{}{1999}]{}
Gieles M., Lamers H.~J.~G.~L.~M., Baumgardt H., 2008, IAU Symp. 246, 171
\bibitem[\protect\citeauthoryear{}{1999}]{}
 Gieles M., Athanassoula L., Portegies Zwart S.F., 2007, MNRAS, 376, 809 
\bibitem[\protect\citeauthoryear{}{1999}]{}
 Gieles M., Portegies Zwart S.F., Baumgardt, H. et al., 2006, MNRAS, 371, 793 
\bibitem[\protect\citeauthoryear{}{1999}]{}
 Gieles M., Baumgardt H., 2008, MNRAS, 389, 28 
\bibitem[\protect\citeauthoryear{}{1999}]{}
 Gieles M., Bastian N., Lamers H.J.G.L.M., Mout J., 2005, A\&A, 441, 949
\bibitem[\protect\citeauthoryear{}{1999}]{}
 Giersz M., Heggie D.C., 1994, MNRAS, 270, 298
\bibitem[\protect\citeauthoryear{}{1999}]{}
 Giersz M., Heggie D.C., 1996, MNRAS, 279, 1037
\bibitem[\protect\citeauthoryear{}{1999}]{}
 Gnedin O.Y., Ostriker J.P., 1997, ApJ, 474, 223
\bibitem[\protect\citeauthoryear{}{1999}]{}
 Hillenbrand L.A., Hartmann L.W, 1998, ApJ, 492, 540  
\bibitem[\protect\citeauthoryear{}{1999}]{}
 Hu Y., Liu Q., Deng L., de Grijs R., 2006, in A. Vazdekis, Peletier R.F., 
 eds, IAU Symp. 241, p.347 
\bibitem[\protect\citeauthoryear{}{1999}]{}
 Hurley J.R., Pols O.R., Tout C.A., 2000, MNRAS, 315, 543
\bibitem[\protect\citeauthoryear{}{1999}]{}
Hut P., McMillan S., Goodman J., et al., 1992, PASP, 104, 981
\bibitem[\protect\citeauthoryear{}{1999}]{}
King I., 1958, AJ, 63, 109
\bibitem[\protect\citeauthoryear{}{1999}]{}
 Kroupa P., 2001, MNRAS, 322, 231 
\bibitem[\protect\citeauthoryear{}{1999}]{}
 Kruijssen J.M.D., Lamers H.J.G.L.M., 2008, A\&A, 490, 151
\bibitem[\protect\citeauthoryear{}{1999}]{}
 Kruijssen J.M.D., 2009, A\&A, 507, 1409
\bibitem[\protect\citeauthoryear{}{1999}]{}
Kruijssen J.M.D., Mieske, S., 2009, A\&A, 500, 785
\bibitem[\protect\citeauthoryear{}{1999}]{}
K\"upper H.W., Kroupa P., Baumgardt, H., 2008, MNRAS, 389, 889
\bibitem[\protect\citeauthoryear{}{1999}]{}
Lamers H.J.G.L.M., Gieles M., Bastian N. et al., 2005,  A\&A, 441, 117
\bibitem[\protect\citeauthoryear{}{1999}]{}
  Lamers H.J.G.L.M., Gieles M., 2006, A\&A Letters, 455, L17
\bibitem[\protect\citeauthoryear{}{1999}]{}
  Larsen S., 2004,  A\&A, 416, 537 
\bibitem[\protect\citeauthoryear{}{1999}]{}
  Lee H.M., 2002, IAU Symp. 207, 584
\bibitem[\protect\citeauthoryear{}{1999}]{}
  Leitherer C., Schaerer D, Goldader J.D., et al., 1999, ApJS, 123, 3
\bibitem[\protect\citeauthoryear{}{1999}]{}
    Makino J., Fukushige T., Koga M., Namura K., 2003, Publ. Astron. Soc. Japan, 55, 1163 
\bibitem[\protect\citeauthoryear{}{1999}]{}
  Maraston C., 2005, MNRAS, 362, 799 
\bibitem[\protect\citeauthoryear{}{1999}]{}
  McGrady N., Graham, J.R., Vacca, W.R., 2005, ApJ, 621, 278 
\bibitem[\protect\citeauthoryear{}{1999}]{}
  Ostriker J.P., Spitzer L.Jr., Chevalier R.A., 1972, ApJ, 176, L51
\bibitem[\protect\citeauthoryear{}{1999}]{}
  Portegies Zwart S.F., Hut P., Makino J., McMillan S.L.W., 1998, A\&A 337, 363 
\bibitem[\protect\citeauthoryear{}{1999}]{}
 Scheepmaker R.A., Haas M.R., Gieles M., et al., 2007, A\&A, 469, 925
\bibitem[\protect\citeauthoryear{}{1999}]{}
 Sommariva V., Piotto G., Rejkuba M., Bedin L.R., Heggie D.C., Milone A., Mathieu R.D., 
  Moretti A., 2009, A\&A, 493, 947
\bibitem[\protect\citeauthoryear{}{1999}]{}
 Spitzer, L. 1940, MNRAS, 100, 396
\bibitem[\protect\citeauthoryear{}{1999}]{}
  Spitzer L.Jr., 1958, ApJ, 127, 544
\bibitem[\protect\citeauthoryear{}{1999}]{} 
  Spitzer L.Jr., 1987, Dynamical evolution of globular
  clusters, Princeton University Press, Princeton, N.J.
\bibitem[\protect\citeauthoryear{}{1999}]{}
  Trenti, M., Vesperini, E. Pasquato, M., 2010, ApJ, 708, 1598
\bibitem[\protect\citeauthoryear{}{1999}]{}
  Vesperini E., Heggie, D.C., 1997, MNRAS 298, 898
\bibitem[\protect\citeauthoryear{}{1999}]{}
 Von Hoerner S., 1960, ZsfAp, 50, 184
\bibitem[\protect\citeauthoryear{}{1999}]{}
  Wielen R., 1985, in {\it Dynamics of star clusters}, IAU Symposium 113, p. 449 
\end{thebibliography}


\appendix

\section[]{recipe for predicting the mass evolution of clusters
in different environments}
\label{sec:AppA}

\subsection{Roche-lobe filling clusters}
\label{sec:tidallylimited}

\vspace{0.5cm}
\noindent{\bf a. Initial conditions}\\
The cluster is defined by its initial mass, \Mi, 
initial half mass radius \rh, initial concentration factor $W_0$, metallicity, $Z$,
and stellar IMF. The environment is defined by {\it only one} parameter, \tzero\ for a single cluster 
or \tnref\ for a cluster ensemble, which 
describes the strength of the dissolution processes due to the environment (see below).
Since the evolution depends on the initial cluster parameters, we first describe these.

The tidal radius, $\rt$, is

\begin{equation}
\rt= \Rgal \times \left(\frac{\Mi}{2 M_{\rm Gal}}\right)^{1/3}~ = ~ \left( \frac{G \Mi}{2 \vgal^2}\right)^{1/3} \Rgal^{2/3}
\label{eq:rt}
\end{equation}
where \Mi\ is the initial cluster mass and $\Rgal$, $\vgal$ and $M_{\rm Gal}$ 
are the Galactic radius, the rotation velocity and the mass within $\Rgal$. The last
equality is only valid for galaxies with a constant rotation velocity.
For clusters in elliptical
orbits $\Rgal$ is the perigalactic distance. 
The half mass radius \rh\ of Roche-lobe filling clusters is $0.187~\rt$ if the initial 
concentration factor of the King density profile is $W_0=5$ and $0.116~\rt$ if $W_0=7$. 

It is important to estimate some of the relevant initial time scales of the cluster.
The  half mass relaxation time, \trh\ is

\begin{equation}
\trh = \frac{0.138}{\sqrt{G_{\rm n}}} \rh^{3/2} 
             \left(\frac{\Ni}{\mmean_{\rm i}}\right)^{1/2} \left(\ln \Lambda \right)^{-1}  
\label{eq:trh}
\end{equation}
with $N_{\rm i}= \Mi/ \mmean_{\rm i}$ and $\Lambda \simeq 0.11 N_{\rm i}$ (Giersz \& Heggie 1994), 
where $\mmean_{\rm i}$
is the initial mean stellar mass. 
The gravitational constant is $G_{\rm n}=0.0044985$ pc$^3$ \Msun$^{-1}$ Myr$^{-2}$.
The crossing times at \rh\ and \rt\ are

\begin{equation}
\tcrh= \frac{ \sqrt{8} \rh^{3/2}}{\sqrt {0.5 G \Mi}}
\label{eq:tcrh}
\end{equation}
and 

\begin{equation}
\tcrt= \frac{ \sqrt{8} \rt^{3/2}} { \sqrt{G \Mi}}
\label{eq:tcrt}
\end{equation}
The core collapse time is typically of the order of about $20$ time-dependent relaxation times.
It can also be expressed in terms of the {\it initial} relaxation time.
The BM03 models show that for Roche-lobe filling clusters

\begin{equation}
\tcc \simeq 16.90~\trh^{0.872}
\label{eq:tcc}
\end{equation}
For clusters in elliptical orbits with $\epsilon \ge 0.2$ this expression underestimates the value
of \tcc\ by 20 \% if the value of \trh\ at perigalacticon is used.

One also needs an initial rough estimate of the total lifetime of the cluster
because dissolution
depends on the number of stars in the cluster $N=M/\mmean$ and $\mmean$ depends on the age.
The lifetime \tone\ can be estimated from \Mi\ and \tzero\ 
using Eq. \ref{eq:tone-tmig} with an accuracy of about 10 percent

\begin{eqnarray}
\tone & \simeq & 3.30 \times \{t_0 \Mi^{0.65}\}^{0.864} ~~~ {\rm if}~ W_0=5    \nonumber \\
      &        & 6.27 \times \{t_0 \Mi^{0.80}\}^{0.778} ~~~ {\rm if}~ W_0=7
\label{eq:tone-app}
\end{eqnarray} 
The values of \tzero\ depend on the mean stellar mass, \mmean, which in turn depends 
on \tmig. So in fact \tone\ depends on $\Mi^{0.61}$ if $W_0=5$ and $\Mi^{0.67}$ if $W_0=7$
(Sect. \ref{sec:9}).

\vspace{0.5cm}
\noindent {\bf b. The total mass loss}\\
To calculate the mass history one has to start with mass \Mi\ and then integrate numerically 
the mass loss rate

\begin{eqnarray}
\dmdteq &=& \dmdteveq + \findmax \times \fdelay(t)  \times  \dmdteveq \nonumber \\
      & &  + \fdelay(t)\times \dmdtdiseq 
\label{eq:dmdttotfinal}
\end{eqnarray}
where the first term is the direct mass loss by stellar evolution, the second term is the
loss of stars induced by stellar evolution,
and the last term is the dynamical mass loss (dissolution) due to the tidal field, 
corrected for the delay
in getting started.

\vspace{0.5cm}
\noindent {\bf c. The evolutionary mass loss}\\
The evolutionary mass loss can be calculated by means of 

\begin{equation}
\dmdteveq = M_{\rm lum}(t)~ \dmuevdteq
\label{eq:dmdtevfinal}
\end{equation}
with

\begin{equation}
\mu_{\rm ev} (t) = a_0+a_1 x + a_2 x^2 + a_3 x^3
\label{eq:qevfinal}
\end{equation}
with $x=\log (t/{\rm Myr})$. The values of the coefficients are given in Appendix B  (Table
\ref{tbl:evolfits}) for a range of metallicities and stellar IMFs\footnote{
By applying Eq. \ref{eq:qevfinal} we ignore the change in the stellar mass function
due to the preferential loss of low mass stars. This is allowed because stellar evolution
dominates the cluster mass loss at early ages before complete mass segregation is established
at $t \simeq 0.15 \tone$}. 
 The luminous mass $M_{\rm lum}$
that appears in Eq. \ref{eq:dmdtevfinal} is the total mass $M(t)$ minus the mass in remnants
$M_{\rm remn}$.
The mass in remnants can be derived from the data in Appendix B by integrating $M_{\rm remn}$
from 0 at $t=0$, using  

\begin{eqnarray}
\dmdtremneq &=&   M_{\rm lum}(t) \times [(1-\fkickBH)\frac{{\rm d \muBH}}{{\rm d}t} \nonumber \\
     &+& (1-\fkickNS)\frac{{\rm d \muNS}}{{\rm d}t}  +(1-\fkickWD)\frac{{\rm d \muWD}}{{\rm d}t}]
\label{eq:dmdtremn}
\end{eqnarray}
with ${\rm d}\mu/{\rm d}t >0$ and using the resulting value of $M_{\rm remn}(t)$ 
to find $M_{\rm lum}(t)=M(t)-M_{\rm remn}(t)$.
The parameters for $\mu_{\rm remn}$ are listed in Table \ref{tbl:evolfits} for various metallicities.
The parameter $f_{\rm kick}$ is the kick factor of the remnants. If all neutron stars and 
black holes are kicked out of the cluster then $\fkickBH=\fkickNS=1$. In the recipe 
described here the kick factor can be adopted 
as a free parameter for the calculation of mass history of clusters.
(The BM03 models described above have
$\fkickBH=\fkickNS=\fkickWD=0$, whereas the Roche-lobe underfilling models have $\fkickBH=\fkickNS=0.90$
and $\fkickWD=0$.)

\vspace{0.5cm}
\noindent {\bf d. The evolution-induced mass loss}\\
The delay function for the evolution-induced mass loss is $\fdelay=  1 - \exp (-t/\tdelay)$ 
where the delay time scale is  $\tdelay \simeq 3.0 \times \tcrt$. 

The scaling factor \findmax\ for Roche-lobe filling clusters can be estimated by

\begin{equation}
\findmax = \{-0.86 + 0.40 \times \log(\tmig)\}\times (1-\epsilon)^5
\label{eq:findmaxfinal}
\end{equation}
with a minimum of $\findmax = 0$.

\vspace{0.5cm}
\noindent{\bf e. Dynamical mass loss: dissolution}\\
The dissolution before core collapse can be written as 

\begin{equation}
\left(\frac{{\rm d}M}{{\rm d}t}\right)^{\rm pre-cc} = -\frac{M(t)^{1-\gamma}}{\tzero}
\label{eq:dmdtdispreccfinal}
\end{equation}
with  $\gamma=0.65$ or 0.80 for clusters with $W_0=5$ or 7 respectively. The start of the
dissolution is described by the function $\fdelay (t)$ in Eq. \ref{eq:dmdttotfinal}.

The value of $\tzero$ describes the strength of the 
tidal field and other dissolution processes. So it depends on the environment.\\
\noindent (i) If the main dissolution process is unknown, \tzero\ is a free parameter.\\
(ii) If shocks due to encounters with GMCs or spiral arms are the dominant dissolution
process, then the value of \tzero\ can be derived based on the properties of the
spiral arms and the GMCs (see Gieles et al. 2006 and 2007).\\
(iii) If tidal dissolution in a galaxy with a constant rotation velocity  
is the dominant effect, then \tzero\ can be derived in the following way.

\begin{equation}
\tzero = \tnref \times \mmean^{-\gamma} \left(\frac{\Rgal}{8.5}\right) (1-\epsilon) 
         \left(\frac{\vgal}{220}\right)^{-1}
\label{eq:tzerofinal}
\end{equation}
with $R$ in kpc and $\vgal$ in \kms\ and $\tnref=13.3$ Myr for $W_0=5$ models and               
 $\tnref=3.5$ Myr for $W_0=7$ models.
The mean stellar mass in the pre-core collapse phase can be expressed as

\begin{equation}
\mmean = a \times (\tzero \Mi^\gamma)^b
\label{eq:mmeantmig}
\end{equation}
with $(a,b,\gamma)=(1.528, -0.121, 0.65)$ for $W_0=5$ models with a Kroupa IMF between
0.10 and 15 \Msun, and $(1.230, -0.094, 0.80)$ for $W_0=7$ models.
(Sect. \ref{sec:5.2}). 
Combining Eqs. \ref{eq:tzerofinal} and \ref{eq:mmeantmig}, yields an explicit
expression for \tzero

\begin{eqnarray}
\tzero^{\rm pred1}& =& \left[ \tnref \times \left( \frac{\Rgal}{8.5}\right) (1-\epsilon) 
               \left( \frac{220}{\vgal}\right) \right]^{q} \nonumber \\
               & \times & a^{-\gamma q}~ \Mi^{- \gamma^2 b q}
\label{eq:tzeropred1}
\end{eqnarray}
with $\tzero$ and $\tnref$ in Myr, $\Mi$ in \Msun\ and  $q= (1+\gamma b)^{-1}$.
For clusters with $\tzero^{\rm pred1} \Mi^{\gamma} < 3$ Gyr the value of $\tzero$ is 
slightly smaller

\begin{equation}
\log (\tzero^{\rm pred2}) = \log (\tzero^{\rm pred1}) + 0.25~[ \log (\tzero^{\rm pred1}\Mi^{\gamma}) - 3.50]
\label{eq:tzeropred2}
\end{equation}
This is because clusters with shorter lifetimes contain high mass stars for a larger fraction 
of their lifetime, resulting in a shorter relaxation and dissolution time.
For clusters in elliptical orbits with $\epsilon \ge 0.2$ the values of \tzero\ are smaller
than predicted by Eq. \ref{eq:tzeropred1}. We found that for all models of clusters in elliptical orbits
calculated by BM03, including the ones not shown in this paper, we can approximate 

\begin{equation}
\log (\tzero^{\rm pred\epsilon}) = \log (\tzero^{\rm pred1}) - 0.195~[ \log (\tzero^{\rm pred1}\Mi^{\gamma}) - 3.60]
\label{eq:tzeropredeps}
\end{equation}
with $\tzero^{\rm pred\epsilon}$ slightly larger or smaller than  $\tzero^{\rm pred1}$, depending on the value of 
$\tzero^{\rm pred1}\Mi^{\gamma}$.

The mass loss rate after core collapse can be expressed by a similar expression
as before core collapse, except that it is a broken power law 

\begin{eqnarray}
\left(\frac{{\rm d}M}{{\rm d}t}\right)^{\rm post-cc} &=& - \frac{M(t)^{1-\gammacc}}{\tzero^{\rm cc}} ~{\rm if}~M>10^3 \nonumber \\
                        & & - \frac{M(t)^{1-\gamma_{cc2}}}{\tzero^{\rm cc2}} ~{\rm if}~M<10^3
\label{eq:dmdtdispostccfinal}
\end{eqnarray}
with $\gammacc =0.70$, $\gamma_{cc2} =0.40$  and with
$\tzero^{\rm cc2}=10^{0.90} ~\tzerocc$ for continuity at $M(t)=10^3$.
The value of \tzerocc\ depends on the environment.

(a) For clusters with tidal dissolution as the dominant effect
the value of $\tzeropostcc$ depends on the mean stellar mass at core collapse,
for the same reason as $\tzero$ before core collapse, but in this case

\begin{equation}
\tzeropostcc=~t_{\rm ref}^{cc} ~ \left( \frac{\mmeancc}{\Msun} \right)^{-\gammacc}~\left( \frac{\Rgal}{8.5} \right) (1-\epsilon) 
           \left( \frac{220}{\vgal}\right)
\label{eq:t0postcc}
\end{equation}
with $\gammacc=0.70$ and $t_{\rm ref}^{cc}=7.2$ Myr if $W_0=5$ and $t_{\rm ref}^{cc}=6.2$ Myr if $W_0=7$ and

\begin{equation}
\mmeancc = a~ (\tmig)^b
\label{eq:mmeancc}
\end{equation}
with $(a, b)= (1.585, -0.0984)$ for $W_0=5$ models,
and $(0.893, -0.0691)$ for $W_0=7$ models. Clusters in elliptical orbits have a smaller
mean stellar mass

\begin{equation}
\mmeancc = \mmeancc (\epsilon=0) \times (1-\epsilon)^{0.18}
\label{eq:mmeancc-exc}
\end{equation}
because of their shorter lifetime.
Substitution of Eq. \ref{eq:mmeancc} or \ref{eq:mmeancc-exc} into \ref{eq:t0postcc} yields the value of 
\tzeropostcc.

(b) For clusters in an environment where external effects (other than the tidal field)
dominate, the value of \tzerocc\ can be derived from \tzero\ by applying the same jump in mass loss
at \tcc\ as for dissolving clusters. This implies

\begin{equation}
\frac{\tzerocc}{\tzero} = \frac{t_{\rm ref}^{\rm cc}}{t_{\rm ref}^{\rm N}} ~\frac{\mmeancc^{-\gammacc}}{\mmean^{-\gamma}} 
\label{eq:tzerocc2}
\end{equation}

We now have a full set of equations that describe the mass loss rates due to stellar evolution,
evolution-induced mass loss and dissolution before and after core collapse
of Roche-lobe filling clusters.
With these sets of equations the mass evolution of Roche-lobe filling clusters can be calculated
by means of numerical integration of Eq. \ref{eq:dmdttotfinal}.

\subsection{Roche-Lobe underfilling clusters}

The recipe for calculating the mass history of clusters that are initially Roche-lobe
underfilling proceeds along the same lines as that for the Roche-lobe filling clusters
with a few modifications. The cluster is defined by the same initial parameters
as in Sect. \ref{sec:tidallylimited} plus the underfilling factor $\mathfrak{F}_{\rm W_0}$
defined by Eq. \ref{eq:uffactor}, with $\mathfrak{F}=1$ for Roche-lobe filling clusters.  
Clusters with an initial density concentration of $W_0 \ne 7$
and $\mathfrak{F} < 1 $ quickly expand and reach a density distribution close to $W_0<7$.
For that reason $\mathfrak{F}_7$ is also needed to describe the mass history.

The values of the initial parameters and time scales are the same as described in Sect. 
\ref{sec:tidallylimited}. The calculation of the 
mass loss rate also proceeds along the same lines.
The scaling factor \findmax\ for the induced mass loss is

\begin{equation}
\findmax =  -0.86+ 0.40 \times \log (\tmig)+2.75 ~ \log \mathfrak{F}_5 
\label{eq:findmax-uf-App}
\end{equation}
The delay time of clusters with $-1.0 < \log \mathfrak{F}_5 <-0.20 $ can be estimated by 

\begin{equation}
\tdelay = 4.31~10^{-3}\times  (\mathfrak{F}_5)^{-1.989} ~ \trh^{1.605}
\label{eq:tdelay-uf-App}
\end{equation}
For $\log \mathfrak{F}_5 >-0.20$ the delay time is about the same as for Roche-lobe filling clusters,
$\tdelay = 3.0 \tcrt$.

The dissolution is described by Eq. \ref{eq:dmdtdispreccfinal} but in this case
$\gamma=0.80$ even if the cluster started with $W_0=5$. 
For clusters in an environment where tidal stripping is not the dominant mass
loss mechanism, \tzero\ is a free parameter.
For clusters with tidal stripping in a galaxy with a constant rotation velocity
the dissolution parameter is given by Eq. \ref{eq:tzerofinal} 
with $\tnrefuf=3.50$ Myr  if $\log (\mathfrak{F}_7)> -0.50$ and

\begin{equation}
\tnrefuf ~=~ 7.395 \times (\mathfrak{F}_7)^{0.65} ~~{\rm if} ~ \log (\mathfrak{F}_7)<-0.50
\label{eq:tnrefuf}
\end{equation}

The mean stellar mass in the pre-core collapse phase is given by 

\begin{equation}
\mmean_{\rm pre-cc}=a \times (\tmig)^b \times \mathfrak{F}_7^c
\label{eq:mmeanprecc-uf-App}
\end{equation}
with (a, b, c, $\gamma$)= (1.38, -0.0984,+0.101, 0.80) for a Kroupa mass function 
of $m_{\rm min}=0.10$ and $m_{\rm max}=100$ \Msun. For other values of $m_{\rm max}$
the parameters will be about the same because the most massive stars are lost before
dynamical effects are important.
Combining Eqs. \ref{eq:tzerofinal} and \ref{eq:mmeanprecc-uf-App} gives an 
explicit expression for \tzero

\begin{eqnarray}
\tzero & =& \left[ \tnrefuf \times \left( \frac{\Rgal}{8.5}\right) (1-\epsilon) 
               \left( \frac{220}{\vgal}\right) \right]^{q} \nonumber \\
               & \times & a^{-\gamma q}~ \Mi^{- \gamma^2 b q} ~\mathfrak{F}_7^{-\gamma cq}
\label{eq:t0uf}
\end{eqnarray}
with $q=1/(1+\gamma b)$.

The core collapse time is given by

\begin{equation}
\tcc = 32.0 \times \trh^{0.872} \times \mathfrak{F}_5^{-0.513}
\label{eq:tccuf-App}
\end{equation}
The mass loss rate after core collapse is given by Eq. \ref{eq:dmdtdispostccfinal}.
The value of \tzerocc\ can be derived from $\tzerocc = \tnrefcc \times \mmeancc^{-\gammacc}$
with $\mmeancc = a_{\rm cc} ~(\tmig)^{b_{\rm cc}}~\mathfrak{F}_7^{c_{\rm cc}}$ 
with $(a_{\rm cc}, b_{\rm cc}, c_{\rm cc}) = (1.507, -0.0984, 0.207)$ (see Eq. \ref{eq:mmeancc-uf})
and 

\begin{equation}
\tnrefcc = 7.50 \times \mathfrak{F}_5^{0.127}
\label{eq:tnrefccuff-App}
\end{equation}
The value of $\gammacc=0.70$ if $M>10^3$ and $\gammacctwo=0.40$
if $M(t) < 10^3 \Msun$. This change is due to the variation of the Coulomb
logarithm for small numbers of stars (see Sect. \ref{sec:6.2}).
The dissolution parameter changes from $\tzerocc$ to $\tzero^{\rm cc2}=10^{0.90} \tzerocc$
if $M(t)<10^3 \Msun$ (see Eq. \ref{eq:dmdtdispostccfinal}).

With these equations the mass history of initially Roche-lobe underfilling clusters
can be calculated.



\section{ Mass loss and remnant production by stellar evolution}
\label{sec:AppB}
 
In this section we describe power law approximations for 
calculating the different contributions to the mass of a cluster due to stellar evolution
as a function of age for non-dissolving clusters.
This is done for metallicities of $Z=0.0004$, 0.001,  0.004, 0.008 and 0.02 for a
Kroupa-IMF from $\mmin < m < \mmax$ with $\mmax = 100~\Msun$ and $\mmin = 0.1~\Msun$. 
The mean initial mass is $<m>_{\rm i}=0.638~\Msun$. 
We adopted the stellar evolution described by Hurley et al. (2000).

%
\begin{table}
\caption[]{
 The maximum  initial mass of a star that has survived at age $t$ is given by the 
 polynomial:\\
  $\log (m_{\rm max})= a_0 + a_1 y + a_2 y^2 + a_3 y^3 + a_5 y^5$ with $y=\log(t/{\rm Gyr})$.
 }
\centering
\begin{tabular}{l|lllll}
  \hline 
  $Z$ &   $a_0$ & $a_1$ & $a_2$ & $a_3$  & $a_5$ \\ \hline \hline
  0.0004 & 0.2732 & -0.3864 & 0.05628 & 0.01524 & -0.005902 \\ 
  0.0010 & 0.2940 & -0.3892 & 0.05075 & 0.02055 & -0.006962 \\
  0.0040 & 0.3078 & -0.3910 & 0.04709 & 0.02407 & -0.007662 \\
  0.0080 & 0.3273 & -0.3867 & 0.04203 & 0.02740 & -0.008361 \\
  0.0200 & 0.3463 & -0.3789 & 0.03389 & 0.03290 & -0.009552 \\
  0.0500 & 0.3307 & -0.3653 & 0.02304 & 0.03663 & -0.010346 \\
\hline
\end{tabular}
\label{tbl:mmax}
\end{table}

Table \ref{tbl:mmax} gives the relation between age and the mass of the star that ends its life
at that age.
Table \ref{tbl:evolfits} gives the parameters for calculating the variation of several cluster 
quantities by means of polynomial fits of the type $y=a_0+a_1x+a_2x^2+a_3x^3$ with $x=\log(t/{\rm Myr})$.
These quantities are: $\mu(t)= M(t)/\Mi$ is the remaining mass fraction of the cluster; 
$\mu_{\rm BH}(t)= M_{\rm BH}/\Mi$, $\mu_{\rm NS}(t)$ and $\mu_{\rm WD}(t)$ are the mass fraction 
(relative to the initial cluster mass \Mi) of the black holes, neutron stars and white dwarfs; 
$<m>$, $<m>_{\rm BH}$, $<m>_{\rm NS}$ and $<m>_{\rm WD}$ are respectively the mean 
masses of all stars and of the black holes, the neutron stars and the white dwarfs (in units of \Msun) and 
$\log(L(t)/\Lsun)$ is the luminosity of the cluster, normalized to an initial cluster mass of 1 \Msun.
The powerlaw fit for the luminosity is split in two parts for two age ranges.
The resulting power law fits can be used to calculate the changes in the mass functions
of dissolving clusters due to stellar evolution.

From the data of this table one can also derive the number of black holes, neutron stars and
white dwarfs as a function of time

\begin{equation}
N_{\rm co}(t) = \Mi~ \mu_{\rm co}(t) / <m>_{\rm co}(t)
\label{eq:Nco}
\end{equation}
where ``co'' stands for compact object, i.e. BH, NS, or WD. Similarly, the total number of objects,
stars plus compact objects, is $N=\Mi ~ \mu / <m>$.
The number of luminous (non-compact) stars is $N_{\rm lum}= N - N_{\rm BH}- N_{\rm NS}- N_{\rm WD}$ and the 
mean mass of the luminous stars is

\begin{equation}
<m>_{\rm lum}=\Mi ~ (\mu(t) - \mu_{\rm BH}- \mu_{\rm NS}-\mu_{\rm WD})/N_{\rm lum}
\label{eq:mlum}
\end{equation}

If compact objects are lost from the cluster by their kick velocities, then the values can be adjusted 
by applying kick-factors \fkickBH, \fkickNS\ and \fkickWD\ with $\fkickBH=0.90$ if 90\% of the
black holes are lost from the cluster during their formation.  For instance
$N_{\rm BH}=(1-\fkickBH)~ \Mi~ \mu_{\rm BH}/<m>_{\rm BH}$. 
 
For clusters with a Kroupa IMF with a higher lower mass limit, \mmin,
the parameters can be calculated by subtracting the mass of the missing lower part of the IMF
from the normalization to \Mi. This is valid because the low mass stars with
$m < 0.80 \Msun$ do not contribute to the evolution at $t<50$ Gyr. 
For instance, a cluster of a given initial mass \Mi\ and an IMF in the range of  $0.50 < m < 100~\Msun$ 
has 1.091 times as many stars with $m>0.5\Msun$ as a cluster with the same \Mi\ but with an
IMF in the range of $0.5 < m < 100~\Msun$. So the fractions $\mu$, $\mu_{\rm BH}$, $\mu_{\rm NS}$ and
$\mu_{\rm WD}$ will be higher by  1.091.

Similarly, for cluster with a Kroupa IMF, but with a lower value of \mmax\ the mass fractions $\mu$, $\mu_{\rm BH}$,
$\mu_{\rm NS}$ and $\mu_{\rm WD}$ can be derived as follows.  Calculate first the age of the cluster, $t(\mmax)$, 
at the time  when a star of mass \mmax\ ends its life from Table \ref{tbl:mmax} and the value of $\mu(t(\mmax))$
from Table \ref{tbl:evolfits}. 
At any time $t$ subtract the value of $\mu_{\rm BH}(t(\mmax)$ from $\mu_{\rm BH}(t)$ and normalize it to the new
initial mass by dividing the result by $\mu(t(\mmax))$ to find the new value of $\mu^{\rm new}_{\rm BH}$. 
The same method can be used to find the values of $\mu^{\rm new}$, $\mu^{\rm new}_{\rm NS}$ and $\mu^{\rm new}_{\rm WD}$.
The number of compact stars is found by 
$N_{\rm co}= N_{\rm co}(t)-N_{\rm co}(t(\mmax))$ and the mean mass is $<m>_{\rm co}=\Mi~\mu^{\rm new}_{\rm co}/N_{\rm co}$.

\begin{table*}
\caption[]{
 Parameters for calculating various stellar evolution parameters for a cluster with a Kroupa IMF
in the range of $0.10 < m < 100~\Msun$, as function
of metallicity $Z$ and age $t$ by means of a polynomial:\\
  $y = a_0 + a_1 x + a_2 x^2 + a_3 x^3$ with $x=\log(t/{\rm Myr})$.
 }
\begin{center}
\begin{tabular}{llrrrrrrrr}
  \hline 
 $y$  & $Z$  &   $a_0$ &  $a_1$   &    $a_2$  &  $a_3$  & $\sigma$ & $x_{\rm min}$ & $x_{\rm max}$ &  $y(x_{\rm max})$\\ 
\hline \hline
$\mu$        &  0.0004 &  1.0541 & -0.10912 & -0.01082 & 0.00285 & 0.000 & 1.0000 & 4.6990 &  0.5981 \\
$\mu$        &  0.0010 &  1.0469 & -0.10122 & -0.01349 & 0.00306 & 0.000 & 1.0000 & 4.6990 &  0.5909 \\
$\mu$        &  0.0040 &  1.0247 & -0.08307 & -0.01845 & 0.00336 & 0.000 & 1.0000 & 4.6990 &  0.5756 \\
$\mu$        &  0.0080 &  1.0078 & -0.07456 & -0.02002 & 0.00340 & 0.000 & 1.0000 & 4.6990 &  0.5682 \\
$\mu$        &  0.0200 &  0.9770 & -0.05709 & -0.02338 & 0.00348 & 0.000 & 1.0000 & 4.6990 &  0.5536 \\ 
\hline   
$\mu_{\rm BH}$ &  0.0004 & -1.0711 &  2.66839 & -2.07015 & 0.52230 & 0.000 & 1.0000 & 1.1406 &  0.0543 \\
$\mu_{\rm BH}$ &  0.0010 & -1.1254 &  2.83983 & -2.25098 & 0.58431 & 0.000 & 1.0000 & 1.1141 &  0.0525 \\
$\mu_{\rm BH}$ &  0.0040 & -0.7770 &  2.00007 & -1.60085 & 0.42042 & 0.000 & 1.0000 & 1.1388 &  0.0455 \\
$\mu_{\rm BH}$ &  0.0080 & -1.2000 &  3.26274 & -2.86105 & 0.83466 & 0.000 & 1.0000 & 1.1208 &  0.0380 \\
$\mu_{\rm BH}$ &  0.0200 &  0.0274 &  0.00000 &  0.00000 & 0.00000 & 0.000 & 1.0000 & 1.1000 &  0.0274 \\
\hline 
$\mu_{\rm NS}$ &  0.0004 & -0.1343 &  0.25061 & -0.15593 & 0.03497 & 0.020 & 1.1226 & 1.7885 &  0.0152 \\
$\mu_{\rm NS}$ &  0.0010 & -0.1099 &  0.19575 & -0.11522 & 0.02497 & 0.020 & 1.1236 & 1.7803 &  0.0143 \\
$\mu_{\rm NS}$ &  0.0040 & -0.0849 &  0.14542 & -0.08131 & 0.01723 & 0.000 & 1.1136 & 1.7120 &  0.0122 \\
$\mu_{\rm NS}$ &  0.0080 & -0.0367 &  0.04471 & -0.01078 & 0.00071 & 0.022 & 1.0840 & 1.7113 &  0.0118 \\
$\mu_{\rm NS}$ &  0.0200 & -0.0128 & -0.00396 &  0.02480 &-0.00823 & 0.001 & 1.0091 & 1.6395 &  0.0111 \\
\hline
$\mu_{\rm WD}$ &  0.0004 &  0.0064 & -0.03566 &  0.01882 &-0.00034 & 0.122 & 1.7571 & 4.6990 &  0.2191 \\
$\mu_{\rm WD}$ &  0.0010 &  0.0087 & -0.03363 &  0.01680 &-0.00007 & 0.110 & 1.7120 & 4.6990 &  0.2144 \\
$\mu_{\rm WD}$ &  0.0040 &  0.0014 & -0.01864 &  0.01000 & 0.00056 & 0.056 & 1.6300 & 4.6990 &  0.1927 \\
$\mu_{\rm WD}$ &  0.0080 &  0.0031 & -0.01809 &  0.00916 & 0.00058 & 0.048 & 1.6020 & 4.6990 &  0.1805 \\
$\mu_{\rm WD}$ &  0.0200 & -0.0173 &  0.00709 &  0.00025 & 0.00133 & 0.012 & 1.5928 & 4.6990 &  0.1595 \\
\hline
$<m>$        &  0.0004 &  0.6681 & -0.06910 & -0.00686 & 0.00180 & 0.000 & 1.0000 & 4.6990 &  0.3787 \\
$<m>$        &  0.0010 &  0.6635 & -0.06416 & -0.00855 & 0.00194 & 0.000 & 1.0000 & 4.6990 &  0.3745 \\
$<m>$        &  0.0040 &  0.6495 & -0.05264 & -0.01170 & 0.00213 & 0.000 & 1.0000 & 4.6990 &  0.3648 \\
$<m>$        &  0.0080 &  0.6388 & -0.04727 & -0.01269 & 0.00216 & 0.000 & 1.0000 & 4.6990 &  0.3606 \\
$<m>$        &  0.0200 &  0.6193 & -0.03621 & -0.01481 & 0.00220 & 0.000 & 1.0000 & 4.6990 &  0.3504 \\
\hline
$<m>_{\rm BH}$ &  0.0004 & 33.5859 &-16.14886 &  0.00000 & 0.00000 & 0.000 & 1.0000 & 1.0925 & 15.9433 \\
$<m>_{\rm BH}$ &  0.0010 & 31.5692 &-14.57680 &  0.00000 & 0.00000 & 0.000 & 1.0000 & 1.0928 & 15.6397 \\
$<m>_{\rm BH}$ &  0.0040 & 30.2147 &-15.18139 &  0.00000 & 0.00000 & 0.000 & 1.0000 & 1.0770 & 13.8643 \\
$<m>_{\rm BH}$ &  0.0080 & 23.6200 &-11.10258 &  0.00000 & 0.00000 & 0.000 & 1.0000 & 1.0647 & 11.7992 \\
$<m>_{\rm BH}$ &  0.0200 &  9.3469 &  0.00000 &  0.00000 & 0.00000 & 0.000 & 1.0000 & 1.0500 &  9.3469 \\
\hline
$<m>_{\rm NS}$ &  0.0004 & 16.3368 &-26.6599  & 16.46480 &-3.46414 & 0.000 & 1.1121 & 1.7650 &  1.5265 \\
$<m>_{\rm NS}$ &  0.0010 & 13.4717 &-20.6296  & 12.27190 &-2.50150 & 0.000 & 1.1121 & 1.7640 &  1.5367 \\
$<m>_{\rm NS}$ &  0.0040 &  9.8973 &-13.4125  &  7.34369 &-1.37739 & 0.000 & 1.0747 & 1.7315 &  1.5403 \\
$<m>_{\rm NS}$ &  0.0080 & 14.1365 &-22.1270  & 13.19290 &-2.66813 & 0.000 & 1.0747 & 1.7211 &  1.5308 \\
$<m>_{\rm NS}$ &  0.0200 & 13.4243 &-20.6297  & 12.06450 &-2.37325 & 0.000 & 1.0500 & 1.7020 &  1.5601 \\
\hline
$<m>_{\rm WD}$ &  0.0004 &  2.8955 & -1.21969 &  0.25054 &-0.01995 & 0.001 & 1.7847 & 4.6990 &  0.6263 \\
$<m>_{\rm WD}$ &  0.0010 &  3.0223 & -1.33991 &  0.28235 &-0.02259 & 0.001 & 1.7847 & 4.6990 &  0.6167 \\
$<m>_{\rm WD}$ &  0.0040 &  3.0228 & -1.34601 &  0.27457 &-0.02079 & 0.000 & 1.7473 & 4.6990 &  0.6035 \\
$<m>_{\rm WD}$ &  0.0080 &  3.0213 & -1.38321 &  0.29001 &-0.02239 & 0.000 & 1.7099 & 4.6990 &  0.6021 \\
$<m>_{\rm WD}$ &  0.0200 &  2.8737 & -1.25599 &  0.24619 &-0.01755 & 0.000 & 1.6352 & 4.6990 &  0.5869 \\
\hline
$\log~L/\Lsun$&  0.0004 &  4.3507 & -2.72949 &  0.88598 &-0.14060 & 0.000 & 1.0000 & 2.5272 &  0.8419 \\
$\log~L/\Lsun$&  0.0010 &  4.4074 & -2.78762 &  0.89986 &-0.14251 & 0.000 & 1.0000 & 2.2258 &  1.0893 \\
$\log~L/\Lsun$&  0.0040 &  4.9761 & -3.84505 &  1.49603 &-0.25205 & 0.000 & 1.0000 & 1.8173 &  1.4165 \\
$\log~L/\Lsun$&  0.0080 &  4.5921 & -3.19055 &  1.11463 &-0.18185 & 0.000 & 1.0000 & 2.7372 &  0.4807 \\
$\log~L/\Lsun$&  0.0200 &  4.8125 & -3.62327 &  1.30573 &-0.20546 & 0.000 & 1.0000 & 2.8070 &  0.3860 \\
\hline
$\log~L/\Lsun$&  0.0004 &  3.2921 & -1.20074 &  0.13311 &-0.01647 & 0.000 & 2.5272 & 4.6990 & -1.1199 \\
$\log~L/\Lsun$&  0.0010 &  5.4132 & -3.20696 &  0.73195 &-0.07364 & 0.000 & 2.2258 & 4.6990 & -1.1351 \\
$\log~L/\Lsun$&  0.0040 &  4.1575 & -2.04685 &  0.35860 &-0.03425 & 0.000 & 1.8173 & 4.6990 & -1.0962 \\
$\log~L/\Lsun$&  0.0080 &  2.8690 & -1.35525 &  0.27530 &-0.03615 & 0.000 & 2.7372 & 4.6990 & -1.1713 \\
$\log~L/\Lsun$&  0.0200 & 10.4008 & -7.65806 &  1.98530 &-0.18815 & 0.000 & 2.8070 & 4.6990 & -1.2696 \\
\hline 
\end{tabular}
\end{center}
(a) $\sigma$ is the standard devaiation of the fit in the range of $x_{\rm min} < x < x_{\rm max}$.\\
(b) $x_{\rm min}$ and $x_{\rm max}$ indicate the validity range of the approximation.\\
(c)  All values of $\mu$ are expressed in units of $\Mi$, so $\mu=M(t)/\Mi$, $\mu_{\rm BH}=M_{\rm BH}/\Mi$ etc.\\
(d) The values of $\mu_{\rm BH}$ and $\mu_{\rm NS}$
     and the values of $<m>_{\rm BH}$ and $<m>_{\rm NS}$ are constant for $x>x_{\rm max}$ at $y(x_{\rm max})$.\\
(e) $<m>$ is the mean stellar mass, $<m>_{\rm BH}$ is the mean stellar mass of the black holes, etc, in units of \Msun.\\
(f) The values of $\log (L/\Lsun)$ are for a cluster of $\Mi=1\Msun$. The fit is split in two age ranges. 

\label{tbl:evolfits}
\end{table*}


\label{lastpage}
\end{document}